\documentclass[journal]{new-aiaa}

\usepackage[utf8]{inputenc}
\usepackage{textcomp}
\usepackage{graphicx}
\usepackage{amsmath}
\usepackage[version=4]{mhchem}
\usepackage{siunitx}
\usepackage{longtable,tabularx}
\usepackage{setspace}

\usepackage{hyperref}
\usepackage{xcolor}
\definecolor{darkblue}{rgb}{0,0,0.80}
\hypersetup{
    colorlinks=true,
    linkcolor={darkblue},
    citecolor={darkblue},
    filecolor=black,
    urlcolor={darkblue},
    plainpages=false,
}

\usepackage{subfig}

\usepackage{tabularx}
\newcolumntype{Y}{>{\centering\arraybackslash}X}

\usepackage{color}
\newcommand{\ba}{{\boldsymbol a}}
\newcommand{\bu}{{\boldsymbol u}}
\newcommand{\bx}{{\boldsymbol x}}
\newcommand{\by}{{\boldsymbol y}}
\newcommand{\mA}{{\boldsymbol A}}
\newcommand{\mB}{{\boldsymbol B}}
\newcommand{\mC}{{\boldsymbol C}}
\newcommand{\mD}{{\boldsymbol D}}
\newcommand{\mI}{{\boldsymbol I}}
\DeclareMathOperator*{\argmin}{argmin}
\DeclareMathOperator*{\argmax}{argmax}
\usepackage{floatrow}
\usepackage{overpic}

\makeatletter
\newcommand\thefontsize{The current font size is: \f@size pt}
\makeatother

\makeatletter
\let\caption@@@make@ORI\caption@@@make
\def\caption@@@make{%
  \caption@ifundefined\caption@lfmt{\let\caption@lfmt\caption@labelformat}{}%
  \caption@ifundefined\caption@fmt\caption@format\relax
\caption@@@make@ORI}
\makeatother

\title{A Database for Reduced-Complexity Modeling of Fluid Flows}

\author{Aaron Towne\footnote{Assistant Professor, Department of Mechanical Engineering, AIAA Member}}
\affil{University of Michigan, Ann Arbor, MI, 48109}
\author{Scott T. M. Dawson\footnote{Assistant Professor, Department of Mechanical, Materials, and Aerospace Engineering, AIAA Senior Member}}
\affil{Illinois Institute of Technology, Chicago, IL 60616}
\author{Guillaume A. Br\`es\footnote{Director of Operations and Senior Research Scientist, AIAA Member}}
\affil{Cascade Technologies Inc., Palo Alto, CA 94303}
\author{Adri\'{a}n Lozano-Dur\'{a}n\footnote{Draper Assistant Professor, Department of Aeronautics and Astronautics, AIAA Senior Member.}}
\affil{Massachusetts Institute of Technology, Cambridge, MA, 02139}
\author{Theresa Saxton-Fox\footnote{Assistant Professor, Department of Aerospace Engineering, AIAA Member}}
\author{Aadhy Parthasarathy\footnote{Research Assistant, Department of Aerospace Engineering, AIAA Student Member}}
\affil{University of Illinois Urbana-Champaign, Urbana, IL, 61801}
\author{Anya R.~Jones\footnote{Professor, Department of Aerospace Engineering, AIAA Associate Fellow}}
\author{Hulya Biler\footnote{Research Assistant, Department of Aerospace Engineering}}
\affil{University of Maryland, College Park, MD, 20742}
\author{Chi-An Yeh\footnote{Assistant Professor, Department of Mechanical and Aerospace Engineering, AIAA Member}}
\author{Het D. Patel\footnote{Research Assistant, Department of Mechanical and Aerospace Engineering, AIAA Student Member}}
\affil{North Carolina State University, Raleigh, NC 27695, USA}
\author{Kunihiko Taira\footnote{Professor, Department of Mechanical and Aerospace Engineering, AIAA Associate Fellow}}
\affil{University of California, Los Angeles, CA 90095, USA}

\begin{document}
\maketitle

\setstretch{1.15}

\clearpage
%%%%%%%%%%%%%%%%%%%%%%%%%%%%%%%%%%%%%%%%%%%%%%%%%%%%%%%%%%%
% -- Abstract ---------------------------------------------

\begin{abstract}

We present a publicly accessible database designed to aid in the conception, training, demonstration, evaluation, and comparison of reduced-complexity models for fluid mechanics.  Availability of high-quality flow data is essential for all of these aspects of model development for both data-driven and physics-based methods.  The database contains time-resolved data for six distinct datasets: a large eddy simulation of a turbulent jet, direct numerical simulations of a zero-pressure-gradient turbulent boundary layer, particle-image-velocimetry measurements for the same boundary layer at several Reynolds numbers, direct numerical simulations of laminar stationary and pitching flat-plate airfoils, particle-image-velocimetry and force measurements of an airfoil encountering a gust, and a large eddy simulation of the separated, turbulent flow over an airfoil.  These six cases span several key flow categories: laminar and turbulent, statistically stationary and transient, tonal and broadband spectral content, canonical and application-oriented, wall-bounded and free-shear flow, and simulation and experimental measurements.   For each dataset, we describe the flow setup and computational/experimental methods, catalog the data available in the database, and provide examples of how these data can be used for reduced-complexity modeling.  All data can be downloaded using a browser interface or Globus.   Our vision is that the common testbed provided by this database will aid the fluid mechanics community in clarifying the distinct capabilities of new and existing methods.  

\end{abstract}

%%%%%%%%%%%%%%%%%%%%%%%%%%%%%%%%%%%%%%%%%%%%%%%%%%%%%%%%%%%
% -- Introduction -----------------------------------------
\section{Introduction}

Reduced-complexity models play a critical role in the study of fluid mechanics.  Precise equations of motion, e.g., the Navier-Stokes equations, are known for most flows, but solving these equations is often computationally expensive and knowledge of the equations does not necessarily produce insight into the unique physics active in a particular flow.  Similarly, experimental investigations typically require substantial time and resources, and quantities of interest are not always directly measurable.  Extracting knowledge from data from either source requires appropriate tools beyond the data or governing equations themselves.  

Reduced-complexity models, i.e., methods employing simplified governing equations and/or a reduced set of variables, offer a path toward overcoming these challenges.  They can be used to identify and extract key physics within a flow, accelerate simulations compared to full-order models, capture unresolved physics, predict statistics and other quantities of interest, reconstruct unmeasured quantities in an experiment, and provide a tractable plant for flow control or design optimization \cite{Taira:AIAAJ2017,Rowley2017model}.  Given their depth and breadth of potential applications, it is not surprising that reduced-complexity modeling is of great interest to the broader fluids community and has produced some of the most-cited publications in fluid mechanics in recent years \cite{Rowley:JFM2009DMD,Schmid:JFM2010,Ling2016reynolds,Taira:AIAAJ2017,towne2018spectral,Brunton2020machine}.

Data are essential for the development and deployment of reduced-complexity models.  Often, model forms are motivated by observations from data.  Many models require training data.  This includes fitting a neural network or other machine-learning algorithms \cite{Brunton2020machine}, supplying data for modal decompositions like proper orthogonal decomposition \cite{Berkooz1993proper, towne2018spectral} or dynamic mode decomposition \cite{Schmid:JFM2010,Schmid2022dynamic}, furnishing mean data for linear stability analysis \cite{SchmidHenningson, Theofilis:ARFM2011}, or supplying sensor data for controller design \cite{Manohar:IEEE18, Proctor:SIAMJADS16, Martini2022resolvent}.   Data are also needed to evaluate the accuracy and cost of a model by providing ground truth against which to compare.  Likewise, data are helpful to demonstrate the properties and capabilities of a new method or to compare several candidate methods for a given application. 

The objective of this paper is to introduce and disseminate a collection of flow data designed to aid in all aspects of reduced-complexity model development and deployment.  We envision that this database will provide a common set of test cases to be used by the community for developing and testing methods, alleviating the aforementioned challenges.  Moreover, it will supplement other recent efforts \cite{Taira:AIAAJ2017, Taira:AIAAJ2020} to lower the barrier of entry into the field of reduced-complexity modeling by offering easily accessible data specifically procured for these applications.  

Since different methods will target different types of flows and different users will have interest in different applications, it is necessary to include a diverse collection of datasets to ensure that such a database is widely applicable.  Based on extensive input from the fluids community, we identified six categories (or regimes) that the database is designed to span.  It includes flows that are: (i) laminar \& turbulent; (ii) transient \& statistically stationary; (iii) tonal \& broadband in their spectral content; (iv) canonical \& application-oriented; (v) wall-bounded \& free shear; and probed using both (vi) simulation \& experiment.

Six datasets were identified that, in aggregate, span these categories.  First, the database contains three canonical shear flows: large-eddy-simulation (LES) data for a turbulent jet, direct-numerical-simulation (DNS) data for a zero-pressure-gradient turbulent boundary layer, and particle-image-velocimetry (PIV) data for the same boundary layer at several Reynolds numbers.  Second, the database contains three aerodynamic flows: DNS data for laminar stationary and pitching flat-plate airfoils, PIV and force data for an airfoil encountering a gust, and LES data for the separated, turbulent flow over a fixed airfoil.  These six datasets are summarized in Table~\ref{tab:datasets}.

\setlength{\tabcolsep}{7.0pt}
\begin{table}
\centering
\renewcommand\arraystretch{1.33}
\begin{tabular}{lllcccccc}
\hline
Sec. & {Flow}  & {Abbreviation} & {Method}  & {L / T}  & {T / SS}  & {T / B} & {C / A}  & {WB / FS}       \\ 
\hline
\ref{sec:jet} & Jet                     & jet & LES   & T &  SS &  both &  C & FS  \\ 

\ref{sec:BL_DNS} & Boundary layer          & BLdns  & DNS   & T & SS  & B  & C  &  WB \\ 

\ref{sec:BL_EXP} & Boundary layer          & BLexp  & EXP   & T & SS  & B  & C  & WB  \\ 

\ref{sec:airfoil_DNS} & Pitching airfoil       & airfoilDNS  & DNS   & L & T  & T  & A  & both  \\ 

\ref{sec:airfoil_EXP} & Airfoil gust encounter  & gustexp  & EXP   & L  & T  & T  & A  & both  \\

\ref{sec:airfoil_LES} & Airfoil wake            & airfoilLES & LES   & T  & SS & both  & A  & both    \\
\hline
\vspace{0.00in}
\end{tabular}
\caption{Overview of datasets within the database.  The columns from left to right indicate: section number; flow description; dataset abbreviation used in the file naming convention; method used to produce the data, i.e., direct numerical simulation (DNS), large eddy simulation (LES), or experimental measurements (EXP); laminar (L) vs. turbulent (T) flow state; transient (T) vs. statistically stationary (SS) flow; tonal (T) vs. broadband (B) spectral content; canonical (C) vs. application-oriented (A) setup; and wall-bounded (WB) vs free-shear (FS) flow.  \label{tab:datasets}}
\end{table}

The database is unique in a number of ways, motivated by its intended use for reduced-complexity modeling.  First, as discussed already, it contains a breadth of flow types to ensure its applicability to a wide variety of models and target applications.  Second, in contrast to other prominent fluids databases that exclusively contain simulation data, we have intentionally included experimental data.  This is important, as the unique challenges that come with using experimental data within the context of reduced-complexity modeling, such as the necessity of rejecting noise and avoiding overfitting, have often been overlooked in the development and deployment of reduced-complexity models.  Third, as much as possible, the database provides lengthy time series, which are often required for data-hungry, data-driven methods.  

The database is hosted within the Deep Blue Data repository at the University of Michigan and is publicly accessible via a web interface or using Globus, a fast, secure, and reliable service for research data management and transfer.  Both transfer options can be accessed at \url{ http://deepblue.lib.umich.edu/data/collections/kk91fk98z}.  All data are stored within hdf5 files, a portable and non-proprietary file format designed specifically for storing large datasets.  Each dataset additionally contains a README file and a Matlab script that exemplifies how the data can be read and manipulated.  Since all the datafiles use the hdf5 format, they can alternatively be read within virtually any other programming environment, e.g., Python, C++, etc.  File and variable naming conventions are consistent between datasets to ensure ease of use, and file names are prepended using the abbreviations indicated in Table~\ref{tab:datasets} to indicate the dataset.  The total size of the database is approximately 8 terabytes; this size has been intentionally held as low as possible without compromising the quality of the data to minimize the computational resources required to use the data.  

The remainder of the paper is organized as follows.  For each dataset (section numbers are indicated in Table~\ref{tab:datasets}), we: (A) introduce the flow setup, e.g., the geometry, dimensionless constants, and methods used to collect the data; (B) describe the data available within the database; and (C) sketch one or more examples of how the data can be used for reduced-complexity modeling.  Finally, in \S\ref{sec:conclusions} we conclude the paper with a brief recap of the database and a discussion of our vision for its future applications and extensions.

%%%%%%%%%%%%%%%%%%%%%%%%%%%%%%%%%%%%%%%%%%%%%%%%%%%%%%%%%%%
% -- Jet Section (Guillaume & Aaron) ----------------------

\section{Turbulent jet large eddy simulation}  % e.g., "Turbulent jet large eddy simulation"
\label{sec:jet}

The first flow included in the database is a turbulent jet.  A jet is a canonical example of a free shear flow, i.e., a flow containing mean velocity gradients but unaffected by walls.  Despite the geometrical simplicity of a jet -- consisting of a fast inner stream that mixes with a slow outer stream -- turbulent jets contain a diverse set of physics.  Large-scale coherent structures created by the Kelvin-Helmholtz instability of the annular shear layer, along with their emitted acoustic radiation, have been studied continuously since the 1960s with increasingly sophisticated tools \cite{Jordan2013wave, Cavalieri2019wave}.  More recently, the important role of other physical mechanisms, including the Orr mechanism \cite{Schmidt2018spectral}, streaks generated by the lift-up effect \cite{Nogueira2019large,Pickering2020lift}, and acoustic waves trapped within the core of the jet \cite{Towne2017acoustic}, have been identified and explored.  This diverse, coexisting set of physics makes a turbulent jet an excellent testbed for reduced-complexity models.  The available dataset contains 10,000 time-resolved snapshots of the jet computed via LES and other quantities derived from these data.

% -- Jet setup ------------------------------------------
\subsection{Setup}

This dataset corresponds to an subsonic turbulent jet issued from a contoured convergent-straight nozzle \cite{BresJFM2018}.  The jet  Mach number is $M_j = U_j/c_j = 0.9$ and the jet is isothermal ($T_j/T_\infty=1.0$), where $U$ is the mean (time-averaged) streamwise velocity, $c$ is the speed of sound, $T$ is the mean temperature, and the subscripts $j$ and ${\infty}$ refer to the jet exit and freestream conditions, respectively.  The Reynolds number is $Re_j = \rho_j U_j D/\mu_j \approx 1\times10^6$, where $D$ is the nozzle exit diameter, $\rho$ is density, and $\mu$ is dynamic viscosity.  The nozzle pressure ratio and nozzle temperature ratio are $NPR = P_0/P_\infty = 1.7$ and $NTR = T_0/T_\infty = 1.15$, respectively, where the subscript $0$ refers to the stagnation (total) property.  All of these parameters match companion experiments performed in the anecho\"ic jet-noise facility of the PPRIME Institute at the Centre d'\'Etudes A\'erodynamiques et Thermiques (CEAT), Poitiers, France \cite{BresAIAA2016}. 

\begin{figure}[t]
    \centering
    \includegraphics[]{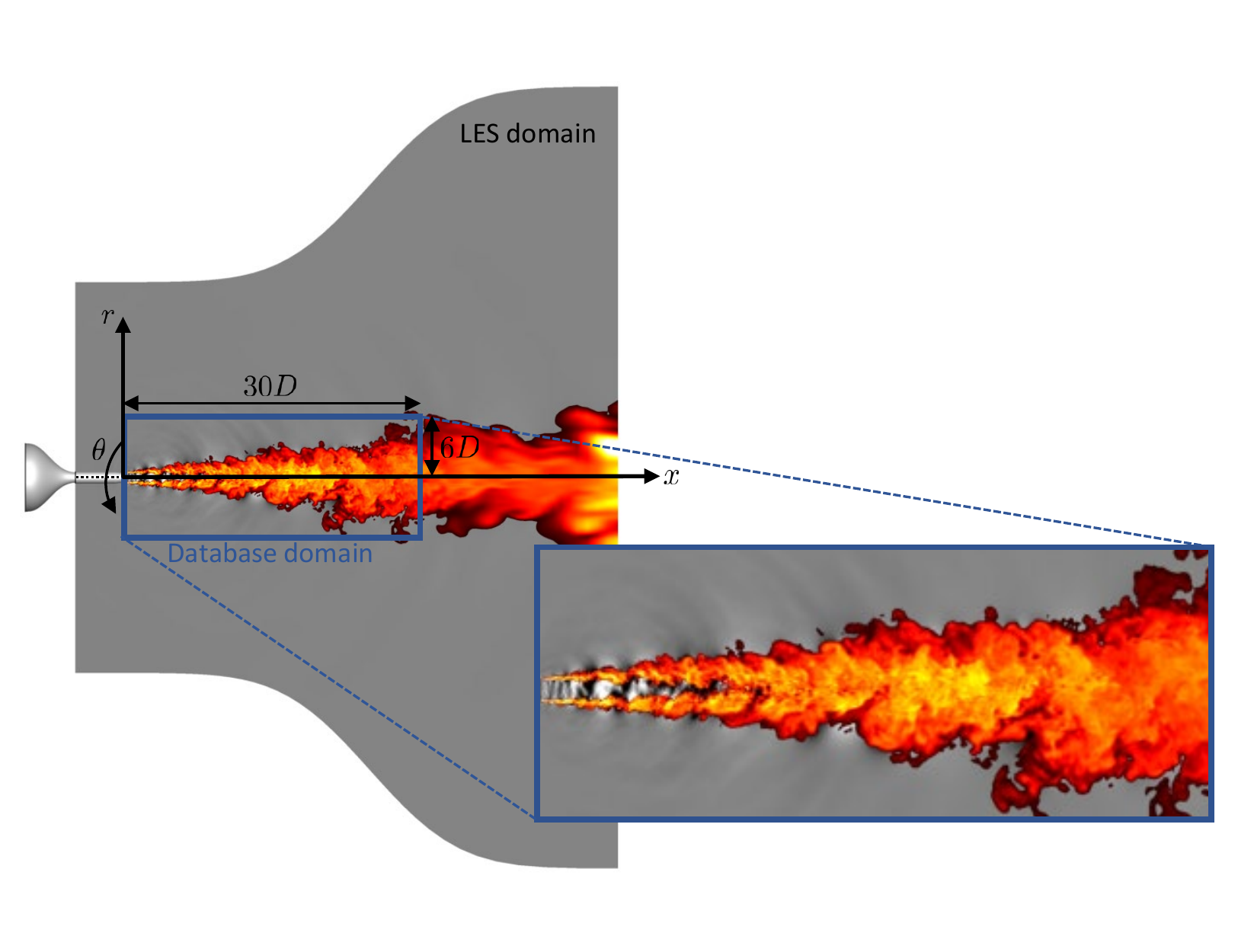}
    \caption{Schematic of the jet LES.  Contours show temperature (color) and pressure (grayscale) fluctuations.}
    \label{fig:jet_schematic}
\end{figure}

This configuration is investigated using high-fidelity large eddy simulation (LES) using the compressible flow solver {\it{CharLES}} developed at Cascade Technologies \cite{BresAIAAJ2017}, which solves the spatially filtered compressible Navier–Stokes equations on unstructured grids using a finite volume method. The round nozzle geometry (with exit centered at the origin) is explicitly included in the axisymmetric computational domain, which extends from approximately $-10D$ to $50D$ in the streamwise direction and flares in the radial direction from $20D$ to $40D$ (see Fig.~\ref{fig:jet_schematic}).  The simulation features localized adaptive mesh refinement as well as synthetic turbulence and wall modeling \cite{Kawai:PoF2012, bodartwm2011} inside the nozzle to match the fully turbulent nozzle-exit boundary layers in the experiments.  All other solid surfaces are treated as no-slip adiabatic walls. A very slow co-flow at Mach number $M_\infty = 0.009$ is imposed outside the nozzle in the simulation to prevent spurious recirculation and facilitate flow entrainment. Sponge layers and damping functions are applied to avoid spurious reflections at the boundary of the computational domain \cite{Freund-97, mani12}. The Vreman \cite{Vreman:2004p754} sub-grid model is used to account for the physical effects of the unresolved turbulence on the resolved flow.  Several meshes were considered as part of a grid resolution study \cite{BresJFM2018}, and the one considered here contains approximately 16 million control volumes.  Comparisons with the experimental measurements show good agreement for the flow and sound predictions, with the far-field noise spectra matching microphone data to within 0.5 dB for most relevant angles and frequencies. The LES methodologies, numerical setup, and comparisons with measurements are described in more detail in Ref. \cite{BresJFM2018}.

% -- Jet data -------------------------------------------
\subsection{Data}

\subsubsection{Time-resolved three-dimensional flow fields}
The database contains 10,000 snapshots of the jet sampled every 0.2 acoustic time units ($t c_{\infty}/D$).  Each snapshot contains cylindrical velocity components, density, and pressure.  To facilitate post-processing, the data have been interpolated from the original unstructured LES grid onto a structured cylindrical output grid that spans $0 \leq x/D \leq 30$, $0 \leq r/D \leq 6$ and contains $(626, 138, 128)$ points in the streamwise, radial and azimuthal directions, respectively.  The streamwise and radial grids approximately mirror the underlying LES resolution, while points are equally-spaced in the azimuthal direction to enable straightforward azimuthal Fourier decomposition.  The Cartesian velocity components computed by the LES were converted to cylindrical velocity components on this grid.  The total size of the resulting database is approximately 2.6 terabytes. Additional details are available in Ref. \cite{BresAIAA2016}.

\subsubsection{Time-resolved azimuthal Fourier modes}
In addition to the three-dimensional time-resolved data described above, the database also contains two quantities derived from the time-resolved data that are precomputed for convenience.  First, the database contains the first three azimuthal modes of each snapshot, computed via a discrete Fourier transform (DFT) in the azimuthal direction. After this transformation, each mode is a function of only the streamwise and radial coordinates.  These azimuthal modes are useful for at least two reasons.  First, the acoustic field emitted by the jet, which is frequently a key quantity of interest, is known to be composed almost entirely of the first three azimuthal modes \citep{Cavalieri2012axisymmetric, Chen2021azimuthal}.  Second, each azimuthal mode can be treated independently within the context of modal decompositions and linear analyses.  For instance, we will focus on the axisymmetric modes (the first azimuthal mode) in both analyses presented in the next section.

\subsubsection{Mean flow field}
Lastly, precomputed mean flow data are provided.  Due to the azimuthal symmetry of the geometry, averages are taken over both time and azimuthal angle, leading to mean quantities that depend on the streamwise and radial coordinates. These mean flow data are useful as input to linear stability analyses.

%In addition to the three-dimensional time-resolved data described above, the database also contains two derived quantities for convenience.  First, mean flow data is provided.  Due to the azimuthal symmetry of the geometry, averages are taken over both time and azimuthal angle, leading to mean quantities that depend on the streamwise and radial coordinates.  Second, the database contains the first three azimuthal modes of each snapshot, computed via a discrete Fourier transform (DFT) in the azimuthal direction.  After this transformation, each mode is, like the mean flow, a function of only the streamwise and radial coordinates.  These azimuthal modes are useful for at least two reasons.  First, the acoustic field emitted by the jet, which is frequently a key quantity of interest, is known to be composed almost entirely of the first three azimuthal modes.  Second, each azimuthal mode can be treated independently within the context of the modal decompositions and linear analyses.  For instance, we will focus on the axisymmetric modes (the first azimuthal mode) in both analyses presented in the next section.

% -- Jet analysis -----------------------------------------
\subsection{Example analyses}

We provide two examples of how the jet data can be used to build, test, compare, and validate reduced-complexity models.

\subsubsection{Modeling coherent structures via spectral POD and resolvent analysis}

Following Towne et al. \cite{towne2018spectral}, we use the jet data to demonstrate a theoretical connection between spectral proper orthogonal decomposition (SPOD) and resolvent analysis.

SPOD is a data-driven method used to identify coherent structures in a flow \cite{Lumley1970stochastic, towne2018spectral}.  The method is derived from a space-time POD problem and yields modes that oscillate at a single frequency and evolve coherently in both space and time, making them an attractive definition for (Eulerian) coherent structures.  SPOD uses data to estimate the cross-spectral density (CSD) tensor, and the eigenvectors of the CSD give a set of modes at each frequency, ordered by the associated eigenvalues that describe their average energetic contribution to the flow.

Resolvent analysis is an operator-based method \cite{Trefethen1993hydrodynamic, McKeon2010critical}.  The Navier-Stokes equations are split into terms that are linear and nonlinear with respect to perturbations to the mean flow, and the nonlinear terms are interpreted as a forcing on the linearized equations \cite{McKeon2010critical}.  The resolvent operator constitutes a transfer function between the forcing and the linear response, and the singular value decomposition of the resolvent operator identifies modes that are most amplified at each frequency.    
An intuitive relationship exists between SPOD and resolvent modes: the most amplified modes (resolvent) are the same as the most energetic modes (SPOD) if the forcing is unbiased, i.e., if it is white noise \cite{towne2018spectral}.  The jet data can be used to explore this theoretical connection.  Specifically, SPOD modes are computed using the time-series data for the axisymmetric azimuthal mode, and resolvent modes are computed using the LES mean flow as input.

\begin{figure}[t]
    \centering
    \includegraphics[]{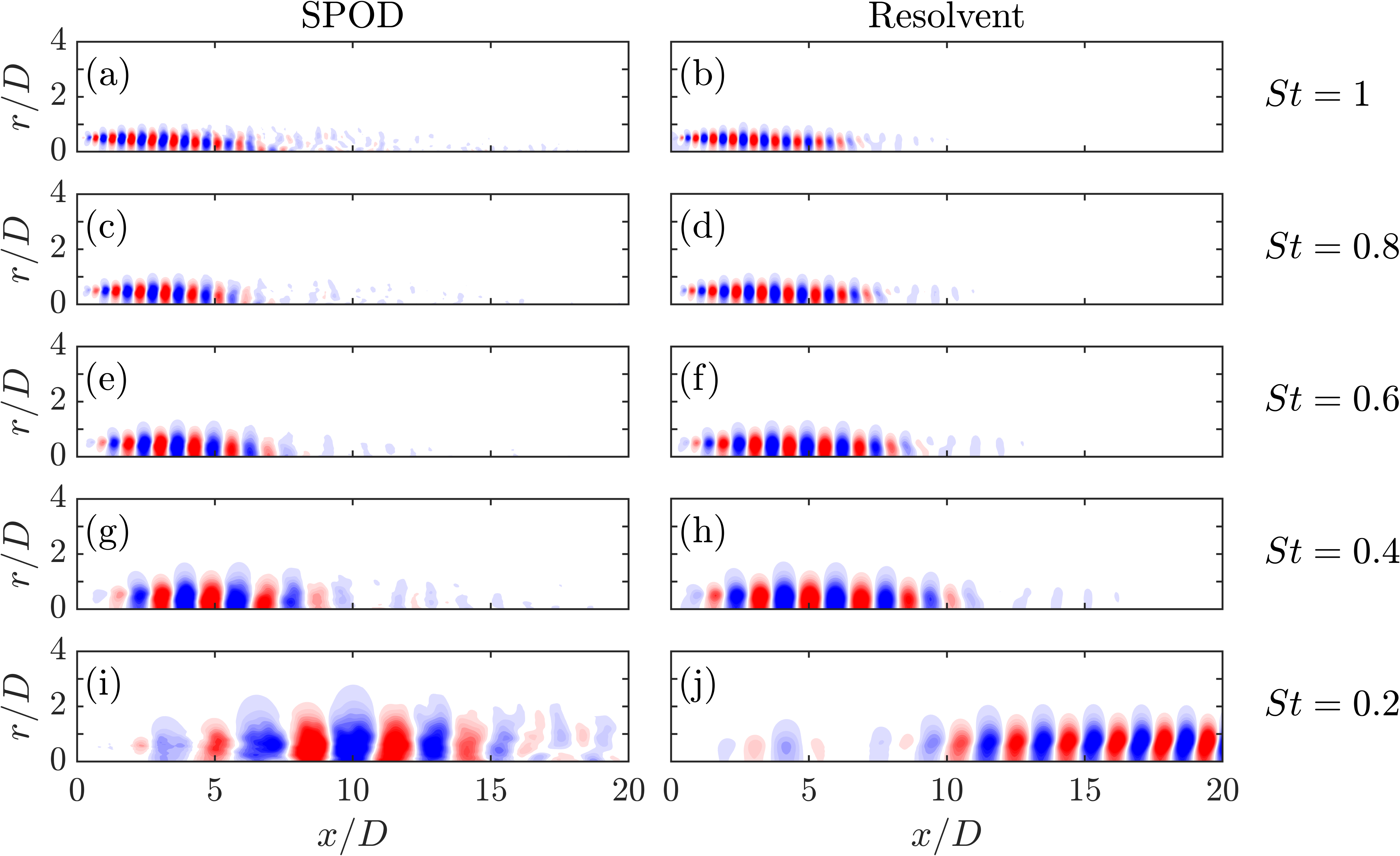}
    \caption{Pressure field from the leading SPOD and resolvent modes at several frequencies.}
    \label{fig:jet_SPOD_RES}
\end{figure}

The leading SPOD and resolvent modes for five frequencies (reported in terms of the Strouhal number $St = f D / U_{j}$) are shown in Fig.~\ref{fig:jet_SPOD_RES}.  A close match is observed for the higher four frequencies, with both methods yielding a coherent wavepacket structure associated with the Kelvin-Helmholtz instability \cite{Jordan2013wave, Schmidt2018spectral}, confirming that white noise is a reasonable forcing model for these frequencies.  On the other hand, significant differences are observed at the lowest frequency.  Here, Orr modes, which are sensitive to the forcing, dominate \cite{Schmidt2018spectral, Pickering2020lift}.  This mismatch, which extends to higher frequencies at higher azimuthal wavenumbers due to the appearance of streaks \cite{Pickering2020lift, Nogueira2019large}, has motivated the investigation of improved forcing models for resolvent analysis of the jet based on analysis of the LES data \cite{Towne2017statistical, Pickering2021resolvent}, sparse sampling strategies \cite{Towne2020resolvent, Martini2020resolvent}, and eddy viscosity models \cite{Pickering2021optimal}.

\subsubsection{Trapped acoustic waves in the jet potential core}

\begin{figure}[t]
    \centering
    \includegraphics[]{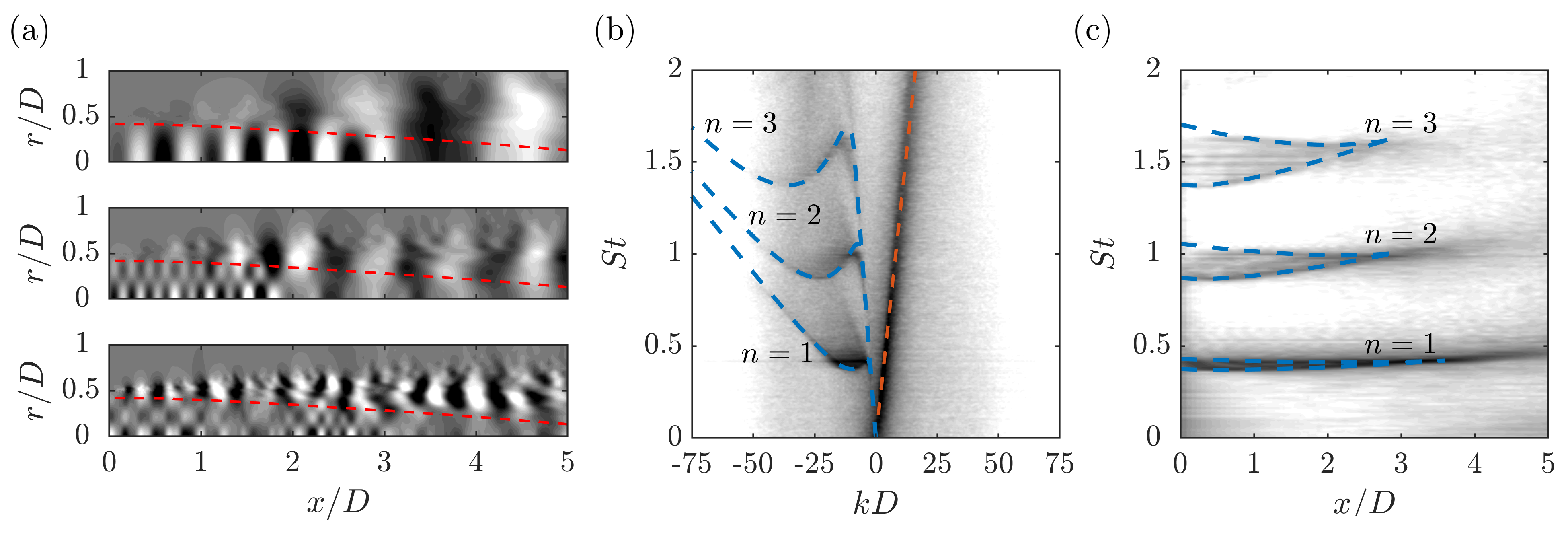}
    \caption{Trapped acoustic waves in the jet potential core: (a) Fourier modes of the pressure field at frequencies $St = 0.39, 0.91$, and $1.52$ (top to bottom) showing the presence of waves in the jet core; (b) dispersion relation along the jet center-line; (c) PSD along the jet center-line.  In (b) and (c), the contours show data from the LES while the lines show predictions from a weakly nonparallel stability theory.  The contour levels span four orders of magnitude, with dark colors showing regions of high energy \cite{Towne2017acoustic}.  }
    \label{fig:jet_trapped_waves}
\end{figure}

The jet data can also be used to investigate waves observed in the potential core of the jet.  These waves are visible in the pressure field of the instantaneous snapshot shown in Fig.~\ref{fig:jet_schematic} and are clearer still in the Fourier modes shown in Fig.~\ref{fig:jet_trapped_waves}a.

The LES data have played an important role in unraveling the mystery of these waves.  First, the mean data have been used as the basis for weakly nonparallel \cite{Towne2017acoustic} and global \cite{Schmidt2017wavepackets} linear stability analyses that revealed the presence of eigenmodes describing acoustic waves that are trapped within the potential core and, at certain frequencies, resonate between the nozzle exit plane and a downstream turning point created by the narrowing of the potential core.  Second, the data have been used to verify that these eigenmodes describe the waves observed in the jet.  Figure~\ref{fig:jet_trapped_waves}b shows an empirical dispersion relation based on the power spectral density (PSD) extracted from the data via streamwise and temporal Fourier transforms overlaid by the theoretical dispersion relation from the weakly nonparallel model. The theoretical curves closely match the energetic (dark-colored) regions of the plot, showing that the corresponding instability waves are present in the flow.  The red line corresponds to the Kelvin-Helmholtz instability, while the blue curves are the trapped acoustic waves.  The existence of both positive and negative slopes, which give the group velocity, for these latter curves indicates that the trapped acoustic waves can travel both upstream and downstream at certain frequencies.  Figure~\ref{fig:jet_trapped_waves}c shows the power spectrum along the jet axis as well as the predicted boundaries of resonance between the upstream- and downstream-traveling trapped acoustic waves from the model.  These curves approximately bound the energetic regions observed in the data, indicating that the trapped waves indeed resonate according to the predictions from the linear model.  The agreement in both cases confirms that the model captures the waves present in the data.  

%%%%%%%%%%%%%%%%%%%%%%%%%%%%%%%%%%%%%%%%%%%%%%%%%%%%%%%%%%%
% -- BL DNS Section (Adrian) ------------------------------

%\section{Time-resolved DNS database of turbulent boundary layers} 
\section{Turbulent boundary layer direct numerical simulations} 
\label{sec:BL_DNS}

Second, we present a zero-pressure-gradient flat-plate turbulent boundary layer, which is the canonical example of an external wall-bounded shear flow. Turbulent boundary layers have been extensively studied; much is now known about their statistics and the coherent structures that populate them. More recently, interest has grown in the nonlinear interactions that occur within the boundary layer, which allow one coherent structure to affect others and generate self-sustaining cycles of interaction.  The dataset contains time-resolved velocity fields computed via direct numerical simulation (DNS) of the Navier-Stokes equations, enabling temporal analysis and reduced-order modeling of the turbulence dynamics.

% -- BL DNS setup -----------------------------------------
\subsection{Setup} 

% Outline:
% 1 Set-up description
% 2 Cases:
% 3 Boundary conditions
% 4 Domain dimensions
% 5 Numerics: resolution and time

\begin{figure}
\begin{center}
\includegraphics[width=0.9\textwidth]{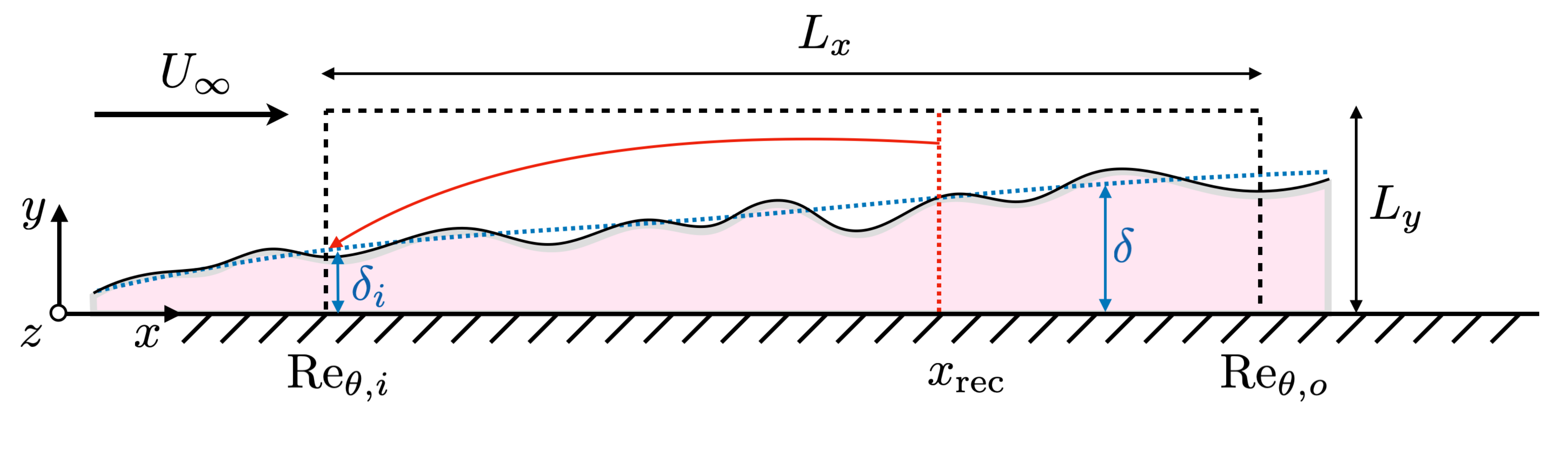}
\caption{Schematic of the numerical setup for the
  zero-pressure-gradient flat-plate turbulent boundary layer.
  ${Re}_{\theta,i}$ and ${Re}_{\theta,o}$ are the 
  momentum thickness Reynolds numbers at the inlet and outlet,
  respectively, $U_\infty$ is the freestream velocity, $\delta$ is
  the boundary layer thickness based on 99\% of $U_\infty$, $L_x$ and
  $L_y$ are the streamwise and wall-normal extent of the computational
  domain, and $x_\mathrm{rec}$ is the streamwise location for
  recycling the inflow boundary condition. The details of the
  simulations are summarized in Table
  \ref{table:cases}.\label{fig:schematic}}
\end{center}
\end{figure}

% 1 Set-up description
Figure \ref{fig:schematic} shows a schematic of the numerical setup for the boundary layer.  The streamwise, wall-normal, and spanwise spatial directions are denoted by $x, y$ and $z$, respectively. The corresponding instantaneous velocities are $u$, $v$, and $w$ and the
velocity vector is denoted by $\boldsymbol{u}=(u,v,w)$. The freestream velocity is $U_\infty$.  The wall is located at $y=0$ and quantities evaluated at the wall are represented by $(\cdot)|_w$. The streamwise distance of the inlet to the origin of the boundary layer is $L_o$. Wall units, denoted by the superscript $+$, are defined in
terms of $u_\tau$, $\nu$, and $\delta$, where 
\begin{equation}
u_\tau \equiv
\sqrt{\left.\nu\frac{\partial \langle u \rangle}{\partial y}\right|_{w}}
\end{equation}
is the friction velocity, $\nu$ is the kinematic viscosity of the flow, $\delta$
is the boundary-layer thickness at 99\% of $U_\infty$, and $\langle \cdot \rangle$ represents the average over $z$ and time. The Reynolds number based on the friction velocity and momentum thickness ($\theta$) are denoted by ${Re}_\tau = u_\tau \delta /\nu$ and ${Re}_\theta = U_\infty \theta /\nu$, respectively. The value of ${Re}_\tau$ and ${Re}_\theta$ at the inlet and outlet of the domain are ${Re}_{\tau,i}$, ${Re}_{\theta,i}$ and ${Re}_{\tau,o}$, ${Re}_{\theta,o}$, respectively.
%
%------------------------------------------------------------------%

% 2 Cases:
Two datasets are presented and labeled as BL1 and BL2. Table \ref{table:cases} offers a summary of the main parameters of the simulations. The dataset for BL1 covers the range of friction Reynolds number from ${Re}_\tau\approx 292$ to $729$, with roughly $10,000$ flow fields stored spanning $26$ eddy turnover times (after transients) based on
$\delta_\mathrm{avg}/u_{\tau,\mathrm{avg}}$, where the subscript `$\mathrm{avg}$' denotes the average along $x$. The time separation between consecutive stored snapshots is constant in time and equal to $\Delta t^+\approx 1.5$.
%
%The total storage space for BL1 is $75$ terabytes
%($7.5$ gigabytes per field) with flow fields containing the three
%velocity components and pressure.
%
The second dataset, BL2, ranges from ${Re}_\tau = 481$ to $1024$, with $7500$ flow fields stored every $\Delta t^+ \approx 0.8$ and spanning a physical time of $7.4$ eddy turnovers.
%
%The total storage
%space for BL2 is $30$ terabytes ($30$ gigabytes per field).
%
In both datasets, the time resolution between stored snapshots is smaller than the characteristic lifetime of the smallest energy-containing eddies ($\Delta t^+ \approx 30$)~\citep{Lozano2014}.
\begin{table}
\begin{center}
\begin{tabular}{l*{6}{c}r}
\hline
% \textbf{Case} & $\boldsymbol{L_x/\theta_\text{avg}}$ & $\boldsymbol{L_y/\theta_\text{avg}}$ & $\boldsymbol{L_z/\theta_\text{avg}}$ & $\boldsymbol{{Re}_{\tau,i}$--${Re}_{\tau,o}}$ & $\boldsymbol{{Re}_{\theta,i}$--${Re}_{\theta,o}}$ & $\boldsymbol{x_{\mathrm{ref}}/\theta_\text{avg}}$ \\ \hline
Case & $L_x/\theta_\text{avg}$ & $L_y/\theta_\text{avg}$ & $L_z/\theta_\text{avg}$ & ${Re}_{\tau,i}$--${Re}_{\tau,o}$ & ${Re}_{\theta,i}$--${Re}_{\theta,o}$ & $x_{\mathrm{ref}}/\theta_\text{avg}$ \\ \hline
BL1 & 480     & 47     & 70    & 292--729      & 832--1982   & 384 \\
BL2 & 469     & 53     & 79    & 481--1024     & 1272--2870  & 375 \\ \hline
\end{tabular}
\end{center}
\begin{center}
\begin{tabular}{l*{6}{c}r}
\hline
% \textbf{Case} & $\boldsymbol{\Delta x^+}$  & $\boldsymbol{\Delta y^+_{\mathrm{min}}}$  & $\boldsymbol{\Delta z^+}$ & $\boldsymbol{\Delta t^+}$ & $\boldsymbol{T u_{\tau,\mathrm{avg}}/\delta_{\mathrm{avg}}}$\\
Case & $\Delta x^+$  & $\Delta y^+_{\mathrm{min}}$  & $\Delta z^+$ & $\Delta t^+$ & $T u_{\tau,\mathrm{avg}}/\delta_{\mathrm{avg}}$\\ \hline
BL1 & 15.1   & 0.42  & 8.8  & 1.5  & 26.1 \\
BL2 & 10.2   & 0.66  & 6.9  & 0.8  & 7.4 \\ \hline
\end{tabular}
\end{center}
\caption{Summary of the simulation parameters for BL1 and BL2. The
  size of the computational domain is $L_x \times L_y \times L_z$,
  $\theta_\text{avg}$ is the streamwise-averaged momentum thickness,
  ${Re}_{\tau,i}$--${Re}_{\tau,o}$ is the range of
  ${Re}_{\tau}$ covered from inflow to outflow (similarly for
  ${Re}_{\theta,i}$--${Re}_{\theta,o}$),
  $x_{\mathrm{ref}}$ is the location of the recycling plane for the
  inflow boundary condition, $\Delta x^+$ and $\Delta z^+$ are the
  averaged streamwise and spanwise grid resolution, respectively,
  $\Delta y^+_{\mathrm{min}}$ is the average minimum wall-normal grid
  resolution, $\Delta t^+$ is the time between stored flow fields,
  $u_{\tau,\mathrm{avg}}$ is the averaged friction velocity,
  $\delta_{\mathrm{avg}}$ is the averaged boundary layer thickness,
  and $T$ is the total time simulated after initial
  transients. \label{table:cases}}
\end{table}

% 3 Boundary conditions:
% top plane
At the top boundary of the computational domain ($y=L_y$), the streamwise and spanwise velocities are $u = U_\infty$ and $w=0$, respectively. The wall-normal velocity at the top boundary is estimated from the known experimental growth of the displacement thickness for the corresponding range of Reynolds numbers as $v(x,L_y,z)=U_\infty \mathrm{d}\delta^*/\mathrm{d}x$ with $\delta^*/x = 0.020 {Re}_x^{-1/7}$, where ${Re}_x$ is the Reynolds number based on the distance to the leading edge~\citep{Monkewitz2007, Jimenez2010}.
%
%The value of $v(x,L_y,z)$ controls the average streamwise pressure
%gradient, whose nominal value is set to zero.
%
% inflow plane
The mean velocity profile at the inflow is kept fixed and equal to the mean velocity profile obtained from previous simulations~\citep{Spalart1988, Jimenez2010}.  The turbulence
fluctuations at the inflow are generated by the recycling scheme of \citet{Lund1998}, in which the velocity fluctuations from a reference downstream plane, $x_\mathrm{ref}$, are used to synthesize the incoming turbulence.
% outflow plane
The convective boundary condition, $\partial \boldsymbol{u}/\partial t + U_\infty \partial \boldsymbol{u} /\partial x= 0$, is applied at the outlet~\citep{Pauley1990} and small corrections are applied to enforce global mass conservation \citep{Simens2009}. The spanwise direction is periodic.

% 4 Domain dimensions:
The length, height, and width of the computational domain in both simulations are roughly $L_x = 450\theta_\text{avg}$, $L_y = 50\theta_\text{avg}$ and $L_z = 70\theta_\text{avg}$, where $\theta_\text{avg}$ denotes the momentum thickness averaged along the streamwise coordinate. The dimensions of the domain are similar to those used in previous studies~\citep{Schlatter2010, Jimenez2010, Sillero2013}.
%
%In both BL1 and BL1, the streamwise, wall-normal,
%and spanwise domain size is larger than $1000 \times 50 \times 150$
%the boundary layer thickness $\delta$ at the inlet.
%
The recycling plane is located at $x_\text{ref}/\theta_\text{avg} >
370$ measured from the inlet location to minimize spurious
feedback~\citep{Nikitin2007, Simens2009}.
% 
% snapshot case I
% A visualization of the instantaneous streamwise velocity at two
% different heights is portrayed in Fig.~\ref{fig:snapshot} for BL2.
% %
% %------------------------------------------------------------------%
% \begin{figure}
% \begin{center}
% \includegraphics[width=0.9\textwidth]{./Figures/BL_DNS//snapshot.eps}
% \caption{Instantaneous streamwise velocity at (a) $y^+\approx 15$ and
%   (b) $y/\delta \approx 0.5$ for BL2 centered at ${Re}_\tau
%   \approx 800$. \label{fig:snapshot}}
% \end{center}
% \end{figure}

% 5 Numerics: resolution and time
The solutions are computed by DNS of the incompressible Navier-Stokes equations. The spatial discretization is a staggered second-order central finite difference scheme~\citep{Orlandi2000}. Time advancement is achieved by a third-order Runge-Kutta scheme~\citep{Wray1990} combined with the fractional-step method~\citep{Kim1985}.
%
%The Poisson solver uses the
%cosine transform to take into account the non-periodic boundary
%conditions in the streamwise direction. The code is parallelized using
%Message Passing Interface with a global transpose from $y$-$z$ to
%$x$-$y$ planes. All computations were run with constant time step
%such that CFL<0.5.
%
The code has been validated in previous studies in turbulent channel
flows \citep{Bae2018, Lozano2019d} and transitional boundary layers
\citep{Lozano2018}.

% -- BL DNS data -------------------------------------------
\subsection{Data}

The dataset includes (time-resolved) volumetric data for BL1 and BL2 along with a number of other quantities.  Specifically, the dataset contains the following quantities.

% 1 Time-resolved data
\subsubsection{Time-resolved three-dimensional velocity fields}

\begin{figure}
\begin{center}
\includegraphics[width=0.9\textwidth]{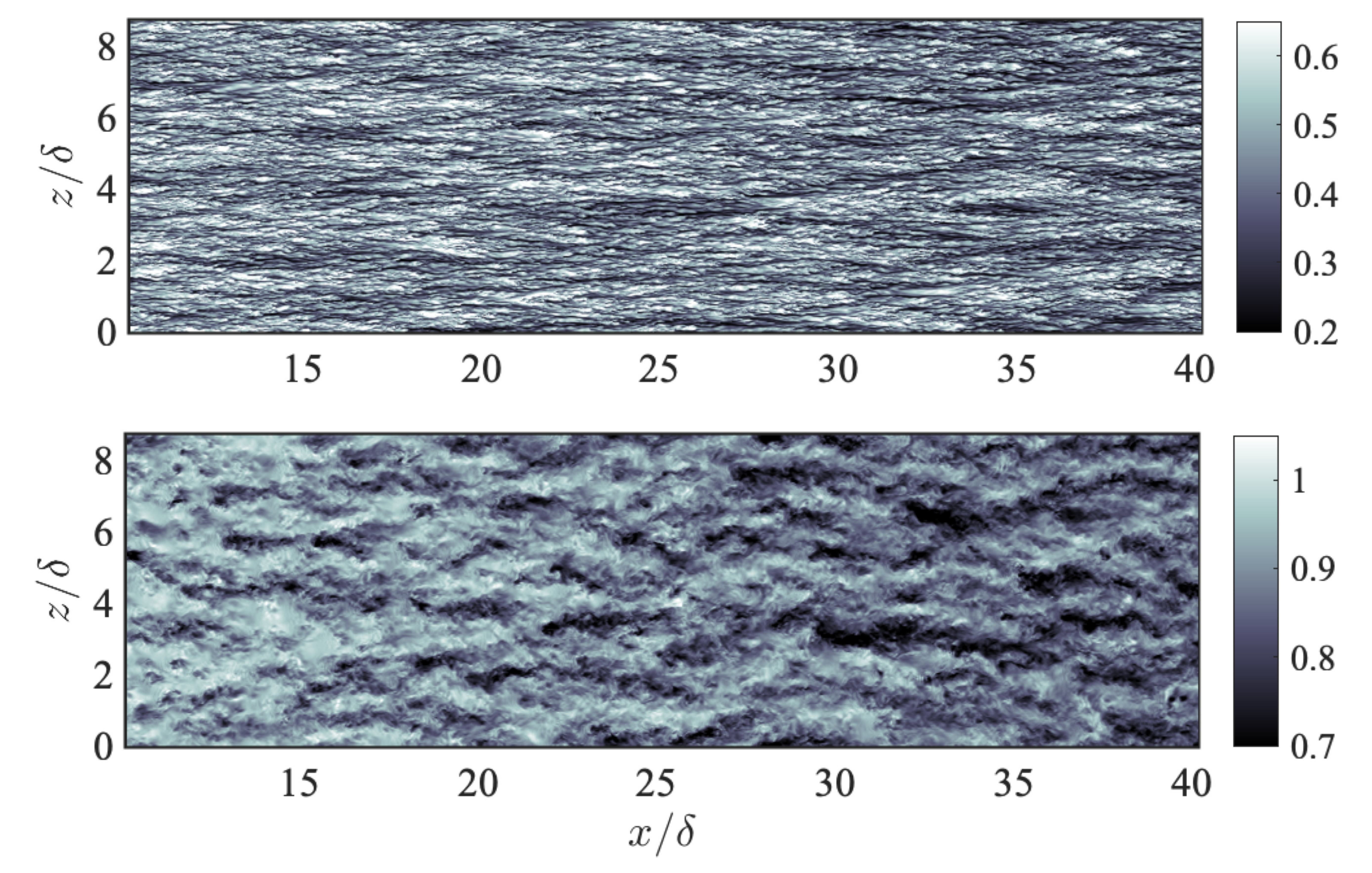}
\caption{Instantaneous streamwise velocity at (a) $y^+\approx 15$ and
  (b) $y/\delta \approx 0.5$ for BL2 centered at ${Re}_\tau
  \approx 800$. \label{fig:snapshot}}
\end{center}
\end{figure}

The publicly available database includes three-dimensional space/time-resolved velocity fields for BL1 and three-dimensional space-resolved fields for BL2. The fields are downsampled by a factor of two in the wall-normal and spanwise spatial directions compared to the original DNS resolution, making the full database 5 terabytes. We anticipate that this downsampled data will be sufficient for most users. The original full-resolution fields for both BL1 and BL2 are available upon request to the authors.  A visualization of the instantaneous streamwise velocity at two different heights is portrayed in Fig.~\ref{fig:snapshot} for BL2.
%
%------------------------------------------------------------------%

% 2 Stresmwise development
\subsubsection{Boundary-layer metrics}

\begin{figure}
\begin{center}
\includegraphics[width=0.7\textwidth]{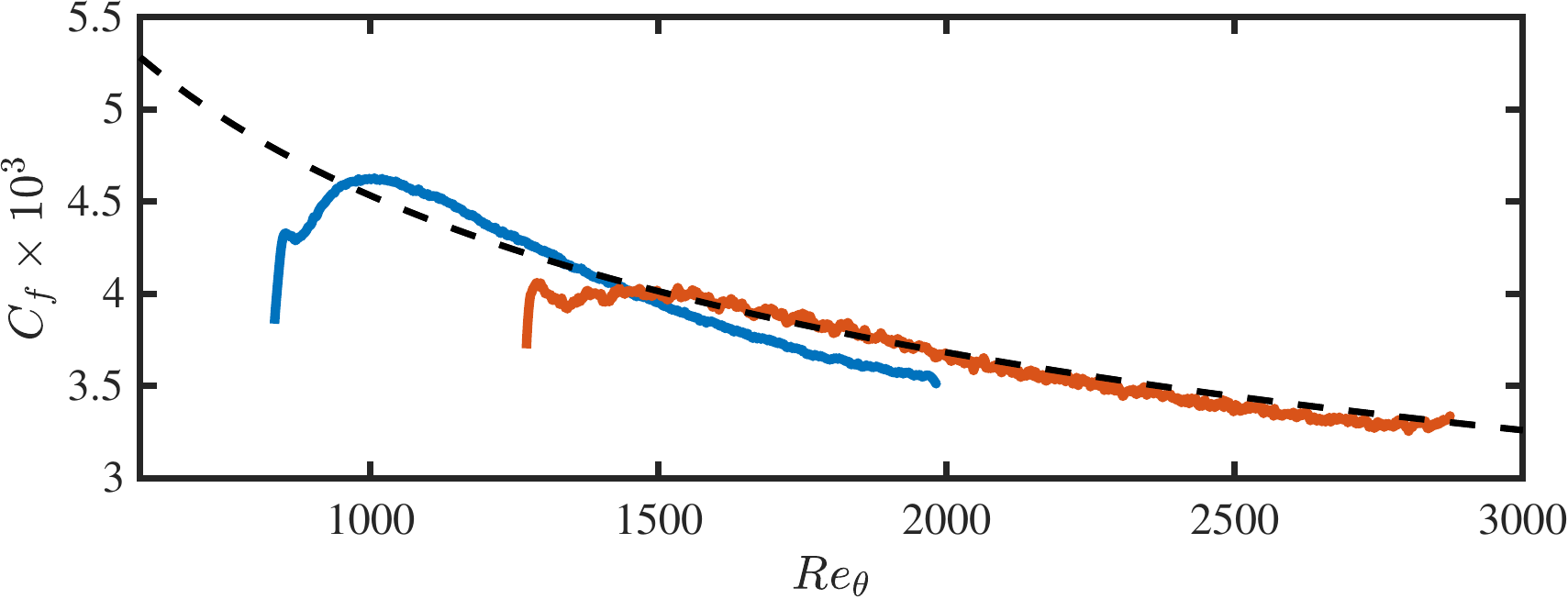}
\caption{Streamwise development of the skin friction coefficient for
  BL1 (blue) and BL2 (red). The dashed line is the empirical fit from \citet{White1974}.\label{fig:TBL:Cf}}
\end{center}
\end{figure}

The dataset also includes several boundary-layer metrics as a functions of $x$: the friction coefficient $C_f$, the boundary layer thickness $\delta$, the momentum thickness $\theta$, the Reynolds number ${Re}_{\theta}$, the friction Reynolds number, ${Re}_{\tau}$, and the friction velocity $u_\tau$.  The streamwise development of the friction coefficient, defined as $C_f = 2\langle \tau_w \rangle/\rho U_\infty^2$, is shown in Fig.~\ref{fig:TBL:Cf} as a function of ${Re}_\theta$. The empirical correlation from \citet{White1974} is included in the plot for reference. The streamwise development of $C_f$ undergoes an initial transient generated by the artificial boundary condition at the inlet. This behavior has also been observed in previous DNS of turbulent boundary layers using flow recycling schemes at the inlet~\citep{Schlatter2010,Jimenez2010, Sillero2013}. After the streamwise transient (${Re}_\theta>1000$ for BL1 and ${Re}_\theta>1500$ for BL2), the flow recovers and approaches the expected empirical value.
%------------------------------------------------------------------%
%

% 3 Mean velocity/vorticity profiles and fluctuations
\subsubsection{Mean and RMS velocity and vorticity profiles}

\begin{figure}
\begin{center}
  \sidesubfloat[]{\includegraphics[width=0.45\textwidth]{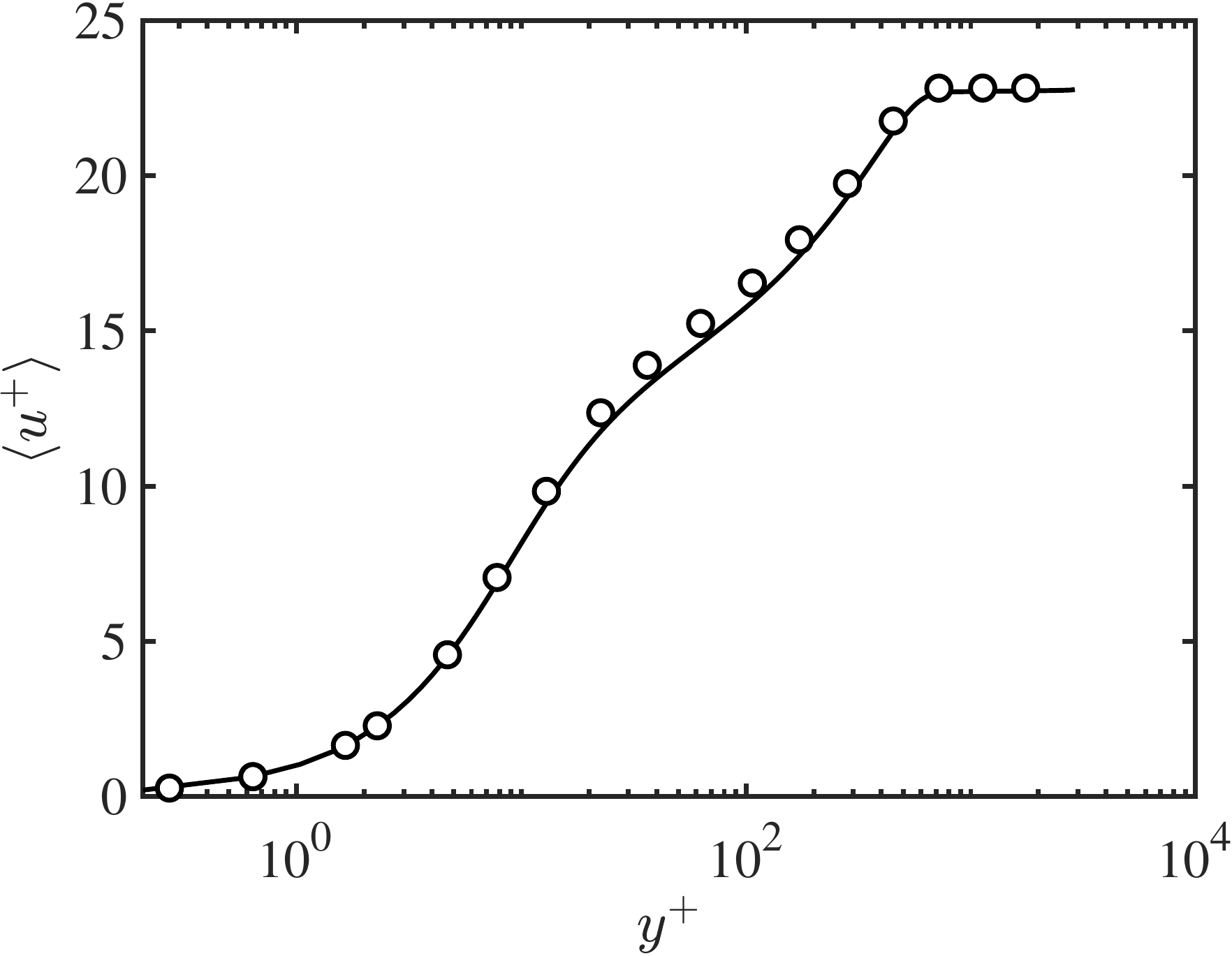}}
  \sidesubfloat[]{\includegraphics[width=0.45\textwidth]{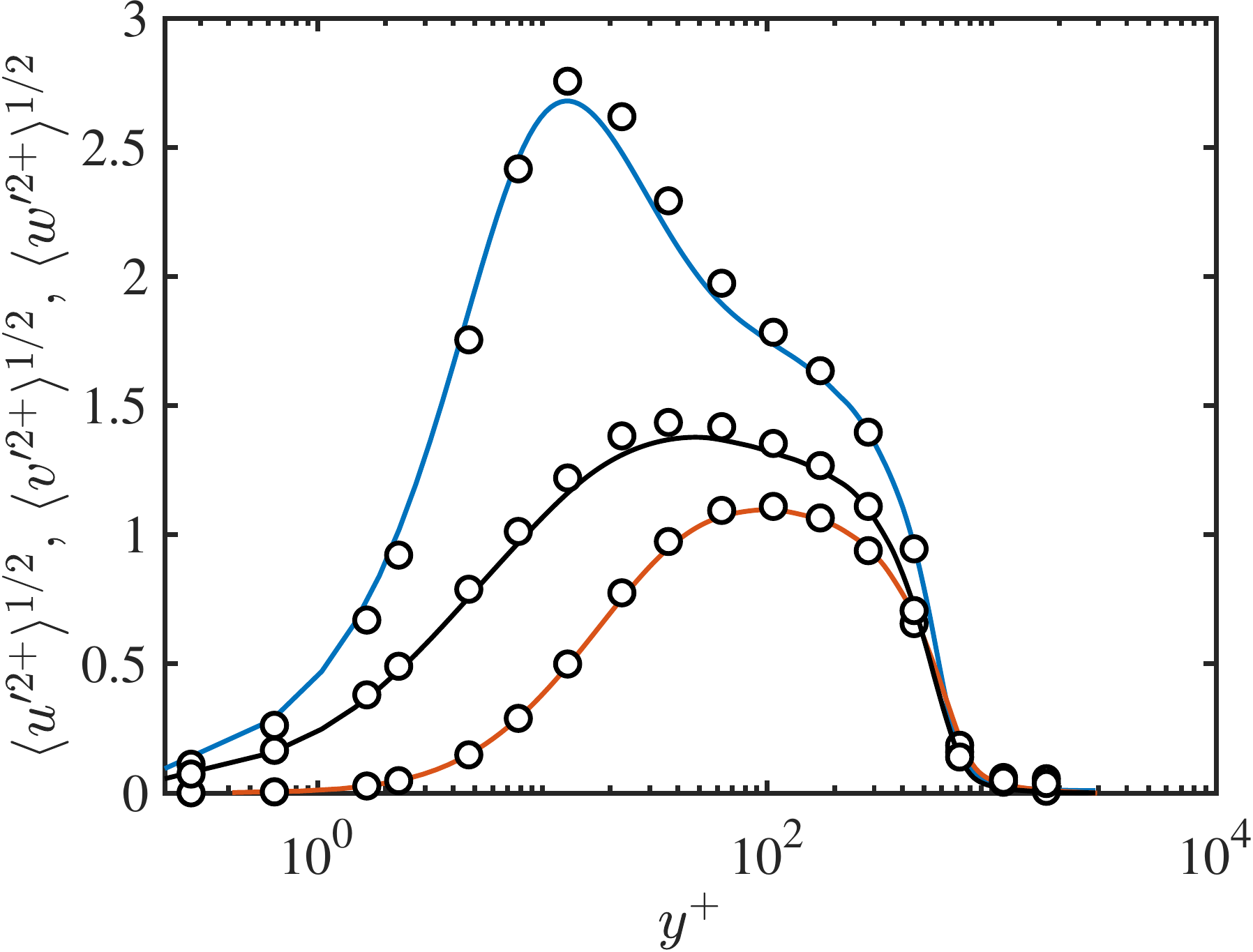}}
\end{center}
\begin{center}
  \sidesubfloat[]{\includegraphics[width=0.45\textwidth]{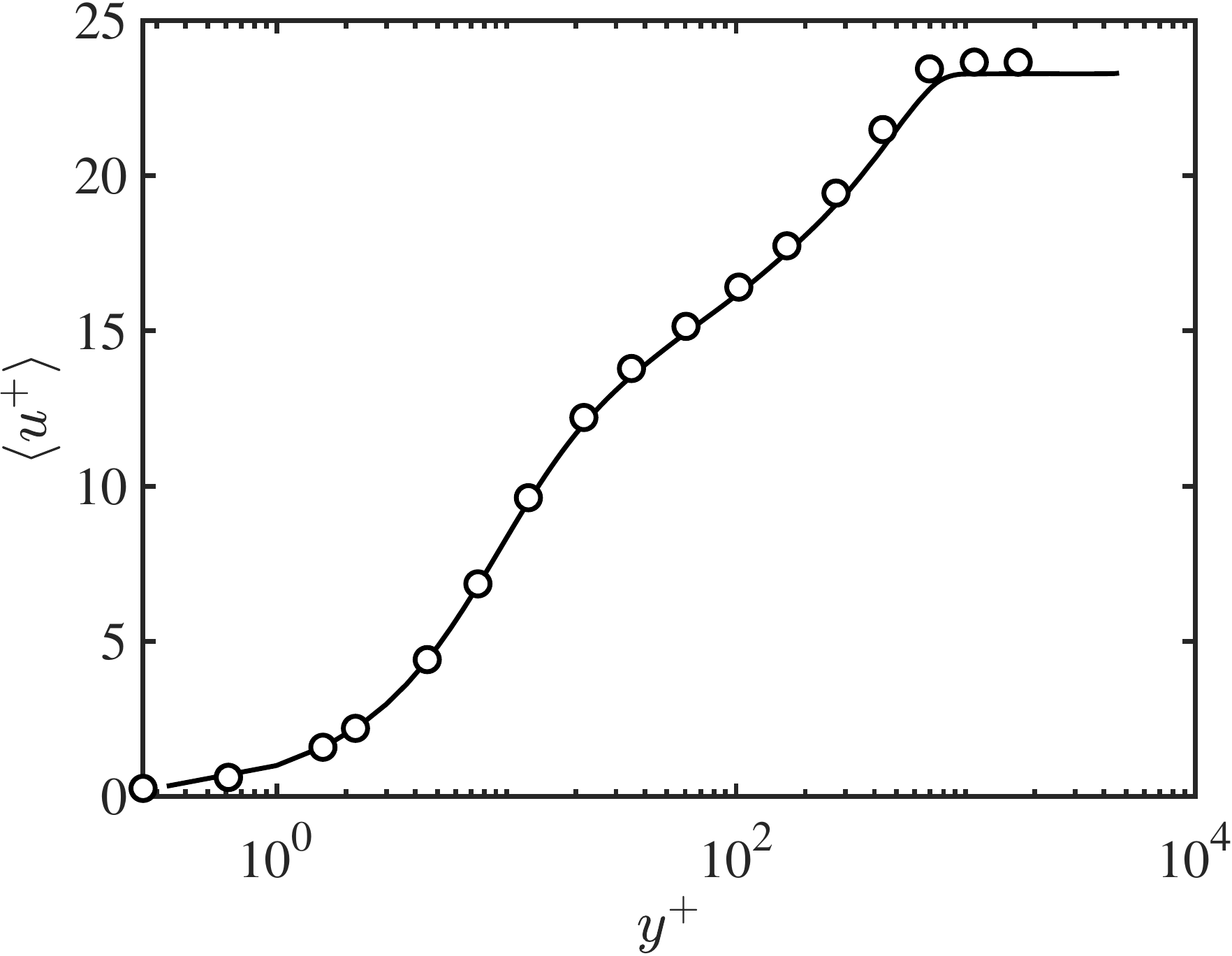}}
  \sidesubfloat[]{\includegraphics[width=0.45\textwidth]{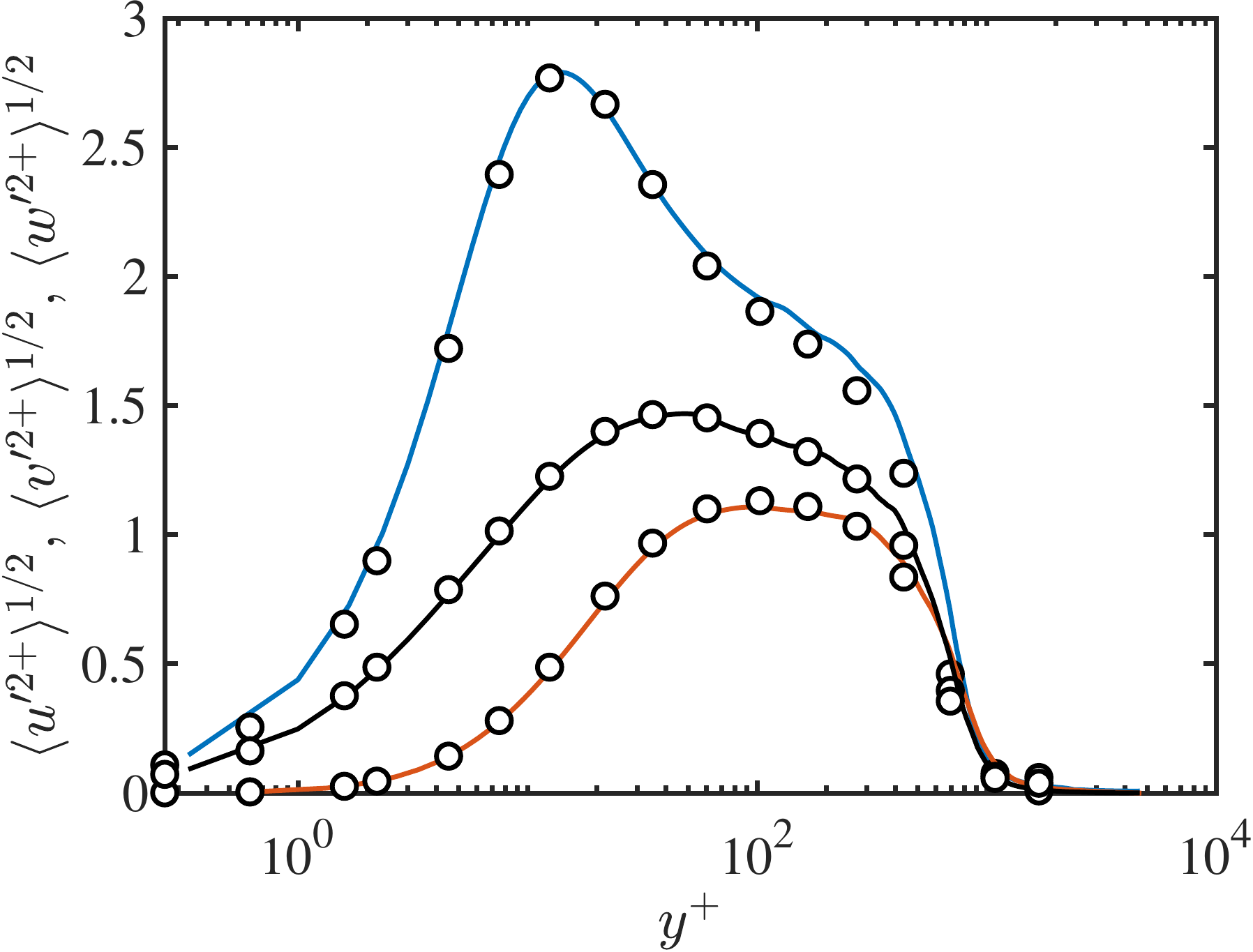}}
\end{center}
\caption{The mean velocities profiles for (a) BL1 at
  ${Re}_\theta=1551$ and (c) BL2 at
  ${Re}_\theta=1968$. The streamwise (blue), wall-normal (red), and spanwise (black) root-mean-squared velocity fluctuations for (b) BL1 at ${Re}_\theta=1551$ and (d) BL2 at ${Re}_\theta=1968$.  Symbols are data from a turbulent boundary layer from \citet{Sillero2013} at a similar ${Re}_\theta$. \label{fig:TBL:profiles}}
\end{figure}

The dataset also includes mean and root-mean-squared (RMS) velocity and vorticty fields as a function of $x$ and $y$.  The streamwise mean velocity profile ($\langle u \rangle$) and the RMS velocity fluctuations $\langle u'^2 \rangle^{1/2}$, where the fluctuations are defined as $u' = u - \langle u \rangle$ (similarly for $v'$ and $w'$) are shown in Fig.~\ref{fig:TBL:profiles} at ${Re}_\theta=1551$ for BL1 and ${Re}_\theta=1968$ for BL2. The profiles are compared to DNS data from \citet{Sillero2013} at similar ${Re}_\theta$ and the agreement is within to 2\% tolerance.
%
%------------------------------------------------------------------%

% 4 Kinetic energy/dissipation correlations
\subsubsection{Two-point correlations}

\begin{figure}
  \begin{center}
    \sidesubfloat[]{\includegraphics[width=0.5\textwidth]{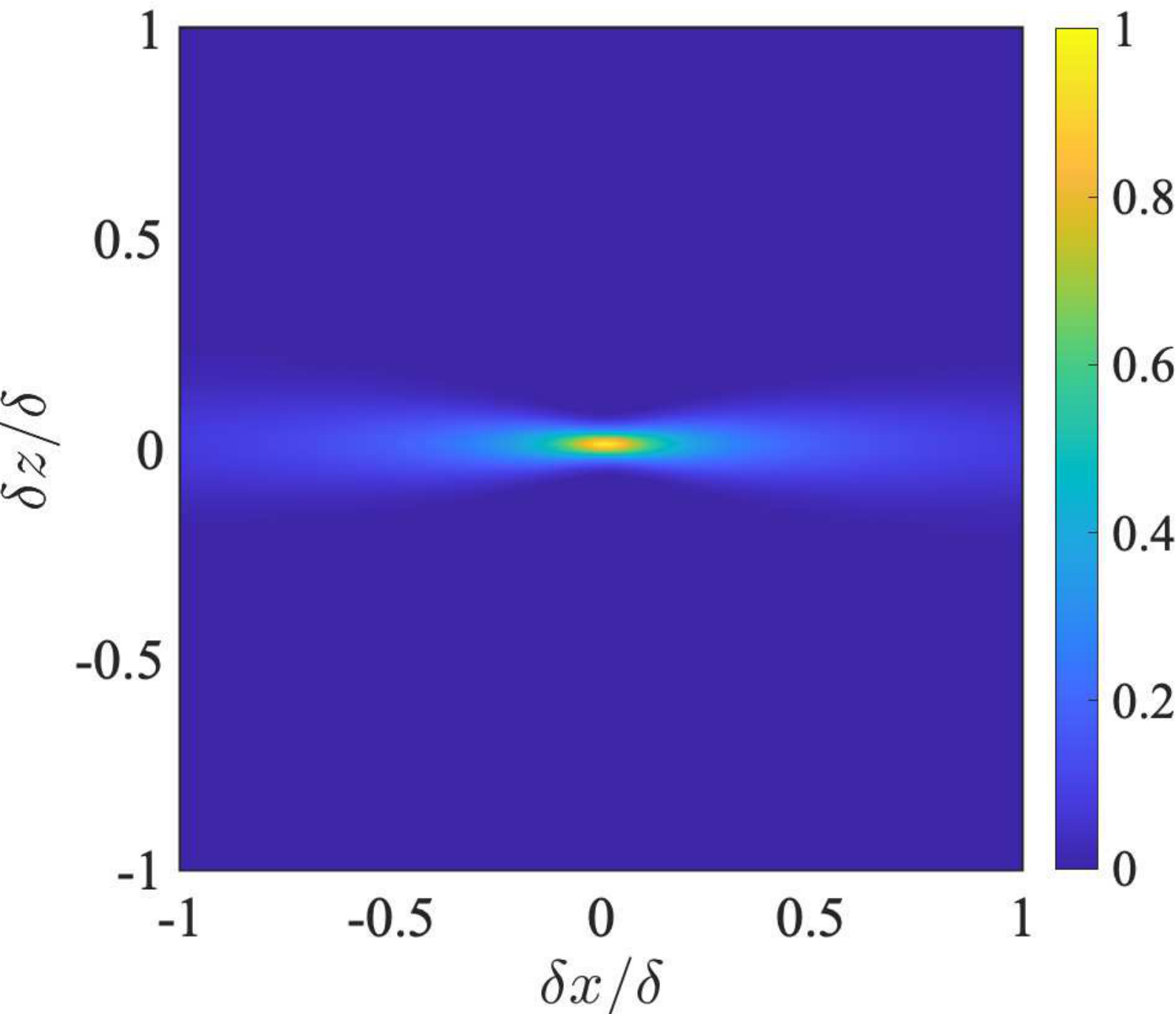}}
    \sidesubfloat[]{\includegraphics[width=0.5\textwidth]{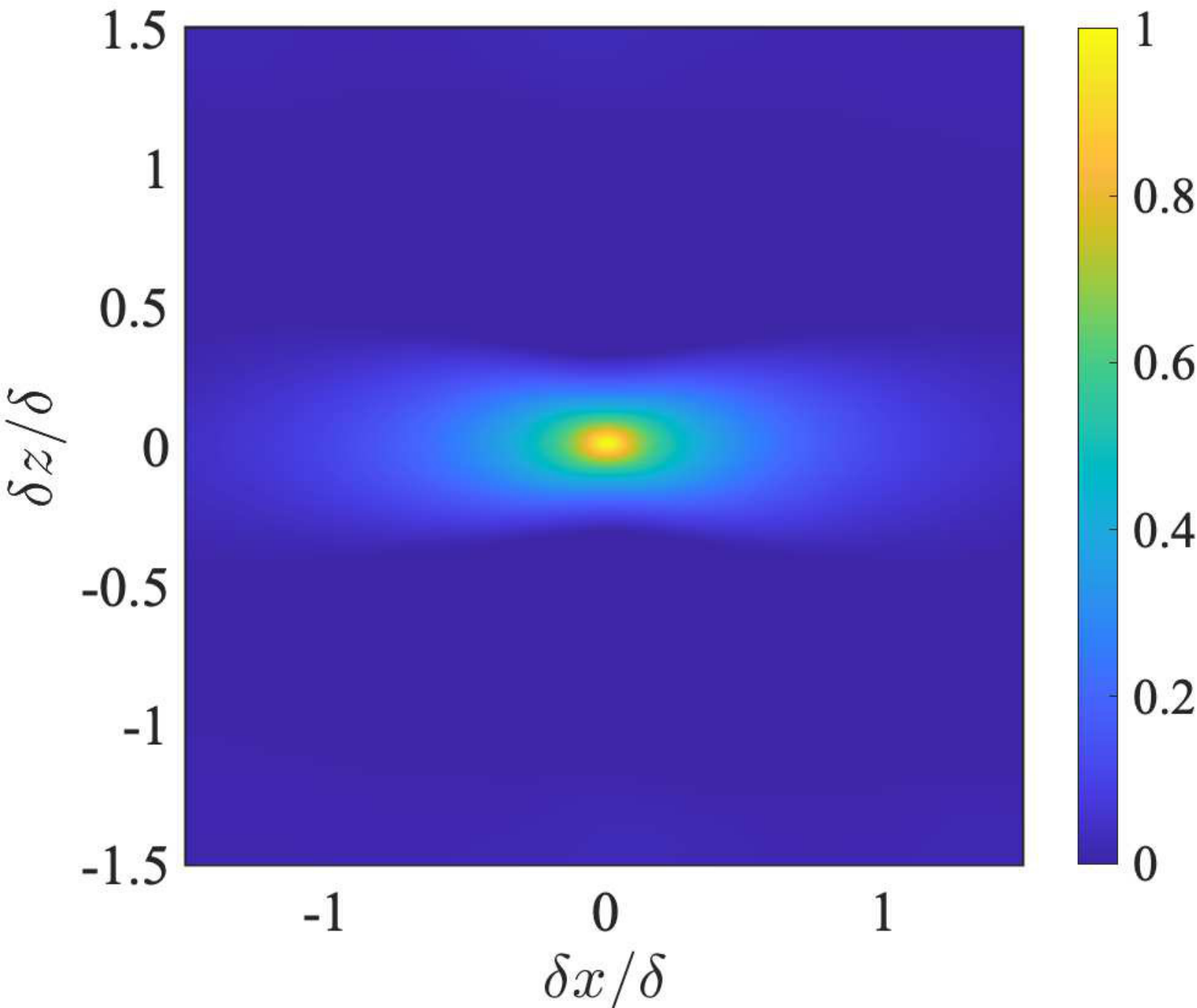}}
\end{center}
  \begin{center}
  \sidesubfloat[]{\includegraphics[width=0.5\textwidth]{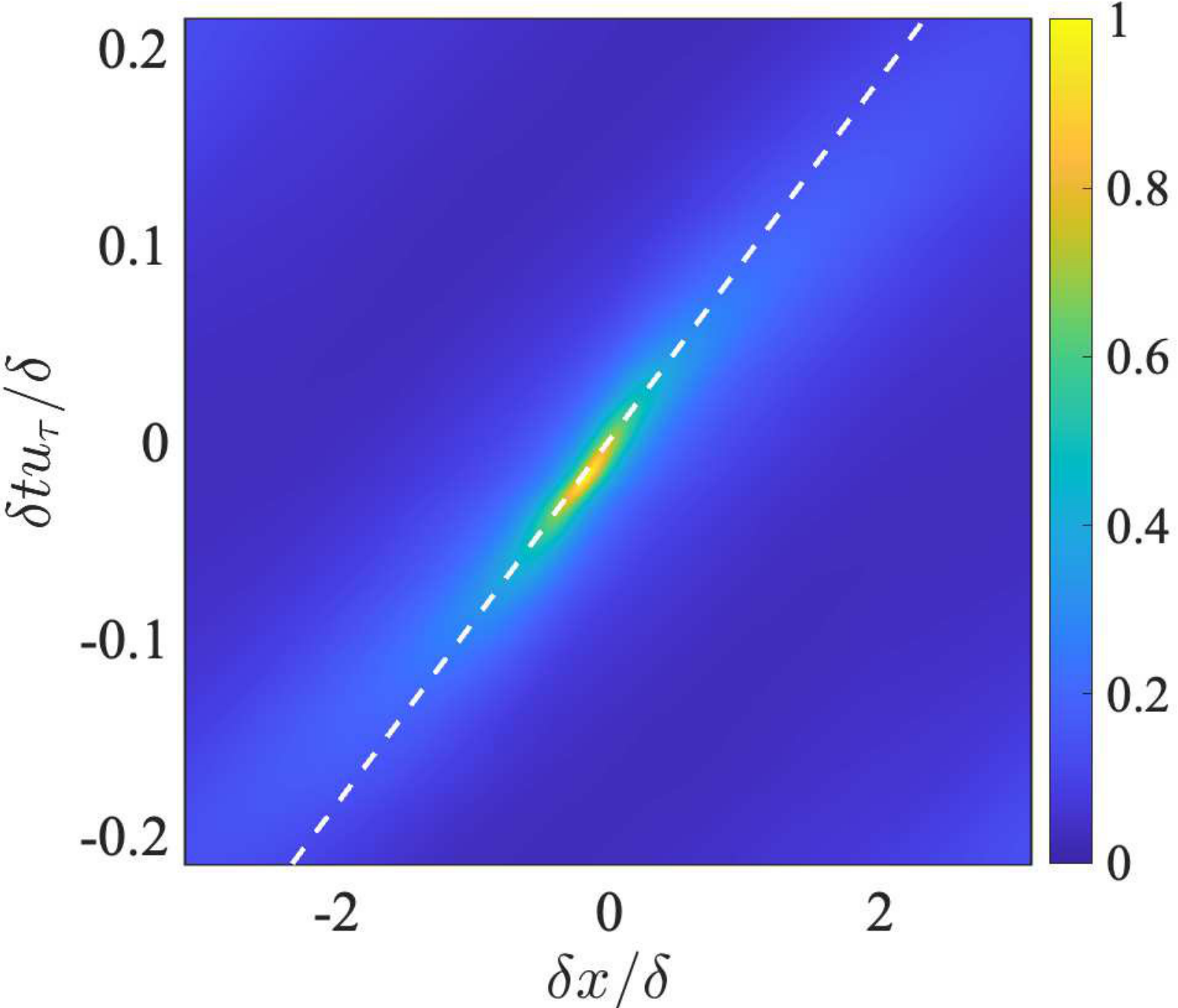}}
  \sidesubfloat[]{\includegraphics[width=0.5\textwidth]{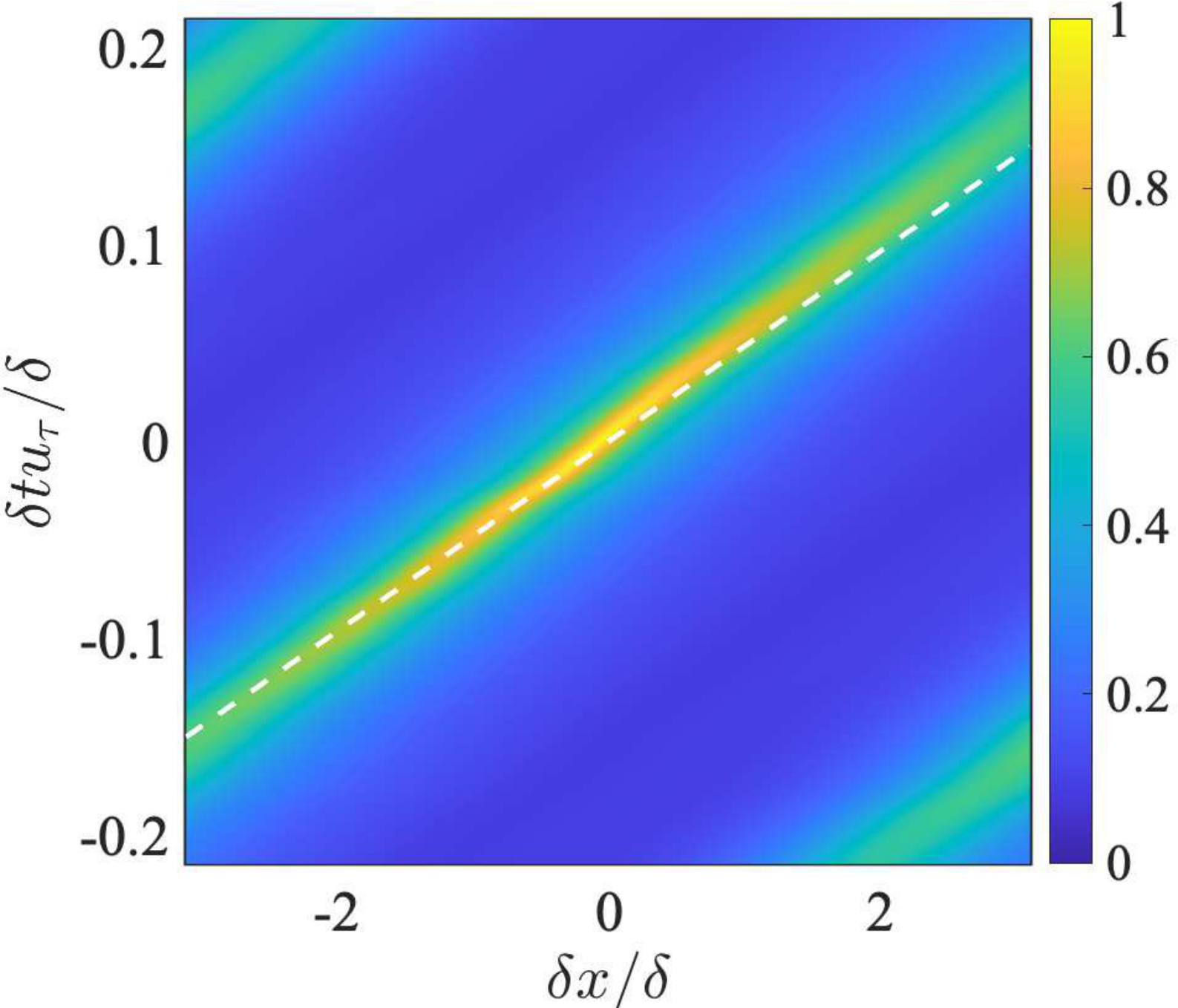}}
\end{center}
\caption{ (a,b) Wall-parallel ($x$-$z$) spacial correlations
  $C^{xz}_{uu}$ and (c,d) space-time ($x$-$t$) correlations
  $C^{xt}_{uu}$ for BL1. The location of $\delta x = 0$ corresponds to ${Re}_\tau=600$. Panels (a) and (c) are for $y_0^+=15$ and the panels (b) and (d) are for $y_0/\delta=0.5$. The white dashed   line in panels (c) and (d) is $\delta x/\delta = \delta t \langle u
  \rangle/\delta$ with $\langle u \rangle$ at the $y_0$ considered.
  \label{fig:TBL:correlation}}
\end{figure}

Several precomputed two-point correlations are also included.  First, spatial correlations in $x$-$z$ planes are provided for each velocity component at $y^+=15$ and $100$ and $y/\delta = 0.1, 0.3,...,0.9$, and $1.1$, and streamwise locations ${Re}_\tau\approx 400, 600$, $700$, and $900$.  For example, the two-point spatial correlation of the streamwise velocity is defined as
\begin{equation}\label{eq:TBL:C}
  C^{xz}_{uu}(\delta x, y_0, \delta z) = \frac{\langle u'(x+\delta x,y_0,z+\delta z,t) u'(x,y_0,z,t)\rangle_{xzt}}{\langle u'(x,y_0,z,t)^2 \rangle_{xzt}},
\end{equation}
where $\langle \cdot \rangle_{xzt}$ represents an average over $x$ (within a streamwise window of size $8\delta$), $z$, and $t$. The correlations for $v$ and $w$ are defined analogously to Eq.~\eqref{eq:TBL:C}. The streamwise velocity correlations are shown in Figs.~\ref{fig:TBL:correlation}a and \ref{fig:TBL:correlation}b for the $x$ region centered at ${Re}_\tau=600$ for BL1. The wall-normal locations selected are $y_0^+=15$ and $y_0/\delta=0.5$, which lie within the inner and outer layer, respectively. The correlations follow the expected trends reported in the literature~\citep{Sillero2014}. Streamwise velocity motions are elongated and flanked by regions of opposite-sign velocity in both the inner and outer layer. Wall-normal and spanwise velocity motions (not shown) are mildly elongated in $x$ within the inner layer and roughly isotropic in the outer layer.
% 
%------------------------------------------------------------------%

Second, space-time correlations in $x$-$t$ planes are provided for each velocity component at $y^+=15$ and $100$ and $y/\delta = 0.1, 0.3,...,0.9$, and $1.1$, and streamwise locations ${Re}_\tau\approx 400, 600$, $700$, and $900$.  The space-time correlation for the streamwise velocity, defined as
\begin{equation}
  C^{xt}_{uu}(\delta x, y_0, \delta t) = \frac{\langle u'(x+\delta x,y_0,z,t+\delta t) u'(x,y_0,z,t)\rangle_{xzt}}{\langle u'(x,y_0,z,t)^2 \rangle_{xzt}},
\end{equation}
is shown in Figs.~\ref{fig:TBL:correlation}c and ~\ref{fig:TBL:correlation}d for BL1. The slope of the band of high correlation in each plot provides a measure of the convection velocity of the fluctuating velocities. At a given wall-normal location, the convection velocity follows approximately the mean velocity profile $\langle u \rangle$ at $y_0$, as marked by the dashed white line in the plots.

% 5: signals
\subsubsection{Planar time-resolved velocity fields}

Lastly, time-resolved velocity fields in the $x-y$ plane (at a fixed spanwise position) are included separately so that they can be accessed without downloading the full volumetric data.  For example, these signals are used in the following causality analysis.

% -- BL DNS analysis -----------------------------------------
\subsection{Example Analysis: cause-and-effect analysis of inner-outer flow motions}

% intro: inner-outer interactions
We leverage the time-resolved nature of the present dataset to study the interactions between the large-scale velocity motions in the outer layer and the near-wall velocity motions.  Previous studies using experimental data of turbulent boundary layer at high Reynolds numbers have revealed the influence of the large-scale boundary-layer motions on the small-scale near-wall cycle~\cite{Hutchins2007, Mathis2009}. To date, the consensus is that the dynamics of the near-wall layer are autonomous~\citep{Jimenez1999}, i.e., large outer-scale motions are not required to sustain realistic turbulence in the buffer layer. Nonetheless, it is agreed that the footprint of the large-scale motions in the inner layer is present via amplitude modulation on near-wall events. Here, we present an alternative approach to investigate the causality between outer and inner scale motions using time-resolved velocity signals and information theory.

\begin{figure}
\begin{center}
\includegraphics[width=1\textwidth]{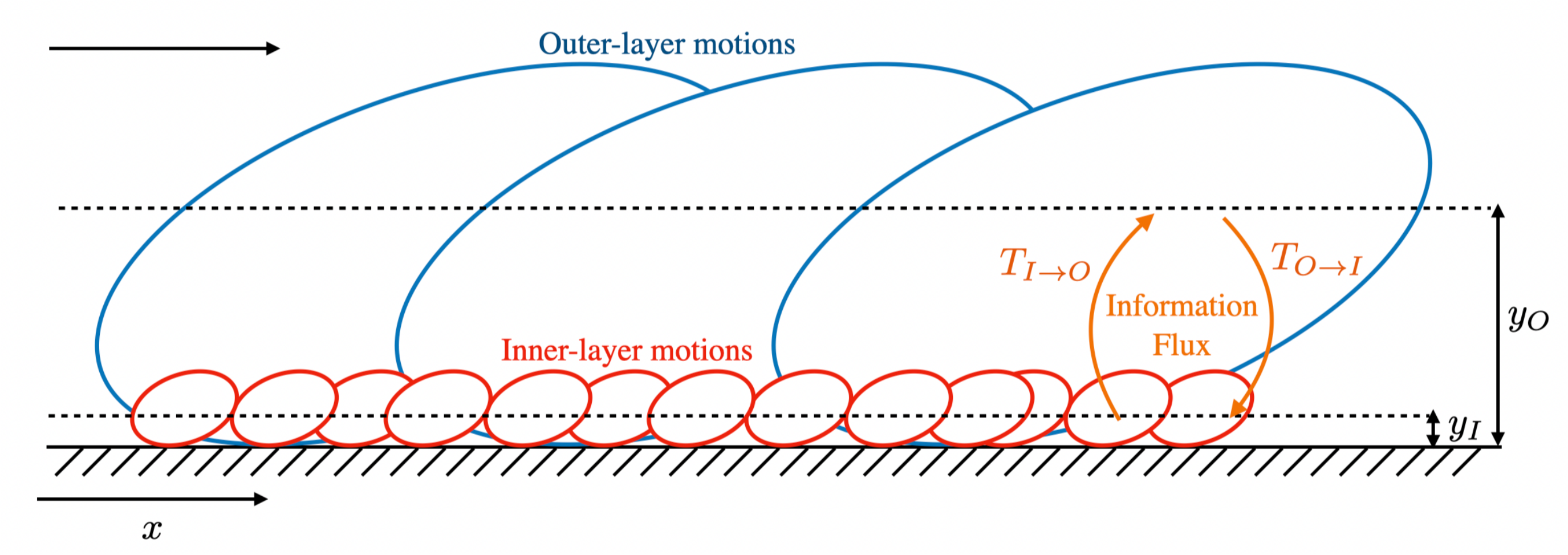}
\caption{ Schematic of outer-layer and inner-layer streamwise velocity
  motions in a turbulent boundary and their interactions via
  information fluxes. \label{fig:TBL:causal_sketch}}
\end{center}
\end{figure}

\begin{figure}
\begin{center}
\includegraphics[width=0.7\textwidth]{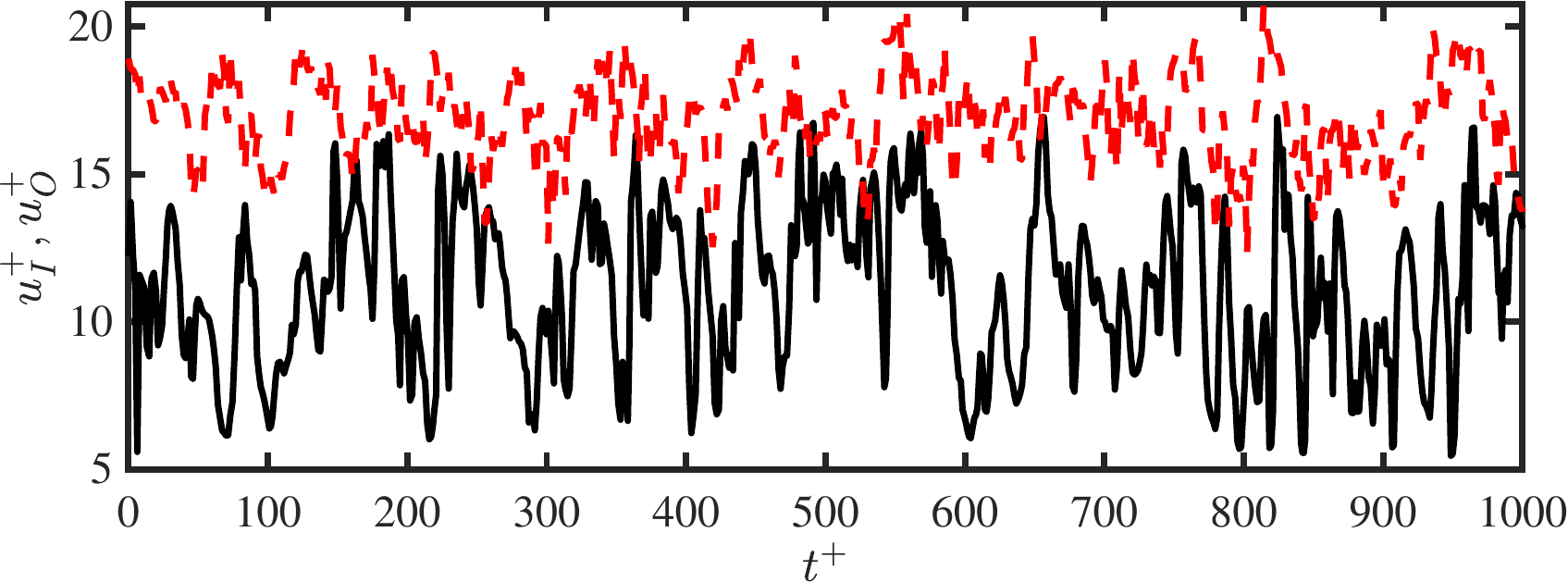}
\caption{ Example of time signals of the streamwise velocity in the
  inner layer $u_I(t)$ at $y^+ = 15$ (solid black) and the outer layer
  $u_O(t)$ at $y/\delta=0.3 (red dashed)$. The signals are extracted at ${Re}_\tau
  \approx 700$ for BL1. \label{fig:TBL:signals}}
\end{center}
\end{figure}

% signal extractions
Figure \ref{fig:TBL:causal_sketch} depicts the setup considered to investigate the causal interactions between outer layer and inner layer velocity motions.
%
%------------------------------------------------------------------%
The time-resolved signals considered are those of the streamwise velocity at the wall-normal heights $y_I^+=15$ (for the inner layer) and $y_O/\delta=0.3$ (for the outer layer) using BL1. The inner and outer layer streamwise velocity signals are denoted by $u_I(t) =
u(x_0,y_I,z_0,t)$ and $u_O(t) = u(x_0,y_I,z_0,t)$, respectively. The streamwise location selected, $x_0$, is such that ${Re}_\tau \approx 700$. Figure \ref{fig:TBL:signals} shows an excerpt of $u_I(t)$ and $u_O(t)$.  Some studies have considered a spatial shift in the streamwise direction between the two signals to increase their correlation~\citep{Howland2018}. Here, we do not apply a spatial shift but instead account for the relative displacement between signals using a time lag.
%
%------------------------------------------------------------------%
%

% tools
The goal is to evaluate the cause-and-effect interactions between $u_I$ and $u_O$. In a broad sense, causality is the mechanism by which one event contributes to the genesis of another~\citep{Pearl2009}.  In the case of turbulence research, causality is usually implied from the cross-time correlation between pairs of time signals representing the events of interest. Here, causality is directly inferred using the information-theoretic framework proposed by \citet{Lozano2022} (see also \citet{Lozano2020}). The approach relies on the quantification of the information flux from the present states of the system to the future states as a proxy for causal inference.  The central quantity of interest is the Shannon information (or entropy)~\citep{Shannon1948} defined for a signal $u(t)$ as
\begin{equation}
H[ u(t) ] = \langle \log_2( p[u(t)] ) \rangle_t,
\label{eq:TBL:H}
\end{equation}
where $p[\cdot]$ is the probability distribution, and $\langle \cdot \rangle_t$ signifies the expected value with respect to time.

% information flux
For the two signals considered, causality from $u_O(t)$ to $u_I(t)$ ($T_{O\rightarrow I}$) is defined as the information flux from the past of $u_O(t)$ to the future of $u_I(t)$~\citep{Lozano2022},
\begin{equation}
T_{O\rightarrow I}(\delta t) = H[ u_I(t) | u_I(t-\delta t) ] - H[
  u_I(t) | u_I(t-\delta t), u_O(t-\delta t) ],
\label{eq:TBL:IF}
\end{equation}
where $\delta t$ is the time-lag to evaluate causality, and $H[ u_I(t) | u_I(t-\delta t) ]$ is the conditional Shannon information. The latter can be interpreted as the uncertainty in the variable $u_I(t)$
given the past of $u_I(t)$ (similarly for $H[ u_I(t) | u_I(t-\delta t), u_O(t-\delta t) ]$).  The Shannon information of $u_I(t)$ conditioned to $u_I(t-\delta t)$ and $u_O(t-\delta t)$ is given by
\begin{equation}
H[u_I(t) | u_I(t-\delta t), u_O(t-\delta t)] =
\langle\log_2( p[u_I(t) , u_I(t-\delta t), u_O(t-\delta t)]\rangle_t -
\langle\log_2( p[ u_I(t-\delta t), u_O(t-\delta t)]\rangle_t.
\end{equation}
Four causal interactions can be defined: the self-induced causality of the signals, $T_{O\rightarrow O}$ and $T_{I\rightarrow I}$, and the cross-induced causality between signals, $T_{I\rightarrow O}$ and $T_{O\rightarrow I}$. The latter represents the interaction between outer and inner layer motions.

\begin{figure}
\begin{center}
  \sidesubfloat[]{\includegraphics[width=0.47\textwidth]{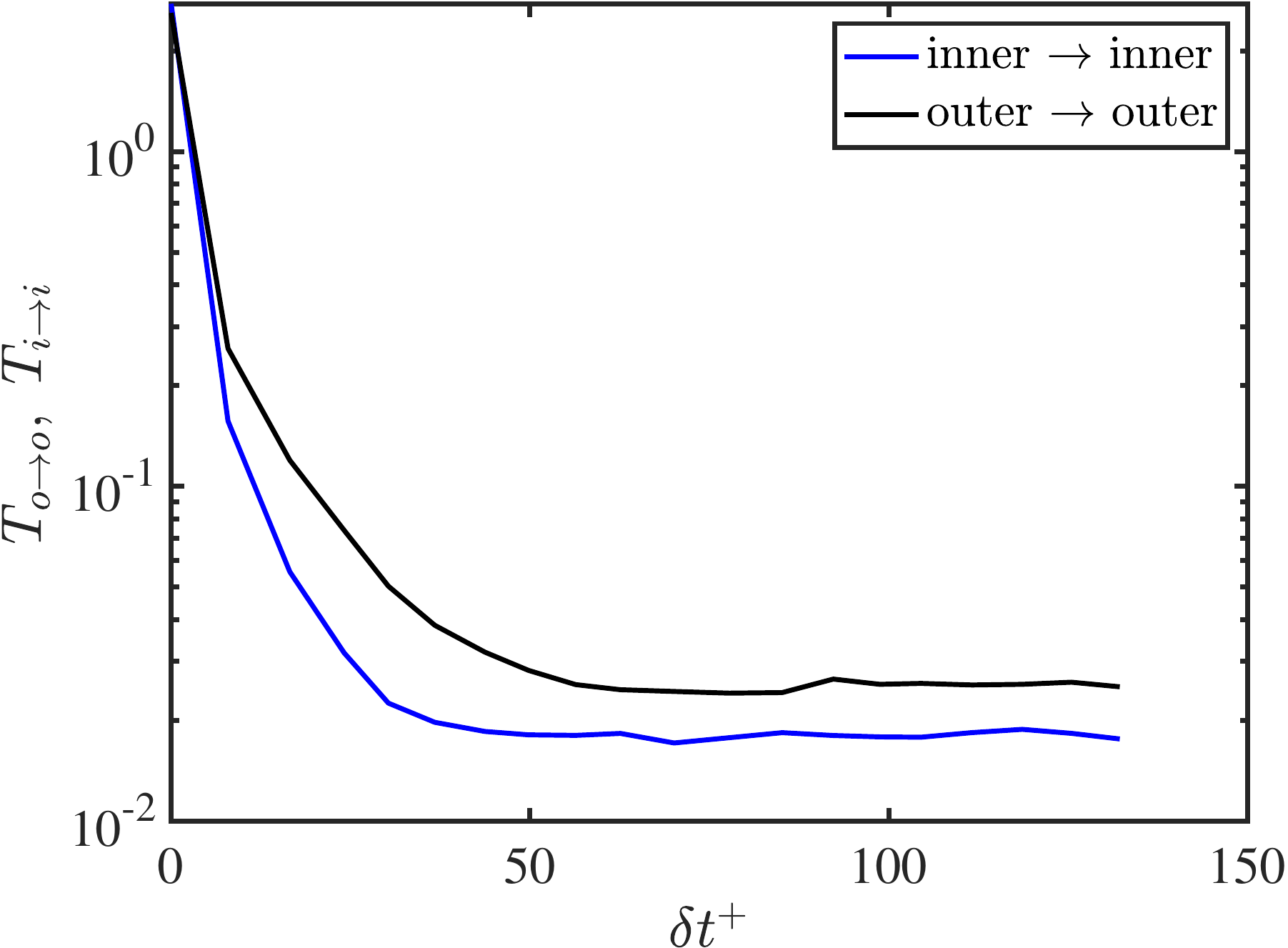}}
  \sidesubfloat[]{\includegraphics[width=0.47\textwidth]{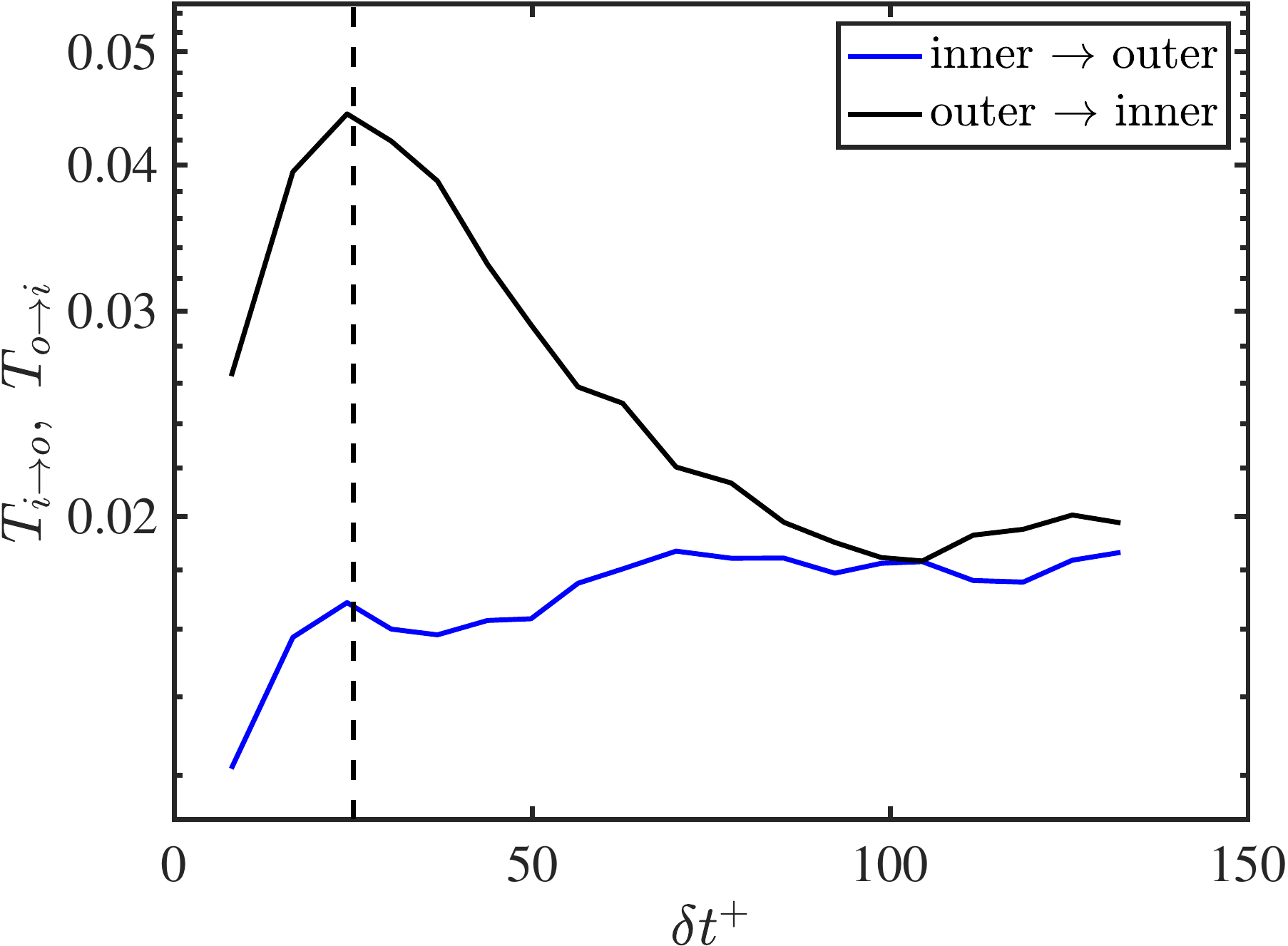}}
\end{center}
\caption{ Causality of inner and outer layer streamwise velocity motions: (a)
self- and (b) cross-information fluxes as a function of the time lag
for causal inference. The vertical dashed line is $\delta t^+ = 25$.\label{fig:TBL:causality}}
\end{figure}

% results
The self- and cross-information fluxes are compiled in Figure \ref{fig:TBL:causality} as a function of the time lag $\delta t$. Self-induced causalities $T_{O\rightarrow O}$ and $T_{I\rightarrow I}$ dominate, especially for short time scales. This is expected, as variables are mostly causal to themselves, i.e., they contain most of the information about their future.  Regarding inner-outer interactions, there is a clear peak in the information flux $T_{O\rightarrow I}$ from the outer motions to the inner motions at $\delta t^+ \approx 25$ which is absent in the reverse direction $T_{I\rightarrow O}$. The result supports the prevalence of top-down interactions: the information flows predominantly from the outer-layer large-scale motions to inner-layer small-scale motions. The outcome is consistent with the modulation of the near-wall scales by large-scale motions reported in previous investigations~\citep[e.g.][]{Hutchins2007, Mathis2009}.
%
%------------------------------------------------------------------%
%

% concluding remarks
Compared to previous investigations, our method is purposely devised to account for the time dynamics of the signals and provides the time scale for maximum causal inference.  Moreover, the present approach only requires the inner/outer time signals without any further manipulation. The information flux is based on probability distributions and, as such, is invariant under shifting, rescaling and, in general, nonlinear $C^1$-diffeomorphism transformations of the signals~\citep{Kaiser2002, Lozano2022}. Hence, our approach unveils in a simple manner the interactions between inner and outer layer velocity motions while minimizing the number of arbitrary parameters typically used in previous studies such as the type of signal transformation, filter width, reference convection velocity, selection of parameters for the extraction of a universal signal, etc.  This is an example of a modern method requiring the time-resolved data provided in this dataset.

%%%%%%%%%%%%%%%%%%%%%%%%%%%%%%%%%%%%%%%%%%%%%%%%%%%%%%%%%%%
% -- BL EXP Section (Tess) ------------------------------

\section{Turbulent boundary layer planar particle image velocimetry}
\label{sec:BL_EXP}

Third, we include an experimental counterpart of the zero-pressure-gradient turbulent boundary layer.  Experimental data present additional challenges, e.g., experimental noise/uncertainties and limited measurements, that are often overlooked when working with simulation data but that, in many cases, should be considered in the design and practical deployment of reduced-complexity models.  Providing both simulation and experimental data for the turbulent boundary layer will provide opportunities for users to explore these issues.  The dataset contains 6000 time-resolved planar velocity fields measured using PIV for five different Reynolds numbers.  The high temporal resolution and large field of view enable the study of convecting large-scale turbulent structures using spatial and temporal data analysis techniques.

% -- BL EXP setup -------------------------------------------
\subsection{Setup}

\begin{figure}[t]
    \centering
    \includegraphics[width=\textwidth]{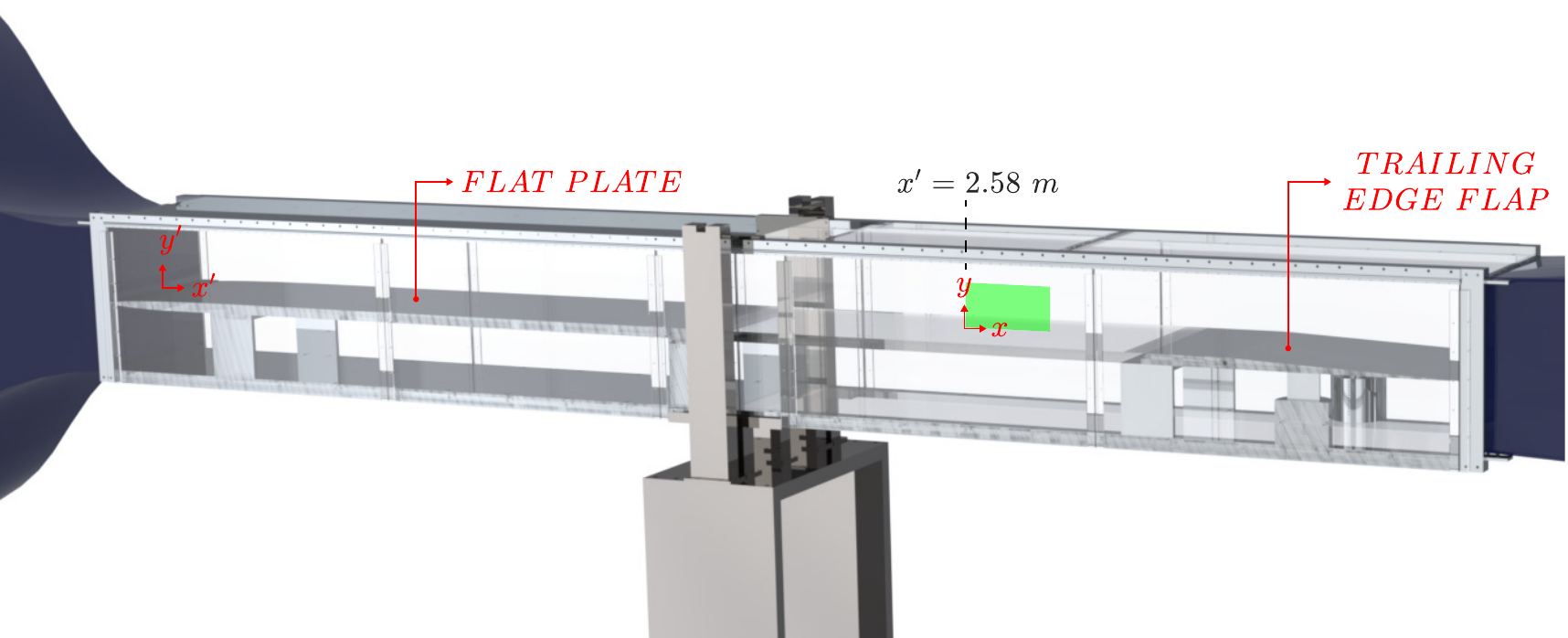}
    \caption{Schematic of the wind tunnel test section. The field of view for PIV is indicated by the green box.}
    \label{fig:testsection}
\end{figure}

Experiments were run in the boundary layer wind tunnel facility at the University of Illinois Urbana-Champaign. The wind tunnel has a 4-inch honeycomb straightener, four turbulence-reducing screens, and a 27:1 contraction area ratio. The test section is 15 inches by 15 inches in cross-section and 12 feet in length. The freestream turbulence intensity is approximately 0.5\%. A turbulent boundary layer was tripped and then developed over 2.5 m on a flat plate centered in the wind tunnel section, as shown in Fig.~\ref{fig:testsection}, before reaching the PIV field of view. The boundary layer developed in a nominally zero-pressure-gradient environment \citep{RodriguezThesis}. The wind tunnel was run at five speeds to yield data at five Reynolds numbers. The velocity field was measured using time-resolved planar PIV in the streamwise--wall-normal ($x$--$y$) plane (Fig.~\ref{fig:testsection}). The flow was seeded with oil particles of approximately 1$\mu$m diameter. A Continuum Terra PIV Nd:YLF 527-80-M laser was used to illuminate the flow. A Phantom VEO 710L camera was used at maximum resolution, fitted with a $50$ mm Nikon lens with $f$/1.4, to capture image pairs at 3.755 kHz. The data were analyzed using DaVis 10.1 software from LaVision. Three passes with an interrogation window of $32 \times 32$ at 50$\%$ overlap, followed by three passes with a final window size of $16 \times 16$ at 75$\%$ overlap were used in computing the vector fields from the particle image pairs. As a result, each instantaneous snapshot contains 319 vectors in the streamwise direction and 150 vectors in the wall-normal direction.  The following flow parameters for each case are provided in Table \ref{tab:parameters}: the 99\% boundary layer thickness, $\delta$, the friction velocity, $u_\tau$, the free stream velocity, $U_\infty$, and the friction Reynolds number, $Re_\tau \equiv \delta u_\tau / \nu$, where $\nu$ is the kinematic viscosity. The $u_\tau$ for each case was computed using the Clauser method \citep{clauser1956}. 

\begin{table}[t]
\setlength{\tabcolsep}{18pt}
\centering
\begin{tabular}{ccccc}
\hline
{Case} & {${Re_\tau}$} & ${\delta}$ {(mm)} & ${u_\tau}$ {(m/s)} & ${U_\infty}$ (m/s) \\
\hline
         1 & 605    & 49    & 0.1827 & 3.71     \\ 
         2 & 987    & 42    & 0.3474 & 7.73     \\ 
         3 & 1373   & 39    & 0.5210 & 12.07    \\ 
         4 & 1782   & 37    & 0.7055 & 16.77    \\ 
         5 & 2227   & 36.7  & 0.8928 & 21.98    \\ \hline
\end{tabular}
    \caption{Velocity parameters for each experimental case: friction Reynolds numbers ($Re_{\tau})$,  99\% boundary layer thickness ($\delta$), friction velocity ($u_{\tau})$, free stream velocity ($U_{\infty}$).}
    \label{tab:parameters}
\end{table}

% -- BL EXP data --------------------------------------------
\subsection{Data}

\subsubsection{Time-resolved planar velocity fields}

\begin{figure}
    \centering
    \includegraphics{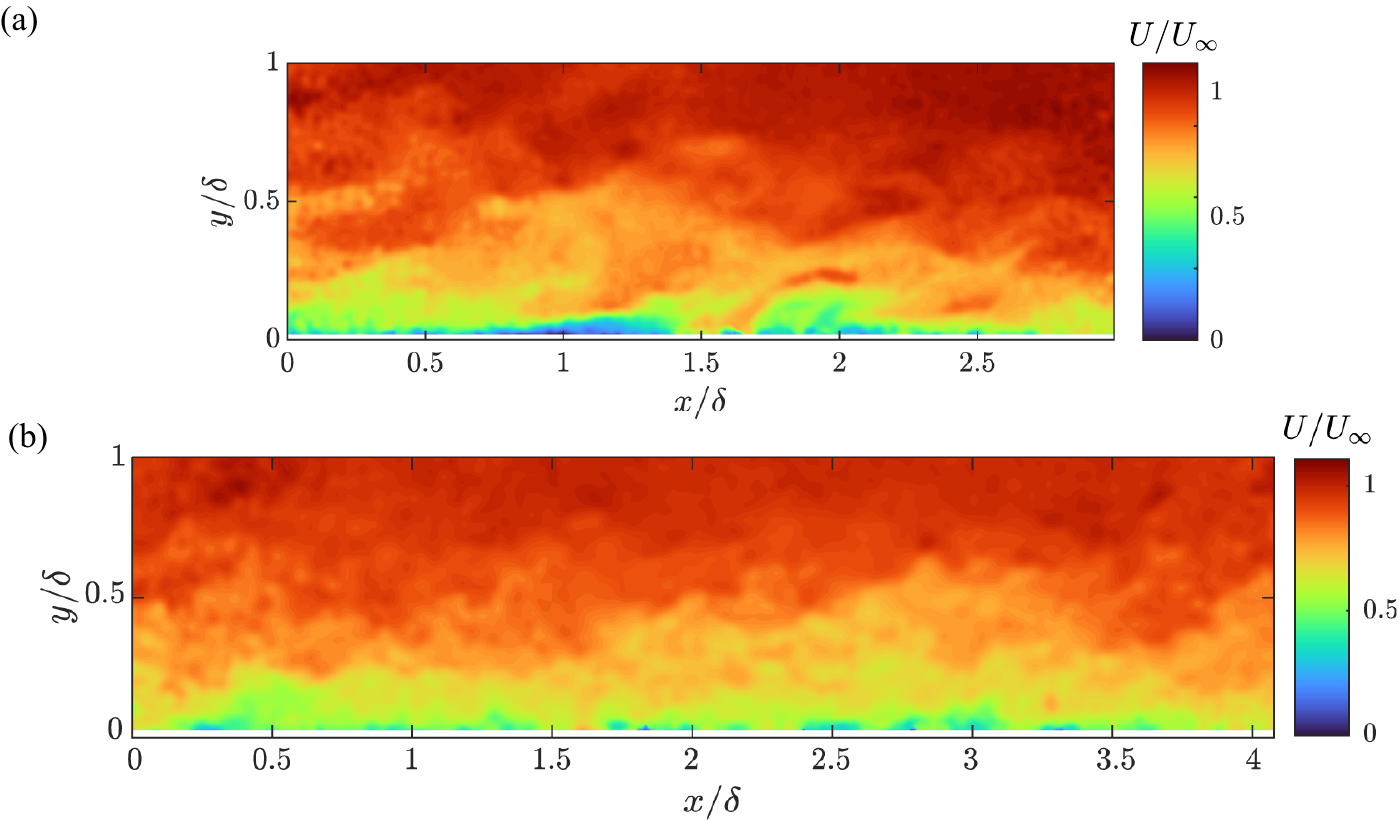}
     \caption{Representative instantaneous snapshot of the streamwise velocity field for (a) $Re_\tau \approx 600$ and (b) $Re_\tau \approx 2200$.}
    \label{fig:snapshotEXP}
\end{figure}

The data include 6000 time-resolved planar snapshots of the streamwise and wall-normal velocity fields for each Reynolds number. Representative snapshots are shown in Fig.~\ref{fig:snapshotEXP} for the $Re_\tau$ = 600 and 2220 cases. The field of view of data collection was chosen to visualize the large-scale motions of the boundary layer while retaining spatial resolution within the boundary layer. The field of view includes 3 to 4 boundary layer thicknesses in the streamwise direction, and the spatial resolution varied from 4 viscous units to 23 viscous units, depending on the Reynolds number. A high temporal resolution was sought to enable spectral analyses of the data. The temporal resolution varied between 0.4 viscous units to 12 viscous units, depending on the Reynolds number. As the freestream velocity increased with Reynolds number, the number of eddy turnover times included in the set of 6000 snapshots provided decreased from 50 to 6. The values for each case are provided in table \ref{tab:data} for the spatial resolution, the field of view, the temporal resolution, and the number of included eddy turnover times. 

\begin{table}[t]
\setlength{\tabcolsep}{18pt}
\centering
\begin{tabular}{cccccc}
\hline
%\textbf{Case} & $\boldsymbol{\Delta l/ \delta}$ & $\boldsymbol{\Delta l^+}$ & \textbf{FOV} $\boldsymbol{/ \delta}$ & $\boldsymbol{\Delta t^+}$ & \textbf{ETT} \\ 
{Case} & ${\Delta l/ \delta}$ & ${\Delta l^+}$ & {FOV} ${/ \delta}$ & ${\Delta t^+}$ & {ETT} \\ 
\hline
         1 & 0.0094  & 4.43     & 3.00 $\times$ 1.40    & 0.44  & 50    \\ 
         2 & 0.011   & 8.87     & 3.49 $\times$ 1.64    & 1.78  & 20    \\ 
         3 & 0.012   & 13.31    & 3.76 $\times$ 1.76    & 4.02  & 12    \\ 
         4 & 0.012   & 18.03    & 3.97 $\times$ 1.86    & 7.36  & 8     \\ 
         5 & 0.012   & 22.82    & 4.00 $\times$ 1.87    & 11.8  & 6     \\ \hline
\end{tabular}
    \caption{Parameters for each experimental case: spatial resolution in outer units ($\Delta l/ \delta$), spatial resolution in inner units ($\Delta l^+$), field of view (FOV), temporal resolution ($\Delta t^+$), and eddy turnover times (ETT).}
    \label{tab:data}
\end{table}

\subsubsection{Mean velocity fields}

\begin{figure}[ht]
    \centering
    \includegraphics{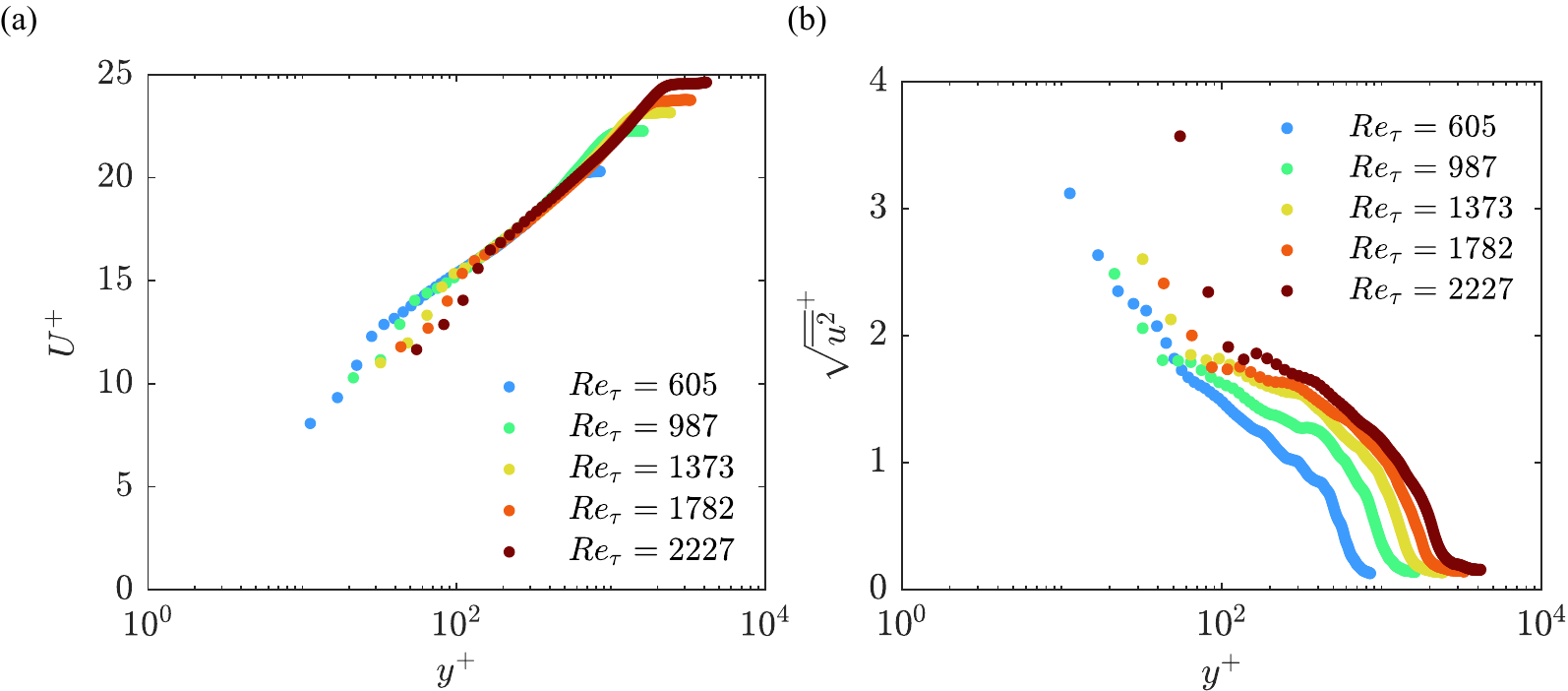}
    \caption{Time-averaged statistics scaled in inner units: (a) mean streamwise velocity and (b) RMS of streamwise velocity.}
    \label{fig:mean}
\end{figure}

The dataset also contains precomputed mean streamwise and wallnormal velocity fields.  The mean streamwise velocity profile at the center of the field of view is shown in Fig.~\ref{fig:mean}a for each Reynolds number. The mean profiles show good agreement in the outer region but show evidence of corruption in the first three data points nearest to the wall due to the reflection of the laser sheet on the surface. In the lowest Reynolds number case, this puts the first trusted data point at $y/\delta = 0.047$ or $y^+ = 28$, while in the highest Reynolds number case, the first trusted data point is at $y/\delta = 0.064$ or $y^+ = 165$. Experimental sources of noise provide opportunities to test data analysis techniques for their robustness to realistic experimental challenges.  Corresponding RMS streamwise velocity profiles are shown in Fig.~\ref{fig:mean}b.  
%The spectral signature of the flow was estimated from the 6000 snapshots of the data using Welch's method and is shown for $Re_\tau$ = 600 and 2220 in Fig.~\ref{fig:spectra}. The spectral signature shows evidence of a growing outer peak at higher Reynolds numbers.

%\begin{comment}
%\end{comment}

% -- BL EXP analysis -----------------------------------------
\subsection{Example analysis}

Next, we provide an example of how this dataset can be used to demonstrate a new reduced-complexity model or, conversely, how the data can be analyzed using a reduced-complexity model.  Specifically, we apply conditional projection averaging (CPA) \citep{saxton2022amplitude}, a form of conditional averaging that allows the user to extract how a behavior at a specified time (and length) scale in the data affects other features of the flow. Here, spectral proper orthogonal decomposition (SPOD) \cite{Lumley1970stochastic, towne2018spectral} is used to generate a model of a structure associated with a single time scale. Using CPA, we then conditionally average on the existence and location of that structure in the instantaneous data. For CPA, one can use any reduced-complexity modeling method that generates modes that have an associated phase, including SPOD, resolvent analysis, and dynamic mode decomposition \cite{saxton2022amplitude}.

Because the interaction between large- and small-scale structures is of interest to the turbulence community, CPA is used here to illustrate a relationship between the phase of a passing large-scale turbulent structure and the local intensity of small-scale turbulence activity. To apply CPA, the instantaneous, fluctuating streamwise velocity data, $u$, are projected onto a model, $\tilde{u}$, representing a large-scale structure. The projection coefficient, $R$, for each snapshot, $t_i$, is recorded at each model phase, $\tilde{\gamma}_j$, 
% \begin{equation}
% R(t_i,\tilde{\gamma}_j) \equiv \left(\frac{u(x,y,t_i)\cdot\tilde{u}(x,y,\tilde{\gamma}_j)}{|u(x,y,t_i)||\tilde{u}(x,y,\tilde{\gamma}_j)|} \right).
% \label{eq:Rgen}
% \end{equation}
\begin{equation}
R(t_i,\tilde{\gamma}_j) \equiv \left(\frac{\langle u(x,y,t_i),\tilde{u}(x,y,\tilde{\gamma}_j)\rangle}{\|u(x,y,t_i)\| \|\tilde{u}(x,y,\tilde{\gamma}_j)\|} \right),
\label{eq:Rgen}
\end{equation}
where $\langle \cdot\, , \cdot \rangle$ is the inner product used to define the projection and $\| \cdot \|$ is the induced norm. If the maximum projection coefficient is above a threshold value, $R_{th}$, then something similar to the model is argued to have been identified in that snapshot of the data. The phase that maximizes the projection coefficient is identified as the approximate phase, $\gamma$, of the identified structure in that snapshot. All snapshots labeled with the same phase are averaged together. For a given field variable, $q$, CPA can be defined as
\begin{equation} 
%\begin{split}
\langle q\rangle_{CPA}(x,y,\gamma)= \frac{1}{N}\sum_{t_i} q(x,y,t_i) \;
\forall t_i \quad \text{  s.t. } \quad  R(t_i,\gamma) > R(t_i,\tilde{\gamma}_j \neq \gamma), \quad 
R(t_i,\gamma) > R_{th},
%\end{split}
\label{eq:phaseaverage}
\end{equation}
where $N$ is the total number of snapshots included in the average \cite{saxton2022amplitude}.

\begin{figure}
    \centering
    \includegraphics{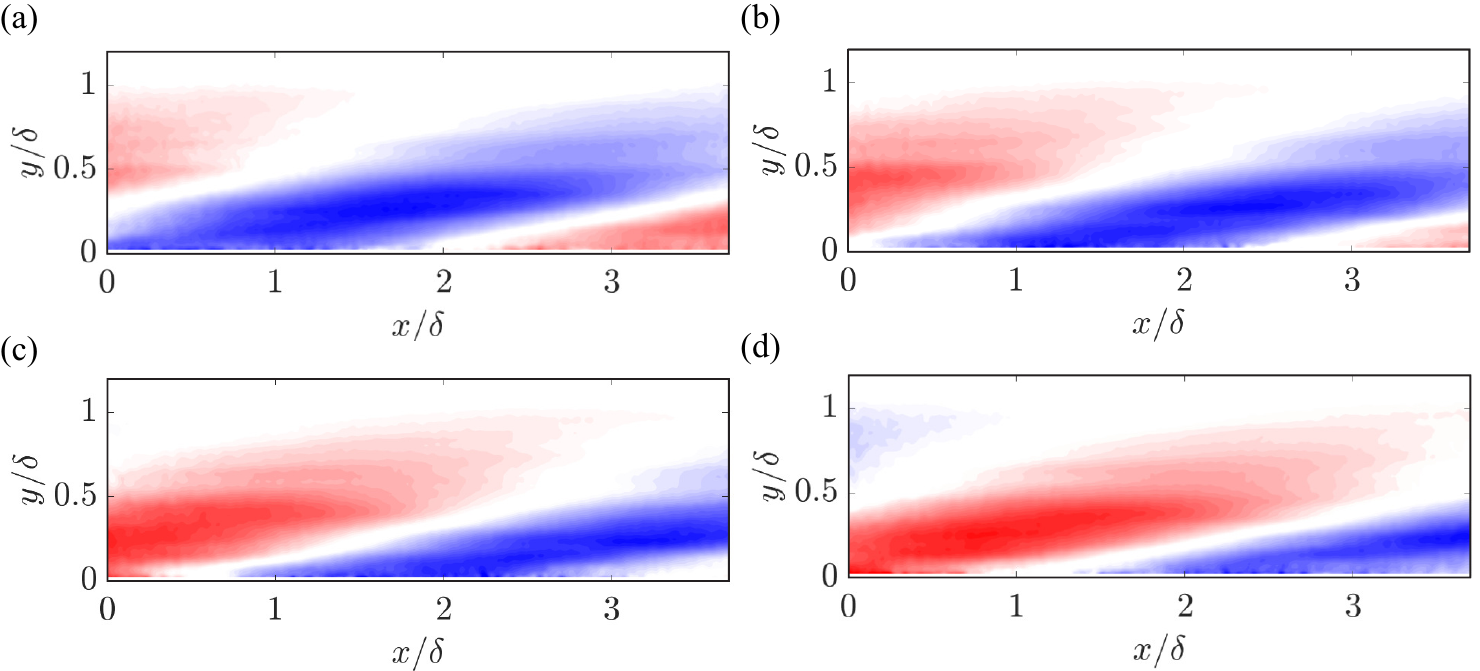}
    \caption{The SPOD mode used as a model for conditional projection averaging at four phases, showing the convection of the large-scale structure.}
    \label{fig:SPOD}
\end{figure}

To generate a model, $\tilde{u}$, the time-resolved data from the $Re_\tau \approx 1370$ case were used to calculate an SPOD mode at a frequency of $f \delta / u_\tau = 4.2$ and an estimated streamwise wavelength $\lambda_x /\delta \approx 6$. This frequency was chosen to approximate the superstructures of the boundary layer \cite{kim1999very, montystatistics}, a relevant large-scale structure that is known to affect small scales locally \cite{Hutchins2007}. The SPOD mode with the largest eigenvalue at that frequency is shown in Fig.~\ref{fig:SPOD} at four different phases. A large, downstream-leaning structure is observed. The instantaneous velocity data were projected onto the mode at eight phases, using Eq.~\ref{eq:Rgen}, with a threshold projection coefficient of $R_{th} = 0.4$. 

\begin{figure}
    \centering
    \includegraphics{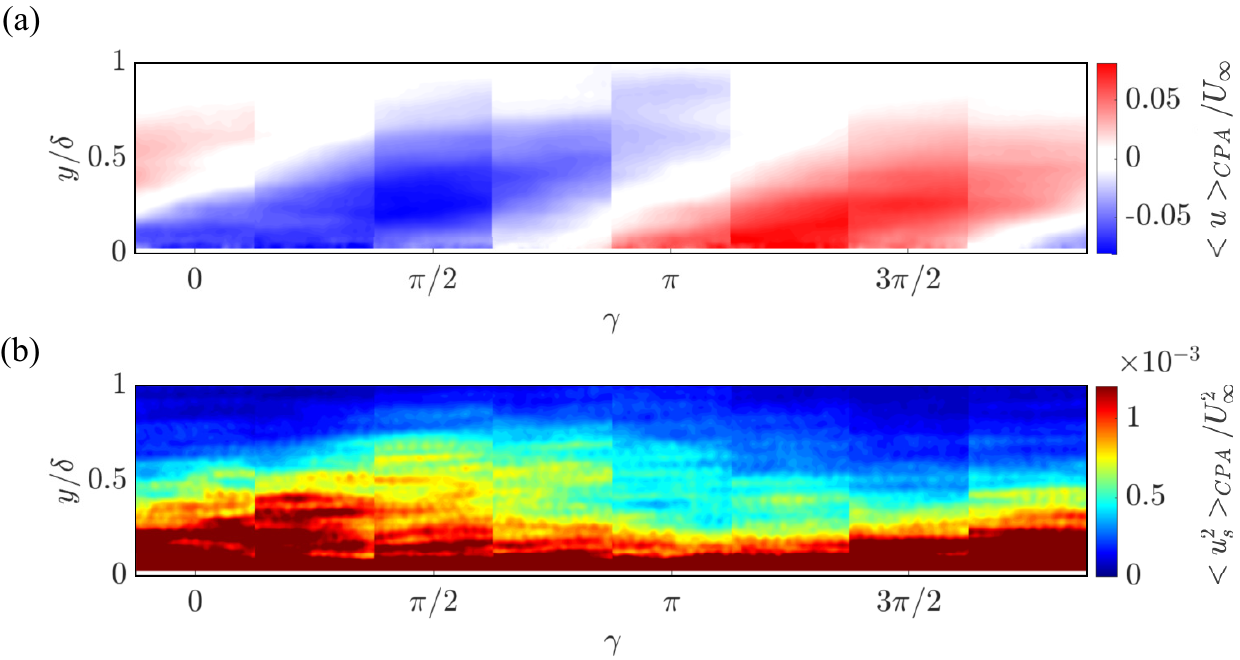}

    \caption{Streamwise fluctuating velocity (a) and small-scale streamwise velocity intensity (b) averaged using conditional projection averaging.}
    \label{fig:CPA}
\end{figure}

  To study the effect of large-scale turbulent structures on small-scale turbulent behavior, the small-scale streamwise velocity field was defined as $u_s(x,y,t) = u(x,y,t)-u_l(x,y,t)$, where $u$ was the fluctuating streamwise velocity field and $u_l$ was defined using a convolution between the fluctuating streamwise velocity field and a Gaussian kernel with a standard deviation of $\sigma_G/\delta = 0.125$ \citep{saxton2022amplitude}. Filtering the velocity field to identify small-scale components is a commonly-used approach when studying the interaction of large- and small-scale turbulent structures \citep{Hutchins2007,ChungMcKeon2010}, and the results have been shown to be fairly insensitive to the details of the filter \citep{gana_hutchins_etal_2012}. The fluctuating streamwise velocity field and the small-scale streamwise velocity intensity were then averaged by setting $q=u$ and $q=u_s^2$ in Eq.~\ref{eq:phaseaverage}, respectively. Note that regardless of the field variable averaged, the definition of the projection always uses the fluctuating streamwise velocity field (Eq.~\eqref{eq:Rgen}).  To enable visualization of the full period of the large-scale phase variation, a portion of each phase average was stitched together to create a single, paneled image. The averaged fluctuating streamwise velocity is shown in Fig.~\ref{fig:CPA}a and the averaged small-scale streamwise velocity intensity is shown in Fig.~\ref{fig:CPA}b. The horizontal axis is the approximate phase of the extracted scale associated with the model.

The averaged fluctuating streamwise velocity field is shown in red and blue contours in Fig.~\ref{fig:CPA}a. The averaged field looks similar to the original mode, suggesting that the projection and averaging successfully isolated the scale of interest at the appropriate phases. The similarity can be quantified by projecting the averaged field of Fig.~\ref{fig:CPA}a onto the mode at the appropriate phase in Fig.~\ref{fig:SPOD}; the average projection coefficient was 0.91, with a maximum across the phases of 0.94 and a minimum of 0.82. The averaged small-scale streamwise velocity intensity is shown in Fig.~\ref{fig:CPA}b and shows a change of intensity depending on the local phase of the large-scale structure. Small scales are observed to be more intense far from the wall when the large scale is in a low-speed phase and more intense near the wall when the large scale is in a high-speed phase. The observed relationship between the large- and small-scale features of turbulence is consistent with previous work averaging using resolvent modes \citep{saxton2022amplitude}, averaging using pointwise conditions \citep{ChungMcKeon2010}, and amplitude modulation statistics \citep{Hutchins2007}. These results provide insight into the spatial organization between large- and small-scale structures in turbulent flows \citep{saxton2022amplitude}.

The CPA analysis demonstrates the nonlinear effect that one scale has on other features in the boundary layer. Spectral proper orthogonal decomposition provides an energetically optimal mode to represent a specific frequency in the data. By projecting onto that mode at different phases and using that information for conditional averaging, CPA identifies the effects that the flow behaviors represented by the SPOD mode have in the boundary layer. CPA can be used to ask questions about how one scale affects other processes in a complex, chaotic flow. 

The data provided here made CPA analysis possible in multiple ways. First, the time resolution permitted the computation of spectral proper orthogonal decomposition modes, which created a representation of the structure of a single frequency in the flow. Second, the large field of view improved the accuracy of the phase estimation from projection. Third, the long time series yielded sufficient convergence to observe expected trends. Finally, the success of SPOD and CPA with this data demonstrates the robustness of each method to experimental noise.

%%%%%%%%%%%%%%%%%%%%%%%%%%%%%%%%%%%%%%%%%%%%%%%%%%%%%%%%%%%
% -- Airfoil DNS Section (Scott) --------------------------

\section{Low-Reynolds-number pitching airfoil direct numerical simulations}  % e.g., "Turbulent jet large eddy simulation"
\label{sec:airfoil_DNS}

Fourth, we consider a set of low-Reynolds-number sinusoidally pitching flat-plate airfoils.  The wings of small  aircraft, both human-made and biological, can be subject to large and rapid motions, leading to substantial fluctuations of aerodynamic forces.  Accurate modeling is essential for reliable and efficient flight of such aircraft. These unsteady aerodynamic flows can in general exhibit a range of distinct phenomena \cite{mccroskey1982annrev}, such as flow separation and reattachment, dynamic stall vortex formation, and instabilities in both the suction surface shear layer and wake.  Accurate reduced-complexity models might typically need to capture the presence and interaction between these phenomena to enable high quantitative accuracy. Periodic (or quasi-periodic) motions can arise due to aeroelastic flutter or prescribed airfoil motion (such as a flapping wing). Considering sinusoidal motions naturally allows for modeling and analysis to be conducted in the frequency domain, a common approach in unsteady aerodynamics dating back to the classical models of \citet{Theodorsen:35} and \citet{sears1941some}.  The dataset contains time series of velocity and vorticity fields and aerodynamic forces for a variety of pitching amplitudes and frequencies computed via DNS.

\subsection{Setup}

The Reynolds number (based on the freestream velocity and  airfoil chord length) is set to 100.  
The dataset focuses on two base angles of attack, $\alpha_0 = 25^\circ$ and $30^\circ$.  In both cases, the flow is entirely separated over the suction surface.  At $\alpha_0 = 25^\circ$, the flow field is stable in the absence of airfoil motion, while at $\alpha = 30^\circ$ the wake is unstable, exhibiting periodic vortex shedding.  We consider airfoil kinematics of the form
\begin{equation}
\label{eq:periodicpitching}
\alpha(t) = \alpha_0 -  \alpha_P \sin(2\pi f_P t),
\end{equation}
where $f_P = f c/U_\infty$ is a dimensionless frequency,  $c$ is the airfoil chord, and $U_\infty$ is the freestream velocity. 

\begin{figure}
\begin{center}
\sidesubfloat[]{\includegraphics[width=0.45\textwidth]{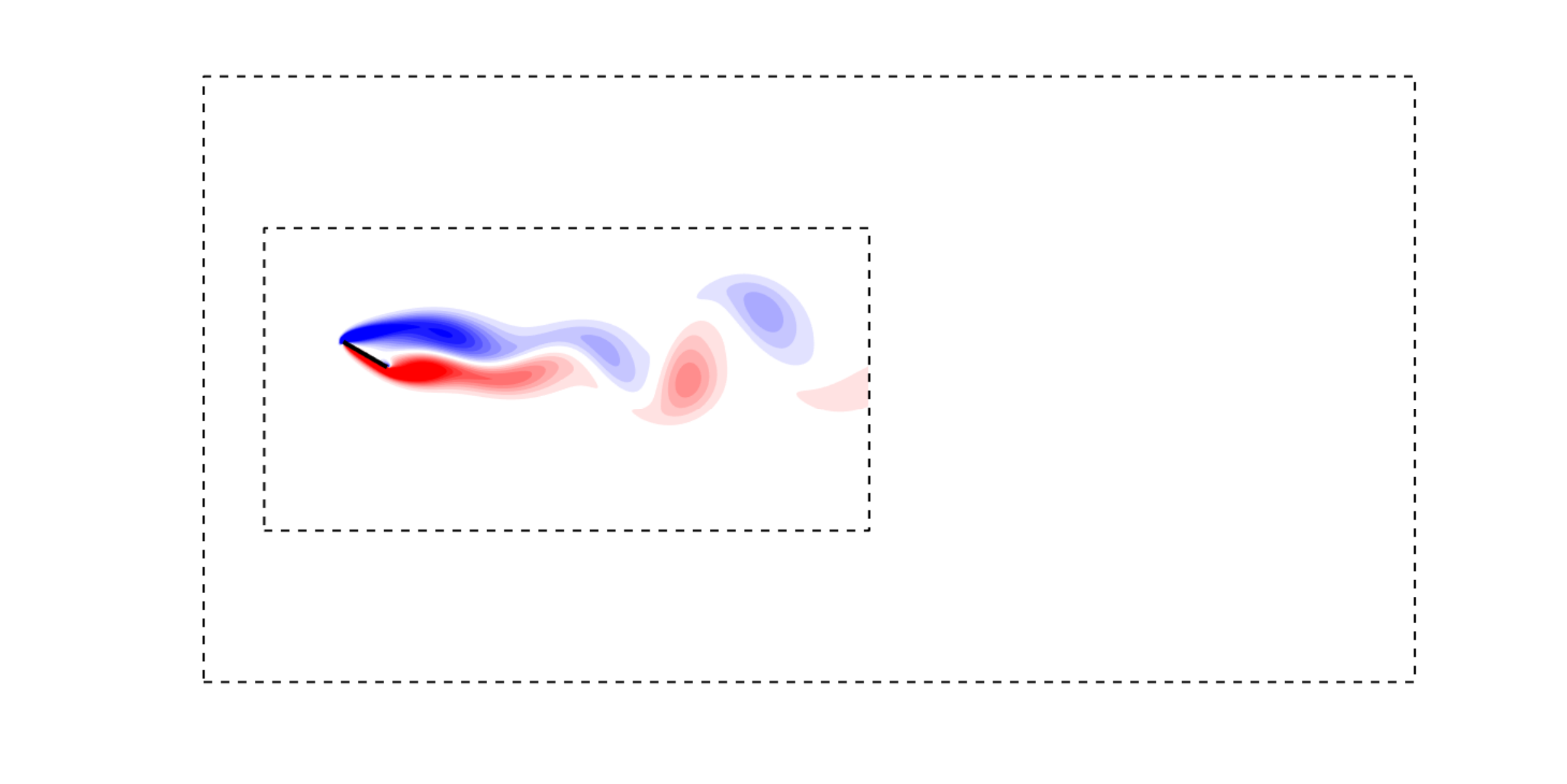}}
 \sidesubfloat[]{\includegraphics[width=0.4\textwidth]{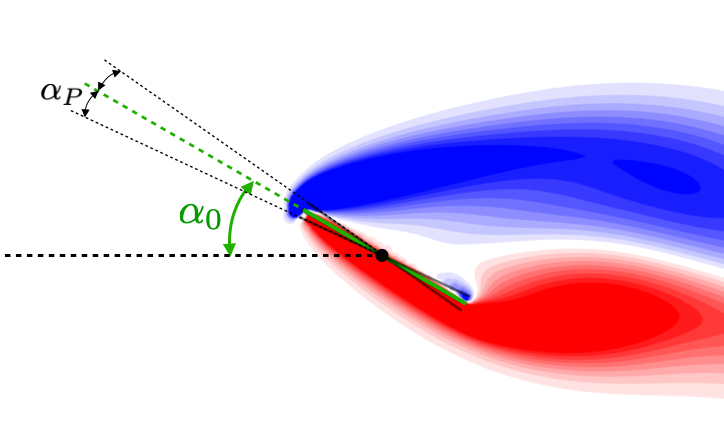}}
\caption{(a) Part of the computational domain used to generate the low Reynolds number airfoil dataset, with vorticity field shown where data is collected. (b) Diagram showing airfoil pitching kinematics.}
\label{fig:AirfoilDomain}
\end{center}
\end{figure}

Direct numerical simulations of the incompressible Navier-Stokes equations are performed using an immersed boundary projection method, following the approaches described in  \citet{taira:07ibfs} and \citet{taira:fastIBPM}.  With this method, the fluid computations are solved on a set of four nested uniform spatial grids, with the finest grid having a spacing of $0.02c$ in both the $x$- and $y$-directions.  Each  grid consists of $600\times300$ grid points, with the spatial extent doubling (and thus the resolution halving) from one grid to the next.  The total computational domain extends 96 and 48 chord lengths in the streamwise and transverse directions, respectively. %Flow field data is provided for the  
Fig.~\ref{fig:AirfoilDomain}a shows the locations of the smallest two grids, with a sample vorticity field plotted within the smallest grid, which is where flow-field data is collected. Fig.~\ref{fig:AirfoilDomain}b shows the range of motion for the pitching airfoil kinematics, as described in Eq.~\eqref{eq:periodicpitching}.

%This computational domain is shown in Fig.~\ref{fig:AirfoilDomain}

%\begin{figure}
%\begin{center}
%(a)\includegraphics[width=0.5\textwidth]{DomainPlotDGDatabaseUpdated}
%\caption[Computational domain used for this study, with the size and location of the airfoil shown]{Computational domain used for the low Reynolds number airfoil dataset, with the size and location of the airfoil shown. Dashed lines represent the borders of each nested grid. flow field data is provided for region within the red dashes.}
%\label{fig:AirfoilDomain}
%\end{center}
%\end{figure}

The boundary conditions of the airfoil are imposed by applying body forces at locations along the surface of the moving airfoil.  These body forces (which are spaced 0.02c apart) are regularized over a small spatial area, and need not conform to the fixed underlying grid.  
Timestepping is performed using Crank-Nicholson and third-order Runge-Kutta time steppers for the linear and nonlinear terms in the governing equations, respectively. A dimensionless timestep of $0.01c/U_\infty$ is used throughout. Simulations are run for 50 convective time units before data is collected, to eliminate transient startup effects. The computational setup and data is similar to that described in Refs.~\cite{dawson2016lift,dawson2017thesis}.

% Timestepping is performed using Crank-Nicholson and third-order Runge-Kutta time steppers for the linear and nonlinear terms in the governing equations, respectively. A dimensionless timestep of 0.01 is used throughout. Simulations are run for 50 convective time units before data is collected, to eliminate transient startup effects.  Data is then collected for 40 convective time units. The computational setup and data is similar to that described in \cite{dawson2016lift}.

\subsection{Data}

This dataset considers pitching amplitudes of $\alpha_P = 0^\circ$ (i.e., a stationary airfoil) and $5^\circ$. In all cases, pitching is about the midchord of the airfoil, which has coordinates $(x,y) = (0.5,0)$. The dimensionless pitching frequency $f_p = f c/U_\infty$ takes values from the set $\{0.05, 0.1, 0.2, 0.25, 0.3, 0.35, 0.4, 0.5\}$.  As a point of reference, the natural vortex-shedding frequency at an angle of attack of $ 30^\circ$ is $f = 0.2418$.

\subsubsection{Time-resolved velocity and vorticity fields}

\begin{figure}
\centering {
\sidesubfloat[]{\includegraphics[width= 0.45\textwidth]{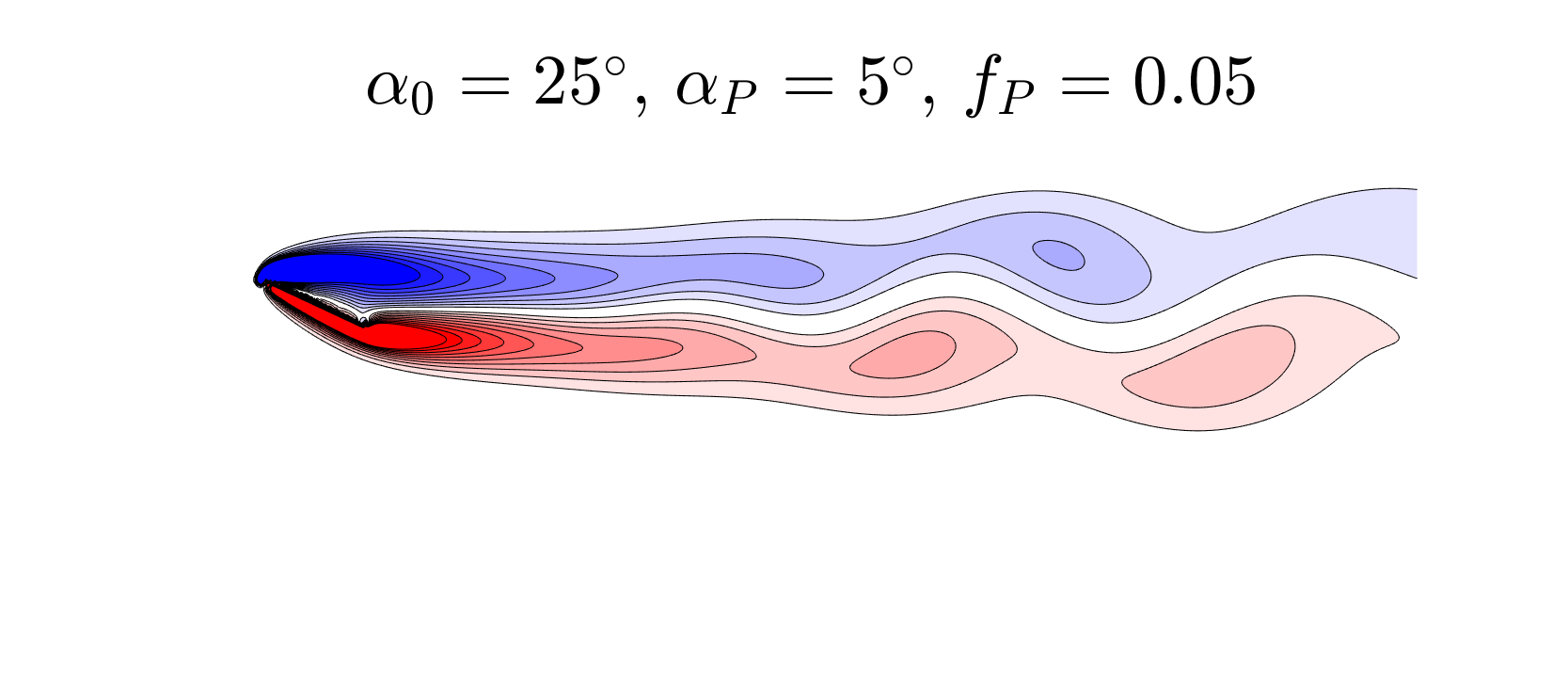}} \ \ 
\sidesubfloat[]{\includegraphics[width= 0.45\textwidth]{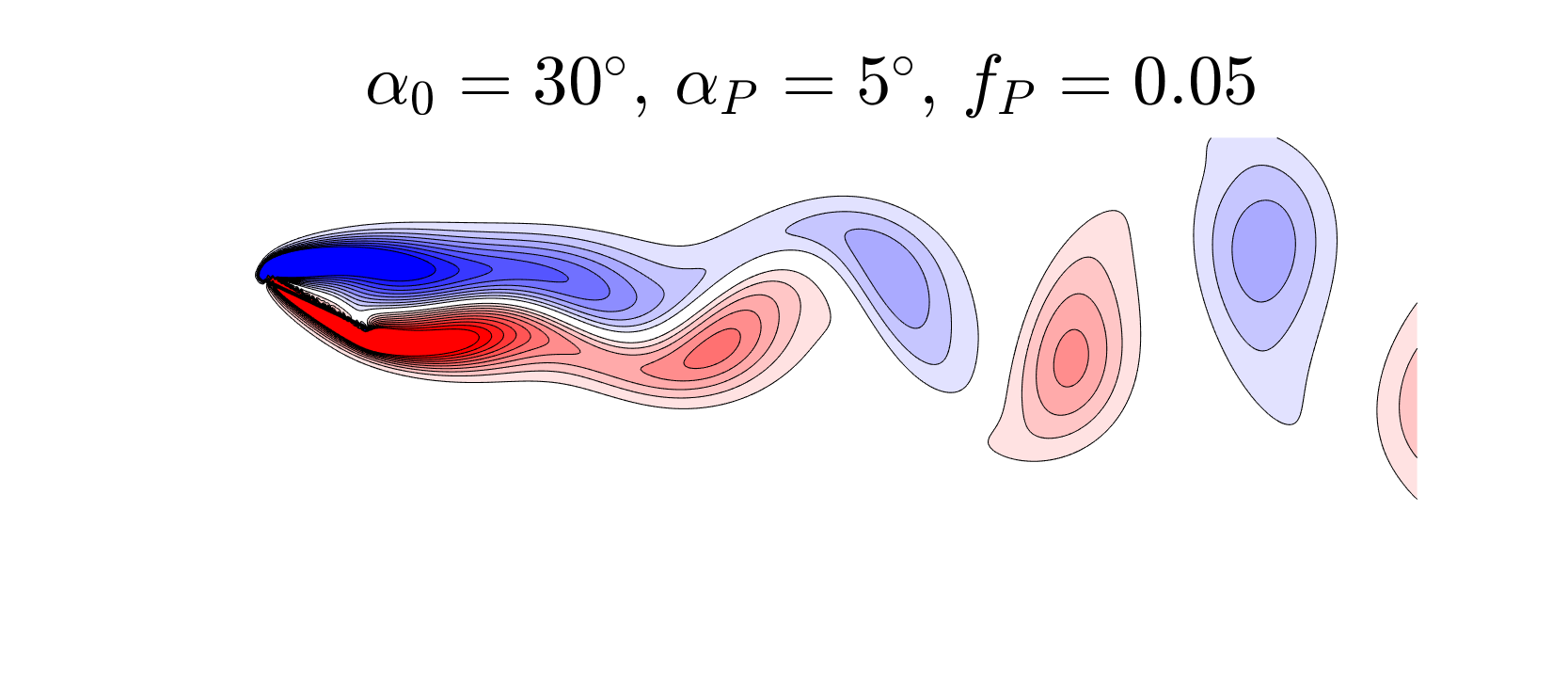}} \\
\sidesubfloat[]{\includegraphics[width= 0.45\textwidth]{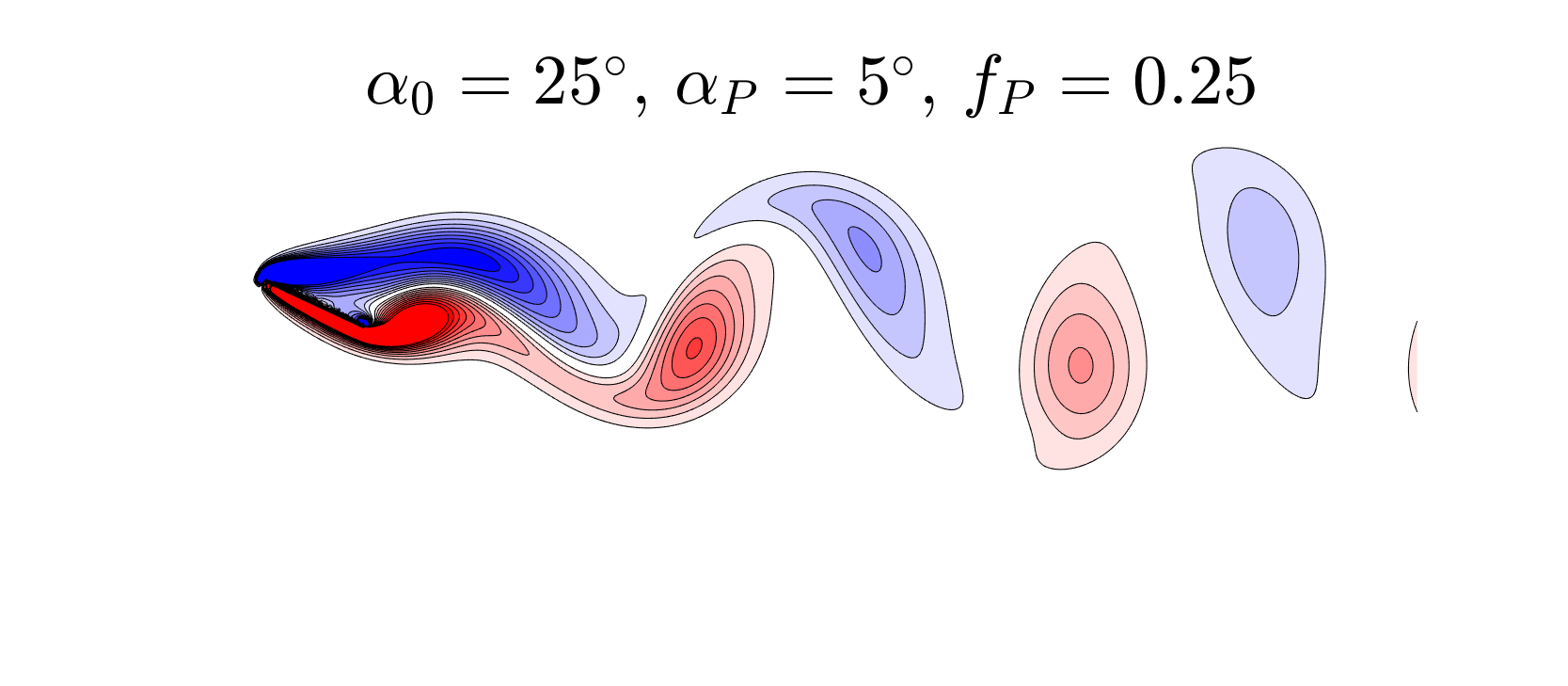}} \ \ 
\sidesubfloat[]{\includegraphics[width= 0.45\textwidth]{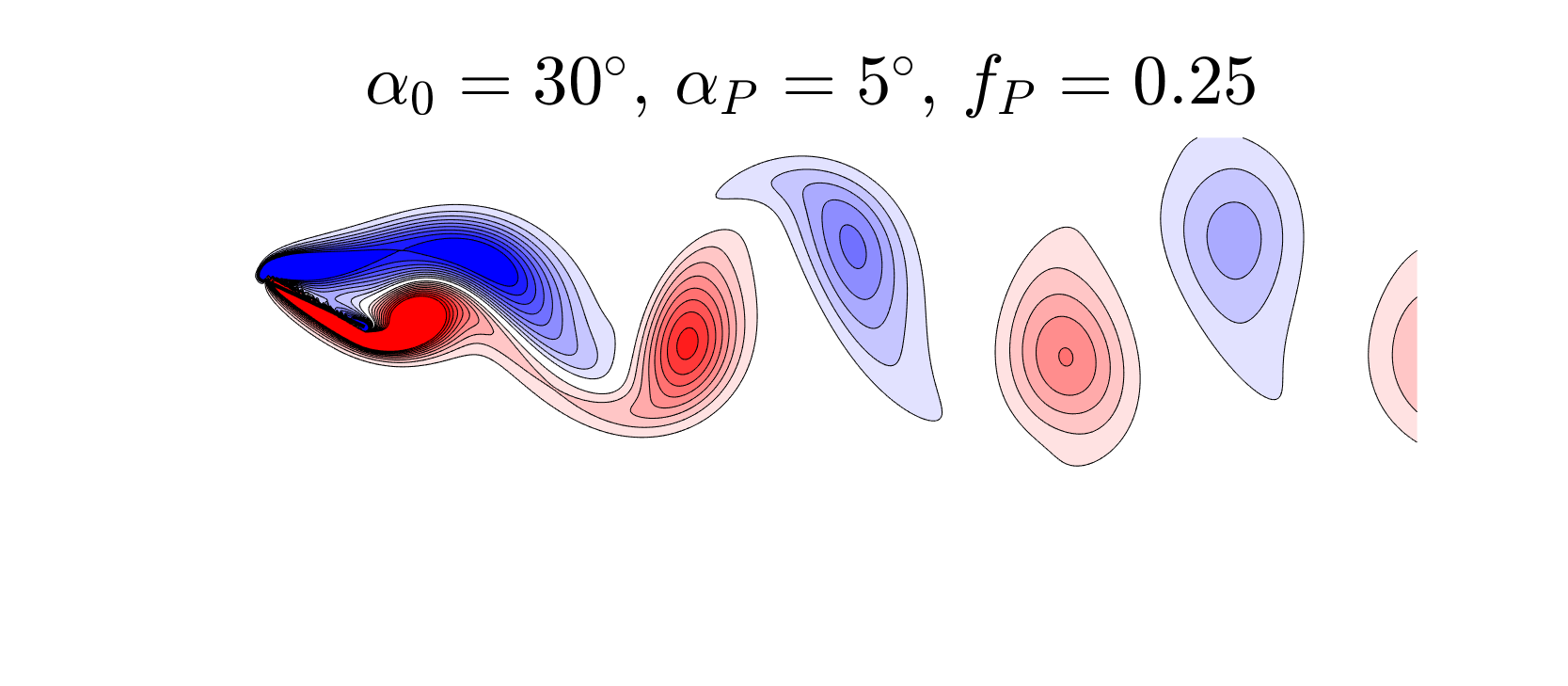}} \\
\sidesubfloat[]{\includegraphics[width= 0.45\textwidth]{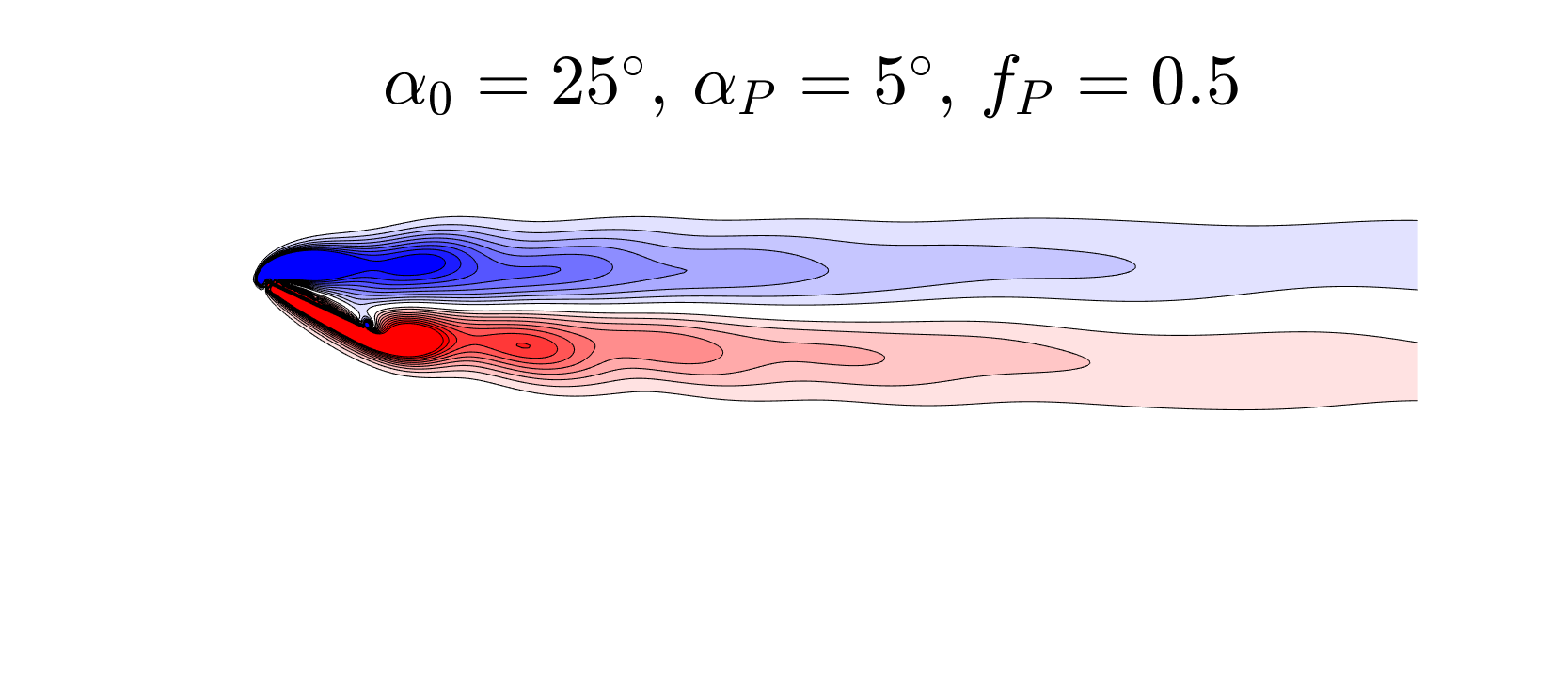}} \ \ 
\sidesubfloat[]{\includegraphics[width= 0.45\textwidth]{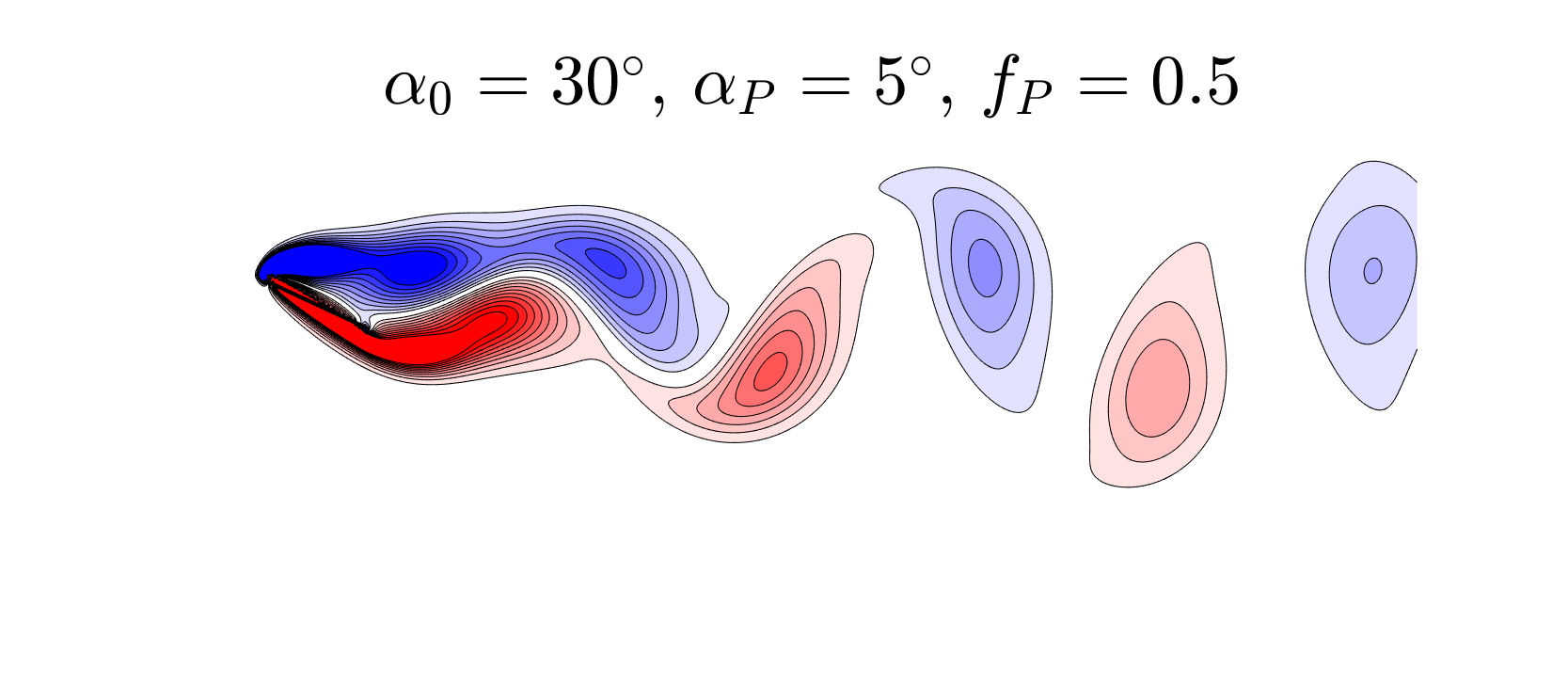}}
}
\caption{Instantaneous vorticity fields for the airfoil pitching at various frequencies.}
\label{fig:vort}
\end{figure}

For each case, the dataset includes time-resolved velocity (streamwise and transverse components) and vorticity fields with a timestep of $0.1c/U_\infty$ within the spatial domain indicated in Fig.~\ref{fig:AirfoilDomain}a. Data is collected over $40 c/U_\infty$  time units for the $(\alpha_0,\alpha_P) = (25^\circ,5^\circ)$ and $(30^\circ,0^\circ)$ cases, and for $100c/U_\infty$  time units for the $(\alpha_0,\alpha_P) = (30^\circ,5^\circ)$ cases. Since the equilibrium state is stable for the $\alpha_0 = 25^\circ$, $\alpha_P = 0\circ$ case, only a single snapshot (of this equilibrium) is included. 
Figure \ref{fig:vort} shows instantaneous vorticity fields for several cases that are discussed further below and further elucidated in the sample analysis in \S\ref{sec:DNSairfoilAnalysis}.

\subsubsection{Time-varying aerodynamic forces and airfoil kinematics}

The dataset also contains time-varying lift and drag coefficients and airfoil kinematics ($\alpha(t)$ and $\dot\alpha(t)$). These quantities are provided with temporal resolution of $0.01c/U_\infty$ time units, corresponding to the timestep used in the simulations.

\begin{figure}
 \centering {
 \sidesubfloat[]{\includegraphics[width= 0.45\textwidth]{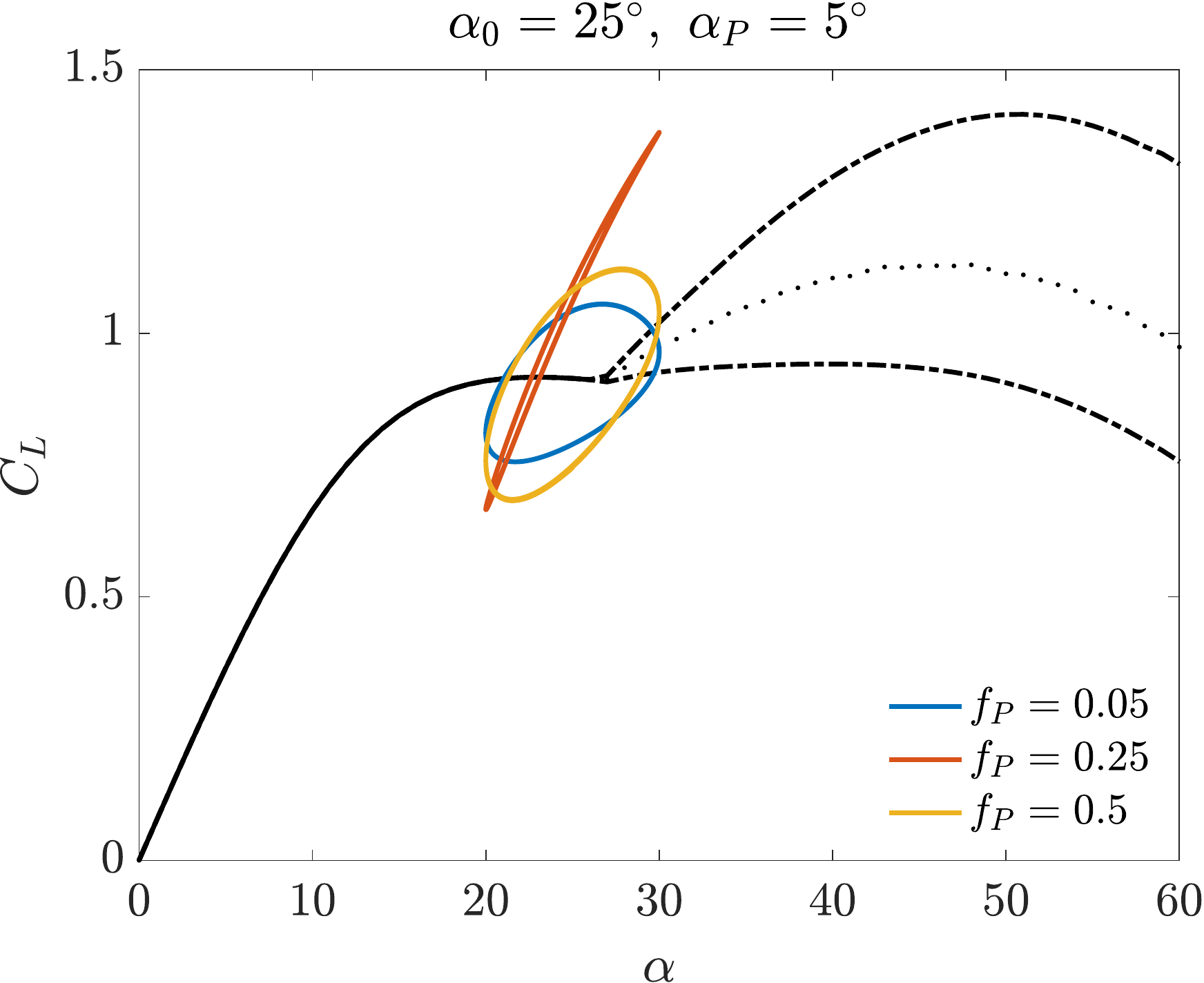}} \ \ 
 \sidesubfloat[]{\includegraphics[width= 0.45\textwidth]{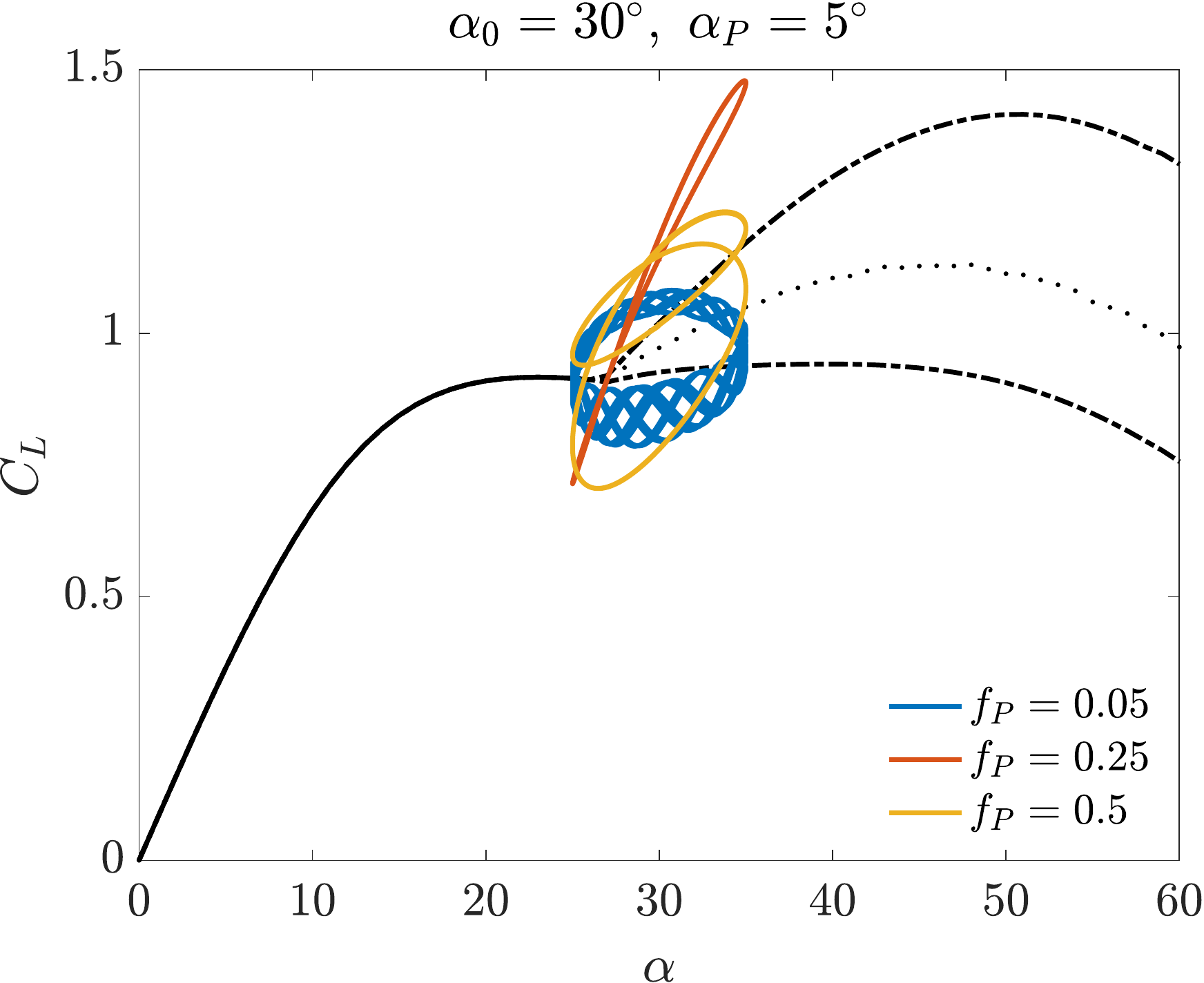}}
}
\caption{Lift trajectories for pitching at various frequencies about a base angle of (a) $\alpha_0 = 25^\circ$ and (b) $\alpha_0 = 30^\circ$. Also shown is the static lift curve, with the equilibrium lift shown when the wake is stable ($\alpha_0 < 27^\circ$), and the maximum, minimum, and average lift shown for cases where the wake behind the stationary airfoil exhibits periodic vortex shedding.}
\label{fig:LiftCurve}
\end{figure}

As an example, Fig.~\ref{fig:LiftCurve} plots the lift coefficient $C_L$ against the angle of attack for a subset of the cases included in the database, with the trajectories for the pitching maneuvers overlaid on the static lift curve/envelope. The lift coefficient $C_L$ is obtained from the lift force $F_y$ via
\begin{equation}
    C_L = \frac{F_y}{0.5 \rho c U_\infty^2}.
\end{equation}
Figure \ref{fig:LiftCurve}a shows cases where the base angle $\alpha_0 = 25^\circ$ is below the bifurcation point where the wake becomes unstable (in the absence of pitching).  In these cases, the lift is periodic with frequency $f_P$.  At $f_P = 0.25$, the pitching frequency excites the least-stable eigenmode in the wake (which ultimately becomes unstable at angles greater than the critical angle of approximately $27^\circ$. This leads to greater amplitude in lift oscillations, a larger average lift, and very little hysteresis of the $C_L$--$\alpha$ curve.

Figure \ref{fig:LiftCurve}b plots the equivalent cases for a base angle of $\alpha_0 = 30^\circ$. While some of the same trends are observed, there is now interaction between the pitching and natural frequencies of the wake. For $f_P = 0.25$, this natural mode locks on to the pitching frequency, again giving a periodic lift response with enhanced mean and fluctuating lift amplitude. At $f_P = 0.5$ (approximately twice the natural vortex-shedding frequency), period-doubling (relative to $f_P$) behavior is observed. At $f_P = 0.05$, the lift response is quasi-periodic, due to the presence of both the pitching and natural frequencies.  

\subsection{Example analyses}
\label{sec:DNSairfoilAnalysis}
Three example analysis methods are described, including analysis of the time-resolved flow field data to determine spectral content, building a linear model for the lift response to pitching, and solving an inverse problem to determine the pitching parameters from limited instantaneous measurements in the wake.

\subsubsection{Modal analysis}
\begin{figure}
 \centering {
 \sidesubfloat[]{\includegraphics[width= 0.3\textwidth]{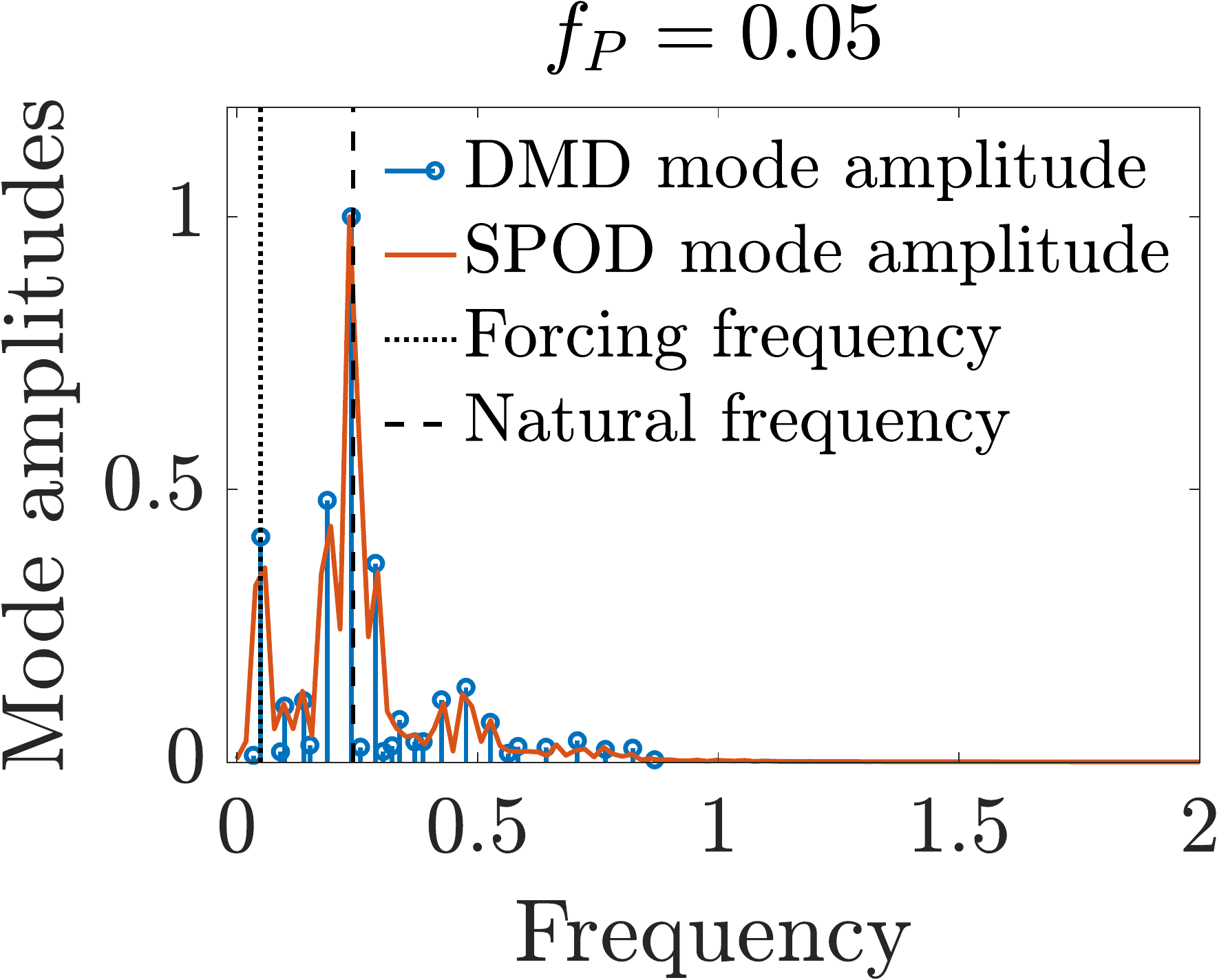}} 
 \sidesubfloat[]{\includegraphics[width= 0.3\textwidth]{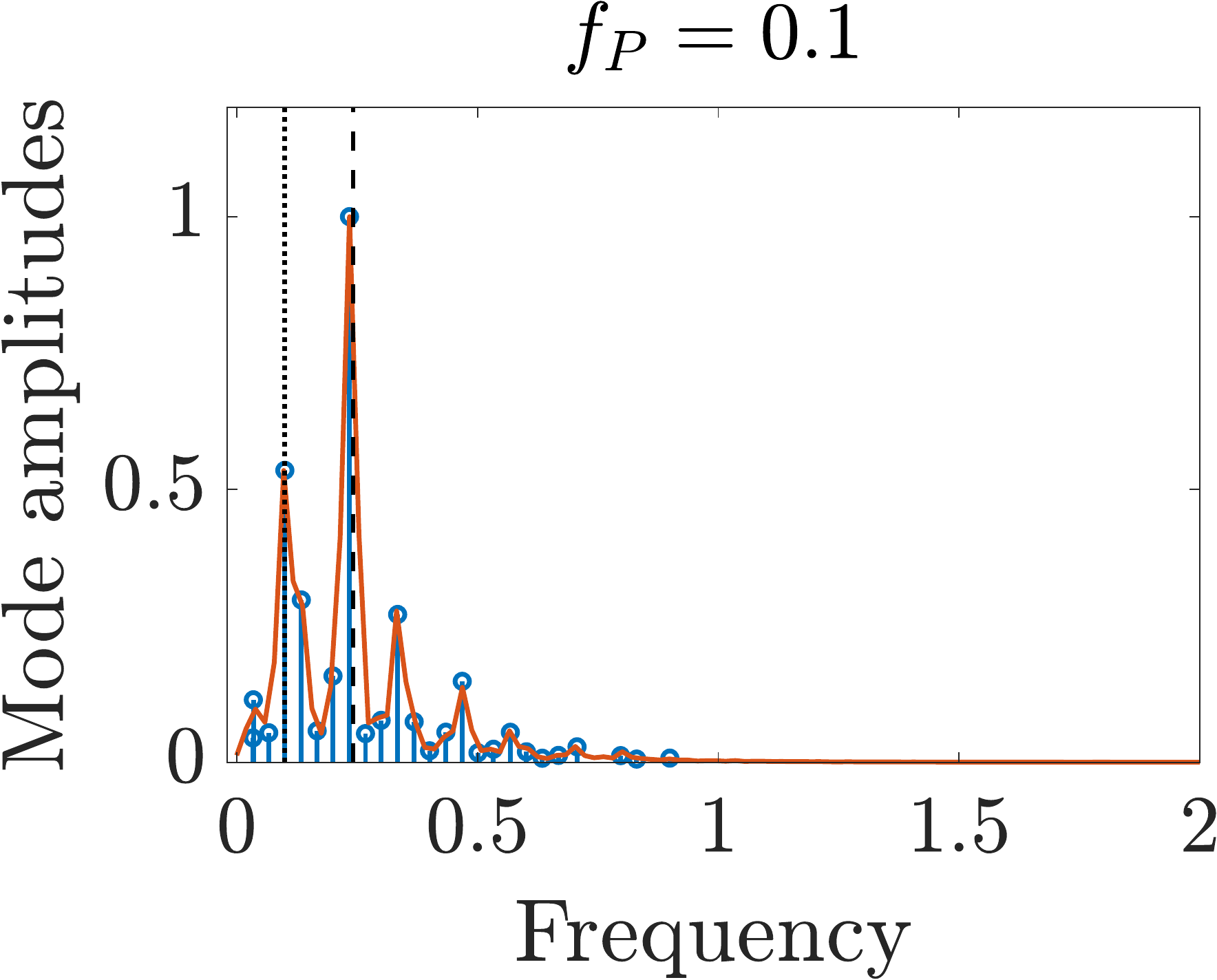}} 
 \sidesubfloat[]{\includegraphics[width= 0.3\textwidth]{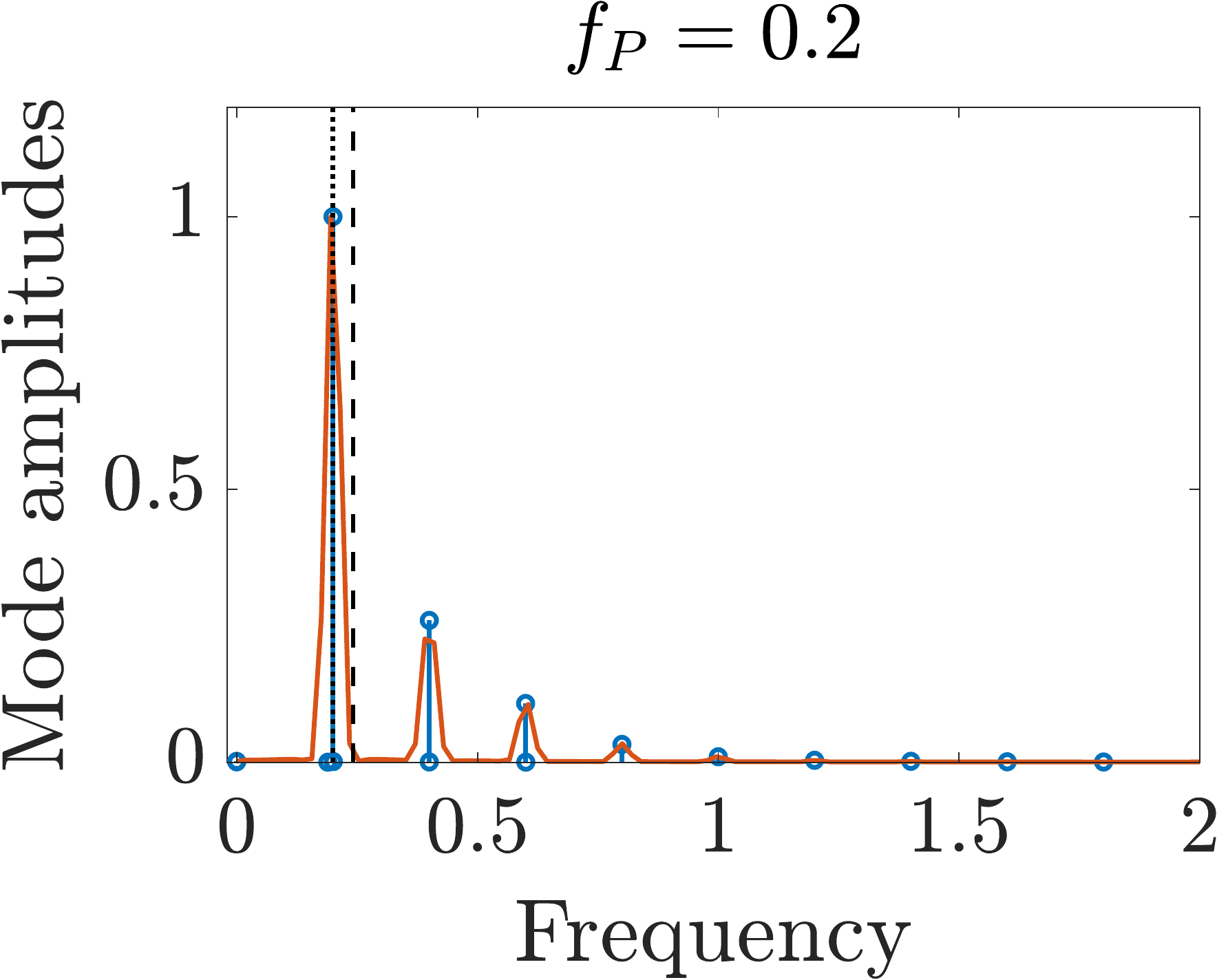}} \\ 
  \sidesubfloat[]{\includegraphics[width= 0.3\textwidth]{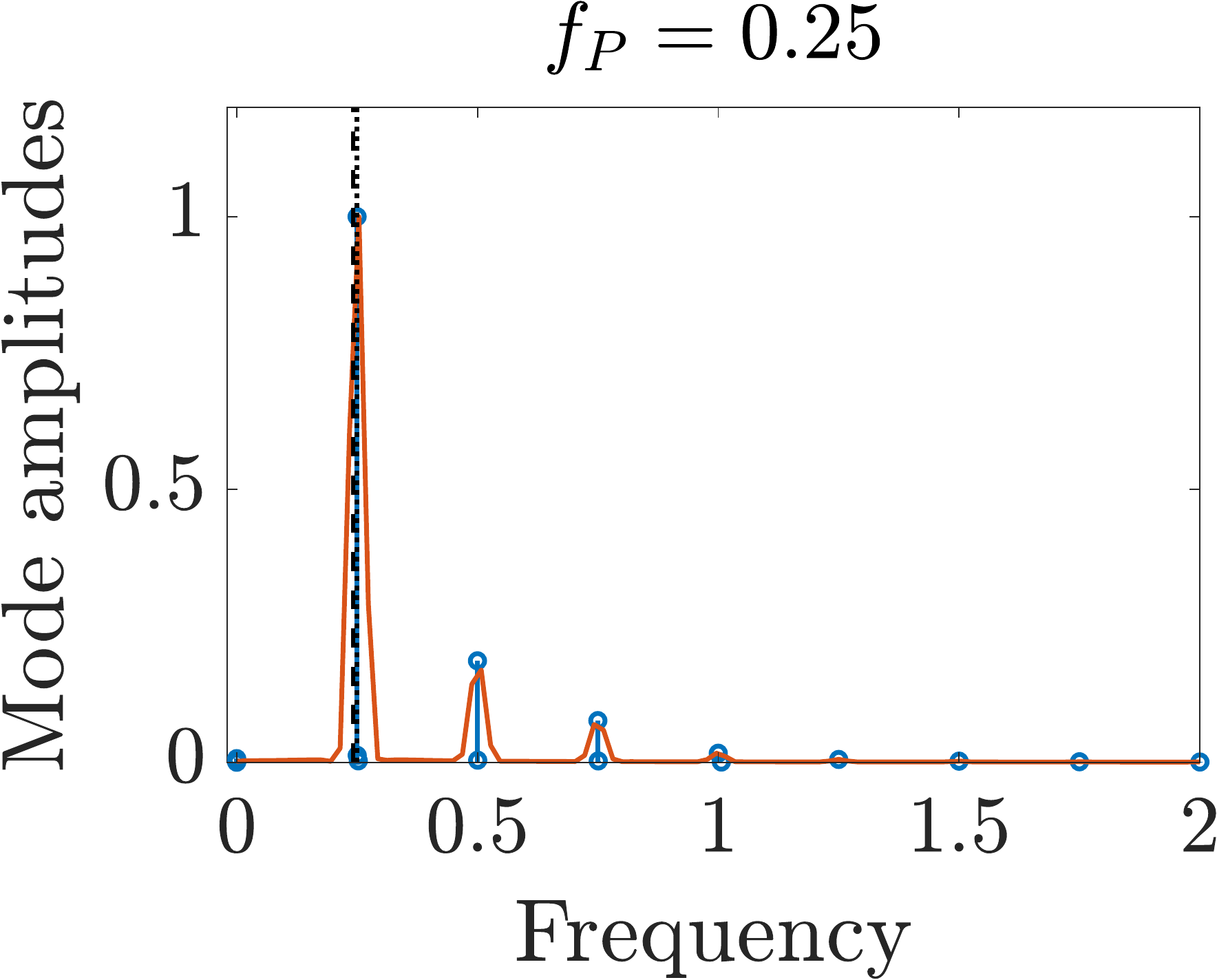}} 
 \sidesubfloat[]{\includegraphics[width= 0.3\textwidth]{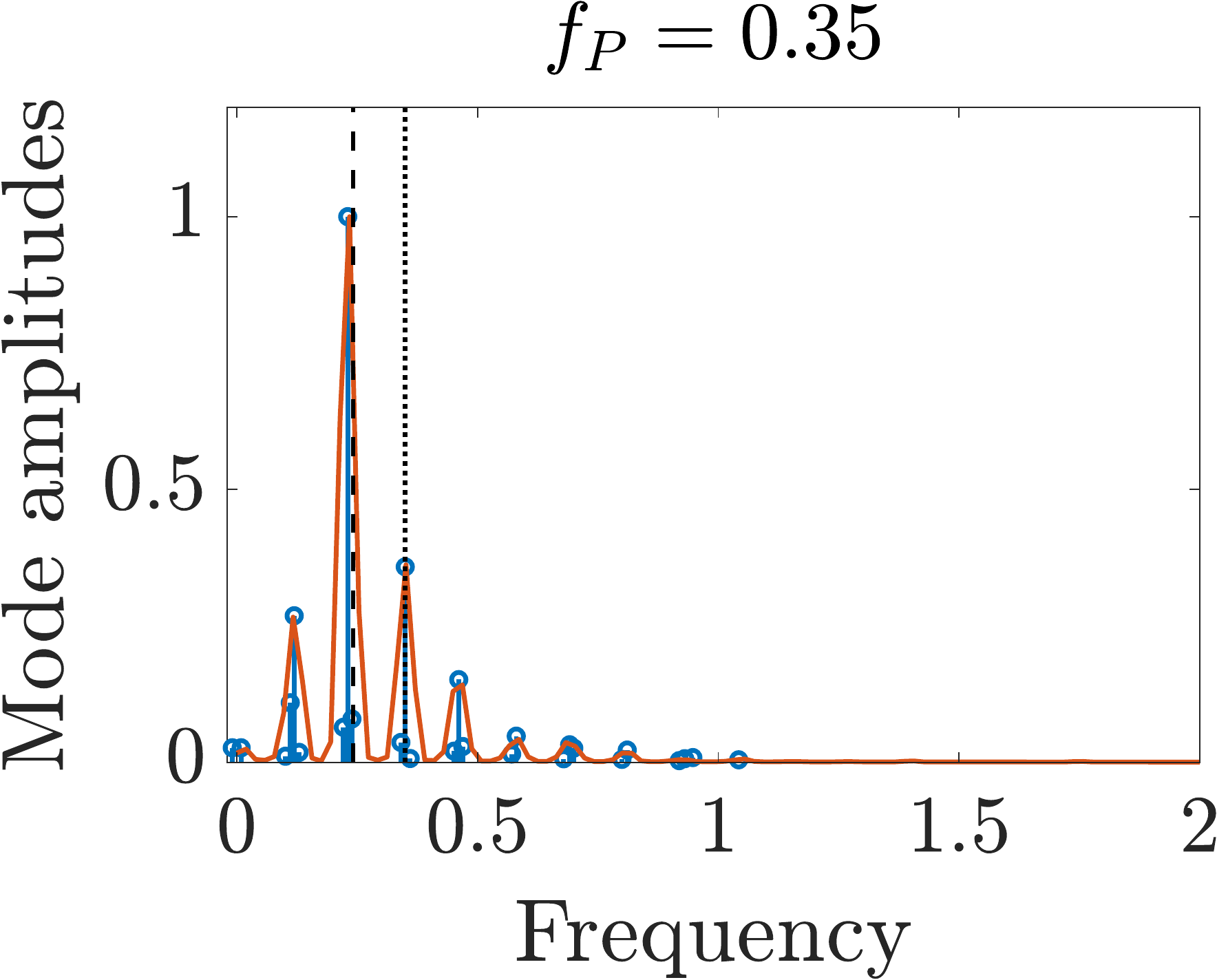}} 
 \sidesubfloat[]{\includegraphics[width= 0.3\textwidth]{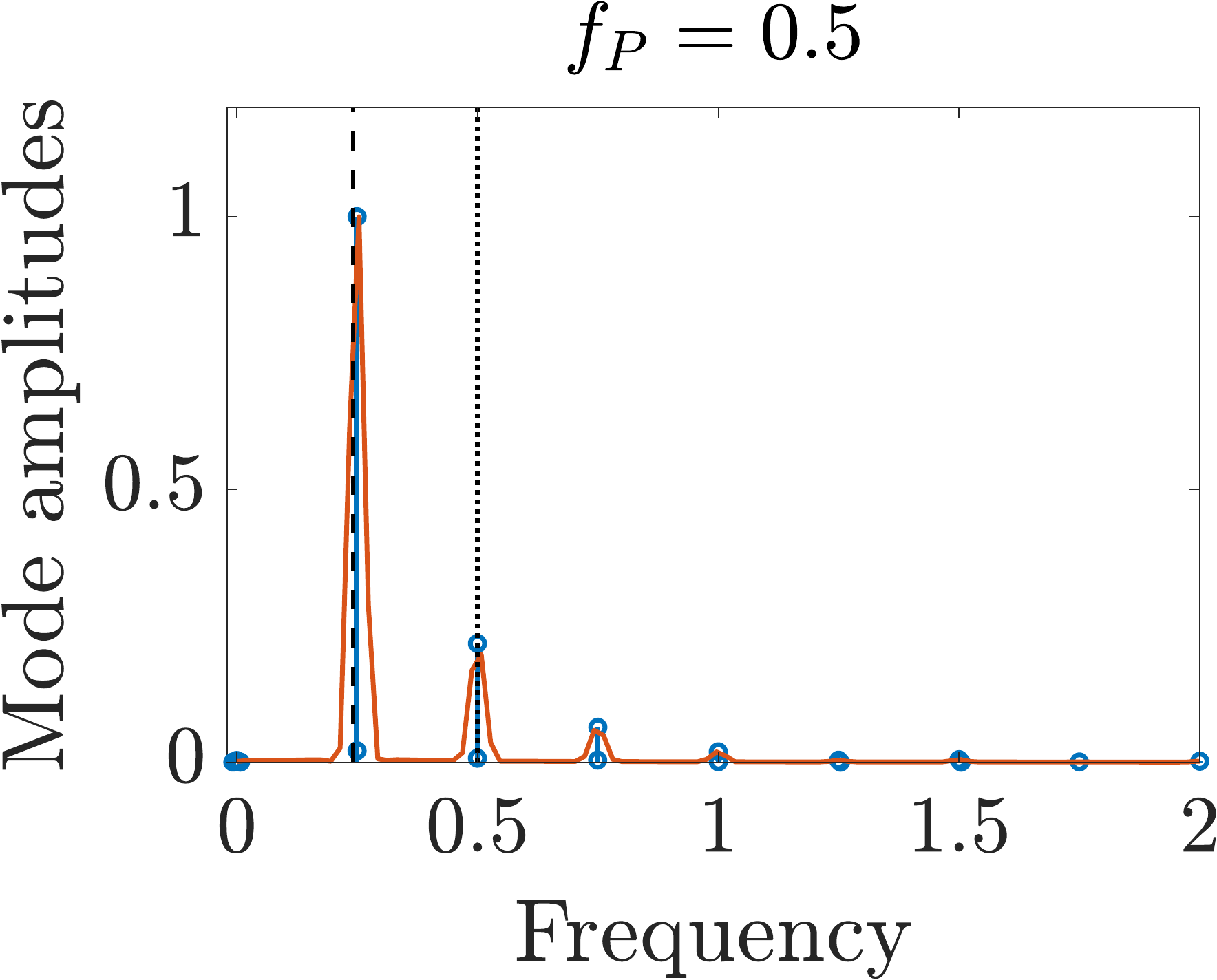}} 
}
\caption{Comparison between DMD and SPOD mode amplitudes for pitching at various frequencies with $\alpha_0 = 30^\circ$  and $\alpha_P = 5^\circ$. }
\label{fig:DMDSPODamps}
\end{figure}

This section considered modal analysis of the flow field data, which reveals the variety of temporal dynamics that emerge across various simulations.  The frequency content of the flow field may be studied through the use of modal analysis methods that are designed to extract such information. 
As an example, shown in Fig.~\ref{fig:DMDSPODamps} are mode amplitudes as functions of frequencies for the airfoil pitching about a base angle of attack $\alpha_0 = 30\circ$ at various frequencies.  Here, we are comparing normalized dynamic mode decomposition (DMD) \cite{Schmid:JFM2010} mode amplitudes (as described, for example, in \cite{Tu:JCD2014}) with normalized spectral proper orthogonal decomposition \cite{Lumley1970stochastic, towne2018spectral} mode amplitudes (these are the square root of SPOD energy levels).  We observe that both methods identify essentially the same frequency content within the data.  This is expected, due to the correspondence between the two methods for non-chaotic flows \cite{towne2018spectral}.

In general, however, there are some practical differences between the two decomposition methods.  For example, SPOD is able to identify multiple modes at each frequency, which is particularly useful for complex, chaotic flows, but typically requires a large number of snapshots to obtain converged results.   
{ On the other hand, DMD can be used for the identification of a linear dynamical system that fits the data, which can be more amenable for prediction and control purposes. } 

Shown in the plots are both the pitching and natural vortex-shedding frequencies of the system.  When these two frequencies are sufficiently close together (as for Figs.~\ref{fig:DMDSPODamps}c and \ref{fig:DMDSPODamps}d), the natural frequency locks on to the imposed frequency, and the peaks in the mode amplitude plots correspond to this frequency and its harmonics. This agrees with the periodic lift response observed for $f_P = 0.25$ in Fig.~\ref{fig:LiftCurve}b.  A more extensive exploration of lock-on between natural and imposed frequencies for low Reynolds number airfoils can be found in \cite{choi2015surging}, for example.

\begin{figure}
\centering{\includegraphics[width= 0.95\textwidth]{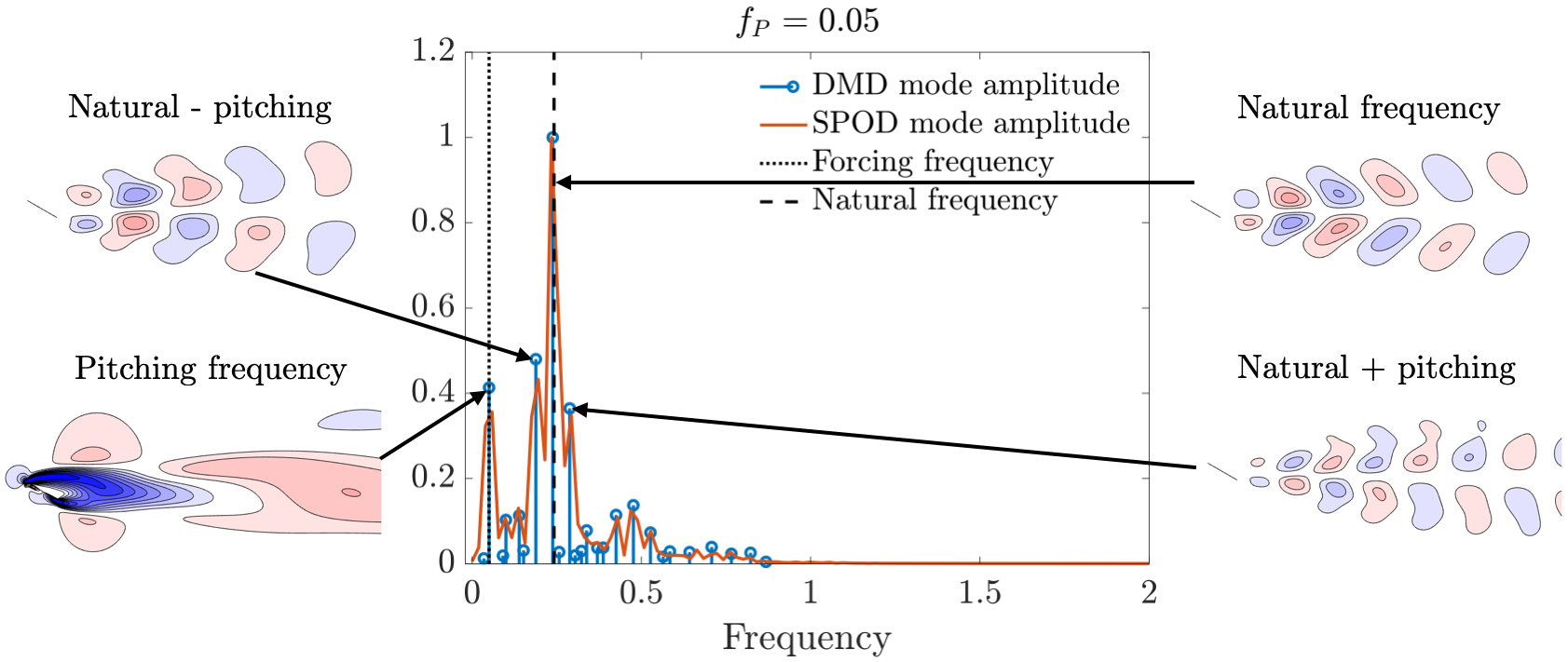}} 
\caption{Mode amplitude plot showing the four largest-amplitude DMD modes (vertical velocity component) identified for  pitching about $\alpha_0 = 30^\circ$ with frequency $f_P = 0.05$.}
\label{fig:DMDmodes}
\end{figure}

Similar behavior is observed for $f_P = 0.5$, except that the dominant frequency is half the pitching frequency. This also agrees with the period-doubling behavior in the lift response observed for this case in Fig.~\ref{fig:LiftCurve}b.  
For other pitching frequencies, there are modes with large amplitude at both the natural and pitching frequencies,  as well as additional frequencies that correspond to sums and differences of these two frequencies and their harmonics.  To show this in further detail, Fig.~\ref{fig:DMDmodes} shows DMD modes with the four largest amplitudes for the $f_P = 0.05$ case.  These frequencies correspond to the natural frequency, pitching frequency, and their sum and difference. 

\subsubsection{Modeling aerodynamic forces}
 We now use the data to develop models that capture the lift and drag response to pitching.  For simplicity, we focus on the $\alpha_0 = 25^\circ$ case, where the response is periodic (with the same period as the input pitching frequency).  In this case, it is reasonable to expect that the data can be well represented by a linear model, i.e. a linear state-space system in the time domain, or equivalently a transfer function in the frequency domain.  
 That is to say, for an input $\bu(t)$ and output $\by(t)$, we can identify a state space model $\{\mA,\mB,\mC,\mD\}$ such that the input and output are related via
  \begin{equation}
\label{eq:ss}
\begin{aligned}
\dot\bx(t)  &= \mA\bx(t) + \mB\bu(t), \\
\by(t)  &= \mC\bx(t) + \mD\bu(t),
\end{aligned}
\end{equation}
 where $\bx$ is an internal system state. Equivalently, in the frequency domain, we have 
 \begin{equation}
Y(s) = G(s) U(s),
\end{equation}
where $Y(s) = \mathcal{L}(y(t))$ and $U(s) = \mathcal{L}(u(t))$ are the Laplace transforms of the inputs and outputs, respectively, and 
\begin{equation}
G(s) = \mC (s \mI - \mA)^{-1} \mB + \mD.
\end{equation} 

Since this dataset consists of the response to sinusoidal pitching, we identify a model in the Laplace (frequency) domain.  A linear system has the property that any sinusoidal input $A_u\sin(\omega t)$ will have a sinusoidal output $A_y\sin(\omega t)$ at the same frequency characterized by a gain
\begin{equation}
\left|\frac{Y(i \omega)}{U(i \omega)}\right| = \frac{A_y}{A_u} = \left| G(i\omega) \right|,
\end{equation}
and phase 
\begin{equation}
\phi = \angle G(i\omega). %= 
\end{equation}
We proceed by identifying models with input $u = \dot\alpha$  and output the mean subtracted lift coefficient $y = C_L-\overline{C_L}$ . The use of  $\dot\alpha$ as an input arises because we expect to have a component of $C_L$ proportional to $\dot\alpha$ (which would be the $\mD$ term in Eq.~\eqref{eq:ss}). If pitching was not about the midchord, we would also expect a term proportional to $\ddot\alpha$. For further discussion of these matters, see Refs.~\cite{brunton:2012a,brunton2013jfm,brunton2014state}.

\begin{figure}
 \centering {
 \sidesubfloat[]{\includegraphics[width= 0.45\textwidth]{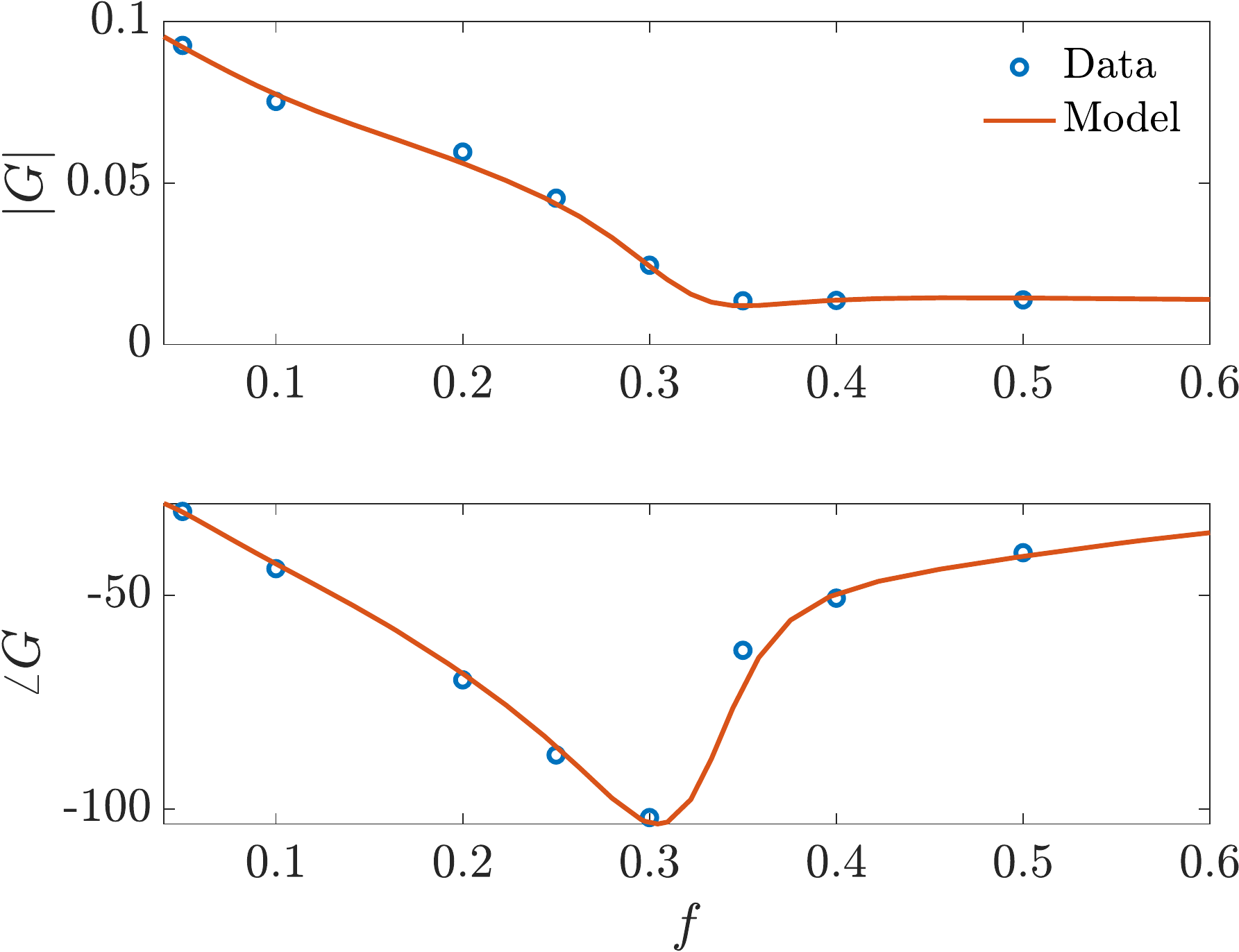}} \ \ 
 \sidesubfloat[]{\includegraphics[width= 0.45\textwidth]{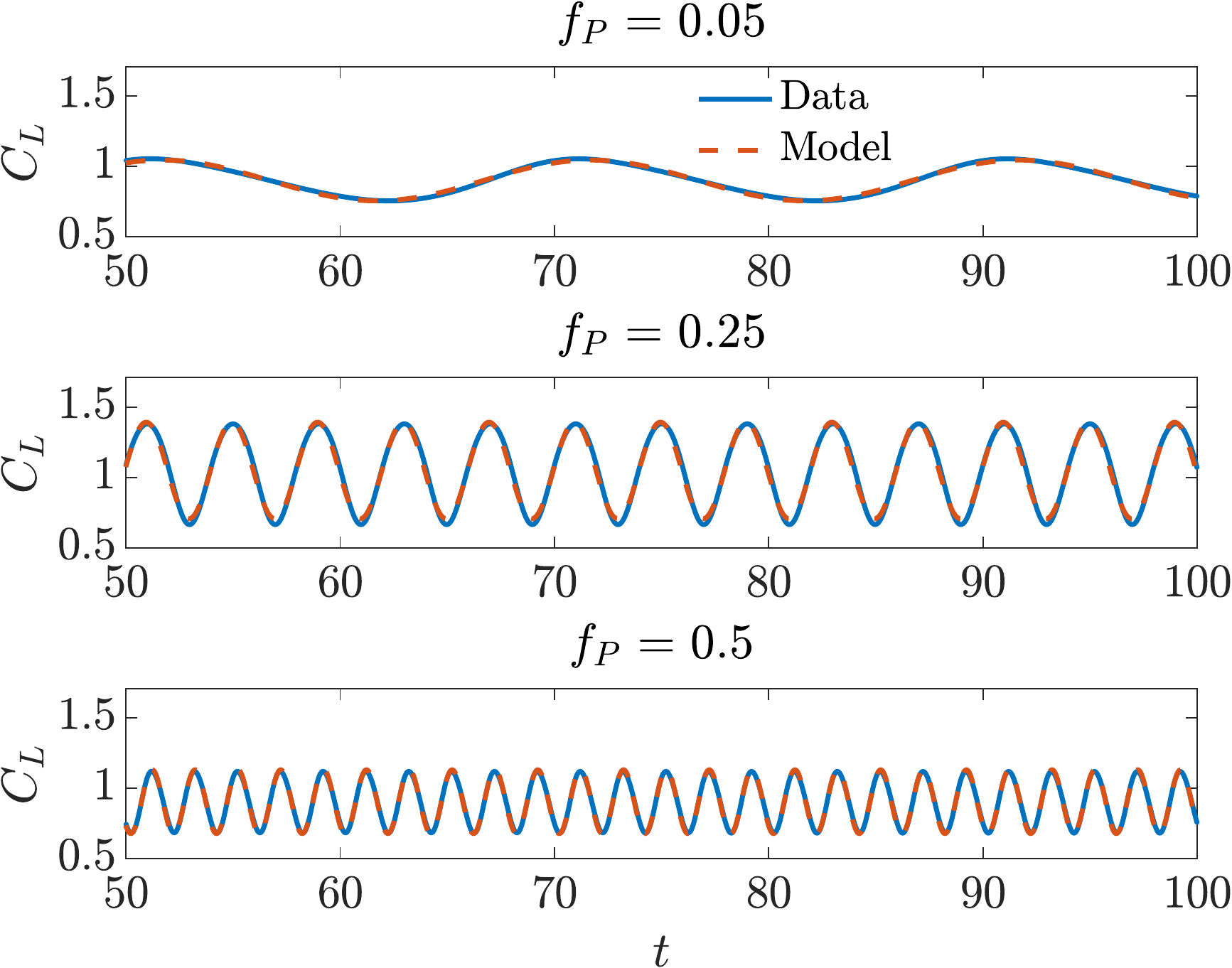}} 
}
\caption{Results for the data-driven aerodynamic forces model: (a) Bode plot showing magnitude and phase of lift ($C_L$) response to pitching at various frequencies (with input defined by $\dot\alpha$), and (b) comparison between true and predicted lift response for pitching at frequencies $f_P = 0.05, 0.25$, and $0.5$.  }
\label{fig:LinModel}
\end{figure}

 We first identify the gain and phase for each frequency by fitting a sinusoidal function to the output data. This frequency response data is next used to identify a transfer function using the \texttt{tfest} command in MATLAB, which utilizes a method developed in Ref.~\cite{drmac2015quadrature}, based on Sanathanan-Koerner iteration \cite{sanathanan1963transfer}. In this process, we enforce that the identified model be stable. 
 Fig.~\ref{fig:LinModel}a shows the magnitude and phase of the data and identified model, as well as a comparison between the original data and model prediction for frequencies. We observe that the model accurately captures both the magnitude and phase of the response. This is further demonstrated in Fig.~\ref{fig:LinModel}b, which compares the predicted and actual response to pitching at three frequencies in the time domain.  Here, to match the data, the model is run for long enough to ensure that all transient dynamics associated with startup have decayed.
The accuracy of this model could likely be improved by using additional data at lower and higher frequencies, or by using time-domain data.  We could also seek to isolate the terms directly proportional to the kinematics ($\alpha$, $\dot\alpha$, and $\ddot\alpha$) as described in Refs.~\cite{brunton:2012a,brunton2013jfm,brunton2014state}, but this is unnecessarily complicated for the purposes of this demonstrative example. Note finally that in this example we computed the mean-subtracted lift using the mean at each frequency. If we instead used the overall mean (or the equilibrium mean), there would be a larger offset between the predicted and actual lift data shown in Fig.~\ref{fig:LinModel}b.
%identify terms proportional to $\alpha$, $\dot\alpha$ and $\ddot\alpha$, as described in Refs.~\cite{brunton2013jfm,brunton2014state}.

The case with $\alpha_0 = 30^\circ$ cannot be accurately represented using a linear model, due to the multiple frequencies being present in the output.  Further modeling efforts could seek nonlinear models capable of reproducing the aerodynamic force responses for this case.
 
 \subsubsection{Flow field reconstruction and discovering airfoil kinematics }
 
As an additional example, we now consider an inverse problem: given some number of measurements of the velocity field made at a single instance of time, we develop a model that can identify the parameters $\alpha_0$ and $f_p$.  We do this by applying a machine learning method based on the ideas that are described and applied to reconstruction and parameter estimation in cylinder flow in 
Refs.~\cite{bright2013compressive,bright2016classification}, as also described in Ref.~\cite{brunton2019data}. %kutz2017leveraging}
We start by performing proper orthogonal decomposition on data from each simulation and assemble a truncated set of $r$ POD modes for each case into columns of a matrix $\Phi^{(j)}_r$, where the superscript  $(j)$ indexes each simulation. For this example, we only consider data from the pitching cases ($\alpha_P = 5^\circ$), and use 401 snapshots from each case. These matrices are concatenated into a global library of modes
\begin{equation}
\label{eq:library}
\Phi_L = \begin{bmatrix} \Phi^{(1)}_r &  \Phi^{(2)}_r & \cdots &  \Phi^{(M)}_r \end{bmatrix},
\end{equation}
where $M = 16$ is the total number of pitching airfoil cases considered.  A given snapshot $\bx$ from simulation $(j)$ may be approximated by $\Phi^{(j)}_r \ba^{(j)}$, where $\ba$ is a vector of POD coefficients.  Equivalently, we can also use the full library to write 
\begin{equation}
\label{eq:xphia}
\bx \approx \Phi^{(j)}_r \ba^{(j)} = \Phi_L \ba_L,
\end{equation}
where $\ba_L$ contains nonzero entries in rows corresponding to the set of columns of $\Phi_L $ containing $\Phi^{(j)}_r$.  

We consider measurements defined by $\by = \mC \bx$, where each of the rows of $\mC$ specifies which value(s) of $\bx$ are used for a given measurement. In this case, we will use point measurements of one velocity component, so each row of $\mC$ will contain zeros and a single entry of 1.  We may use compressed sensing principles \cite{donoho2006compressed,candes2008introduction} to reconstruct the state $\bx$ from measurements $\by$. This is achieved by solving the convex optimization problem
\begin{equation}
\label{eq:compressed}
\hat \ba_L = \argmin{\|\ba_L\|_1},  \ \ \text{such that}  \ \ \| \mC \Phi_L \ba_L - \by\|_2 < \epsilon,
\end{equation}
from which the full state $\bx$ can be obtained from Eq.~\eqref{eq:xphia}.  Minimizing the $l_1$ norm promotes a sparse solution for $\hat \ba_L$, where most of its entries are zero. The use of the inequality $\| \mC \Phi_L \ba_L - \by\|_2 < \epsilon$ is a relaxation of the equality constraint  $\by = \mC \Phi_L \ba_L$, which can account for imperfect measurements. 
Ideally, the nonzero entries  of the reconstructed coefficients $\hat \ba_L$ would correspond to the POD modes associated with the correct simulation parameters, allowing for the frequency and base angle of attack to be identified. To predict the simulation parameters  (defined by the index $j$) from the single-snapshot measurements  $\by$, we may compute
\begin{equation}
\label{eq:kinRecon}
j = \argmax_{j'} \|\ba^{(j')}\|_2.
\end{equation}
That is, we determine the set of modes $\Phi^{(j)}_r$ which have the largest coefficients in the reconstructed data.

\begin{figure}
\centering{\includegraphics[width= 0.6\textwidth]{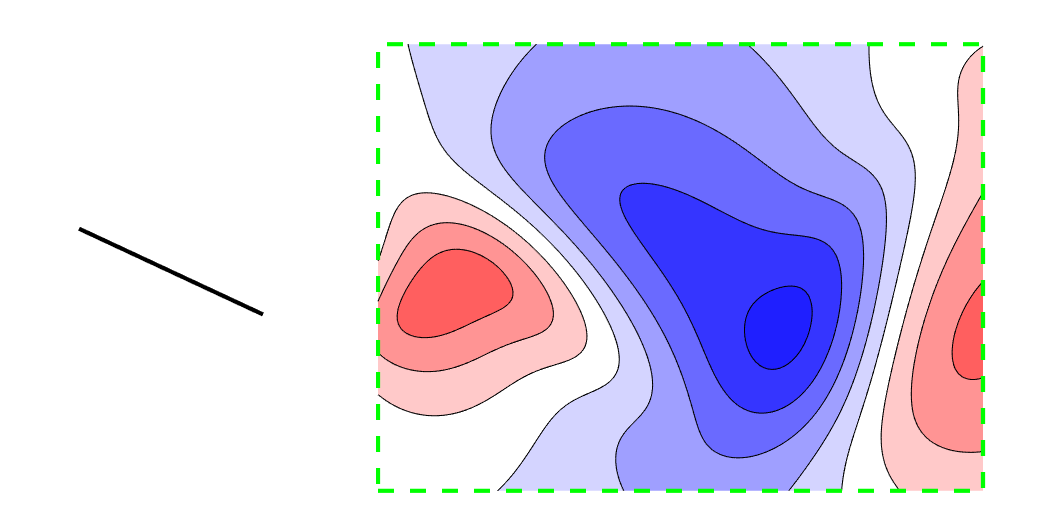}} 
\caption{Window used for sampling and flow field reconstruction, and identification of airfoil kinematics.}
\label{fig:compressedWindow}
\end{figure}

To test this method, we randomly select 50 locations to measure data at a given instance in time of a given simulation and reconstruct the flow field via Eqs.~\eqref{eq:xphia}-\eqref{eq:compressed}.  Eq.~\eqref{eq:xphia}  is solved using the CVX package \cite{grant2014cvx} in MATLAB, with  $\epsilon = 10^{-4}$. For the purposes of this example, we take measurements from, and perform reconstruction of, a limited window downstream of the wake, as indicated in Fig.~\ref{fig:compressedWindow}. This means that the measurements do not have any near-airfoil information to assist in the reconstruction or prediction of parameters. We additionally only consider the vertical ($y$) component of velocity. Of the 401 snapshots available for each simulation, we use the first 351 for model training (i.e., to generate the mode library in Eq.~\eqref{eq:library}) using $r=10$, and the remaining 50 for testing. 

   \begin{figure}
 \centering {
 \sidesubfloat[]{\includegraphics[width= 0.45\textwidth]{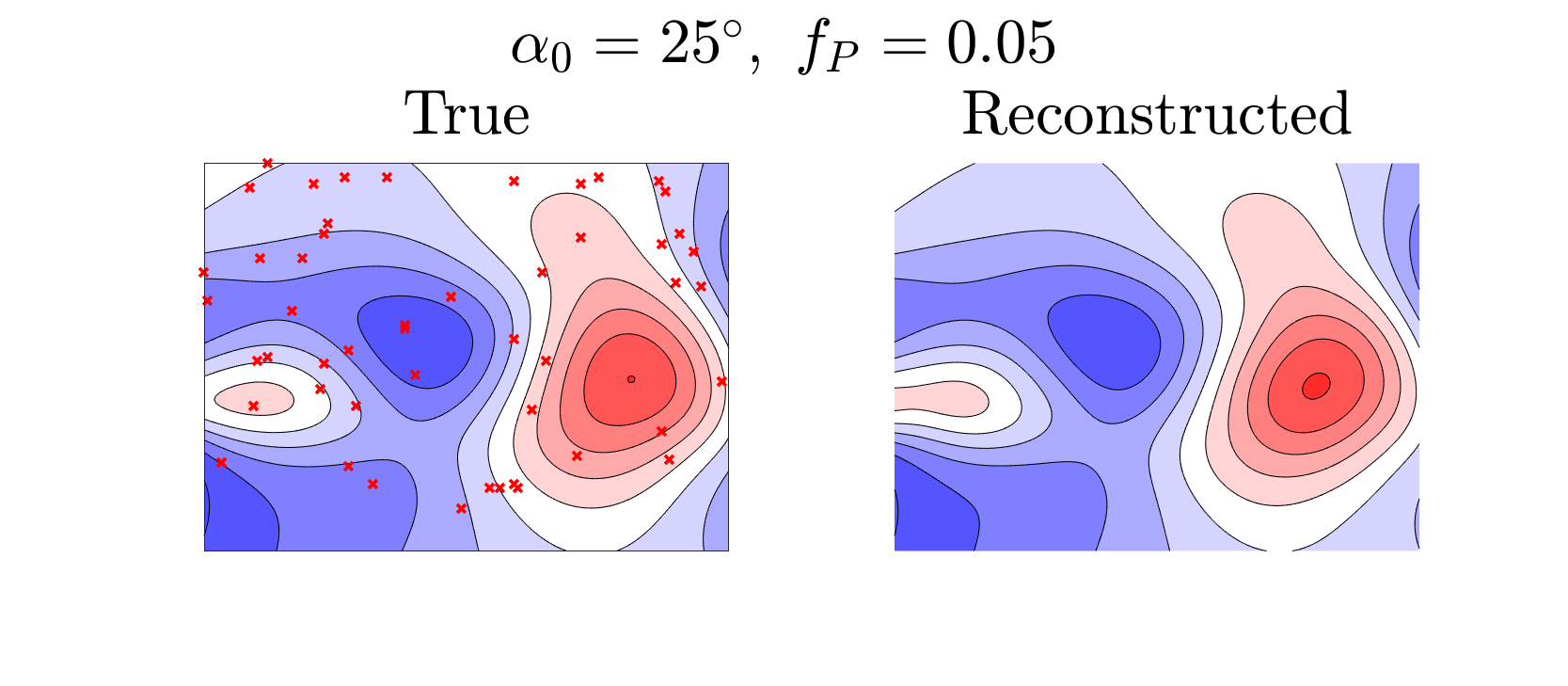}} 
 \sidesubfloat[]{\includegraphics[width= 0.45\textwidth]{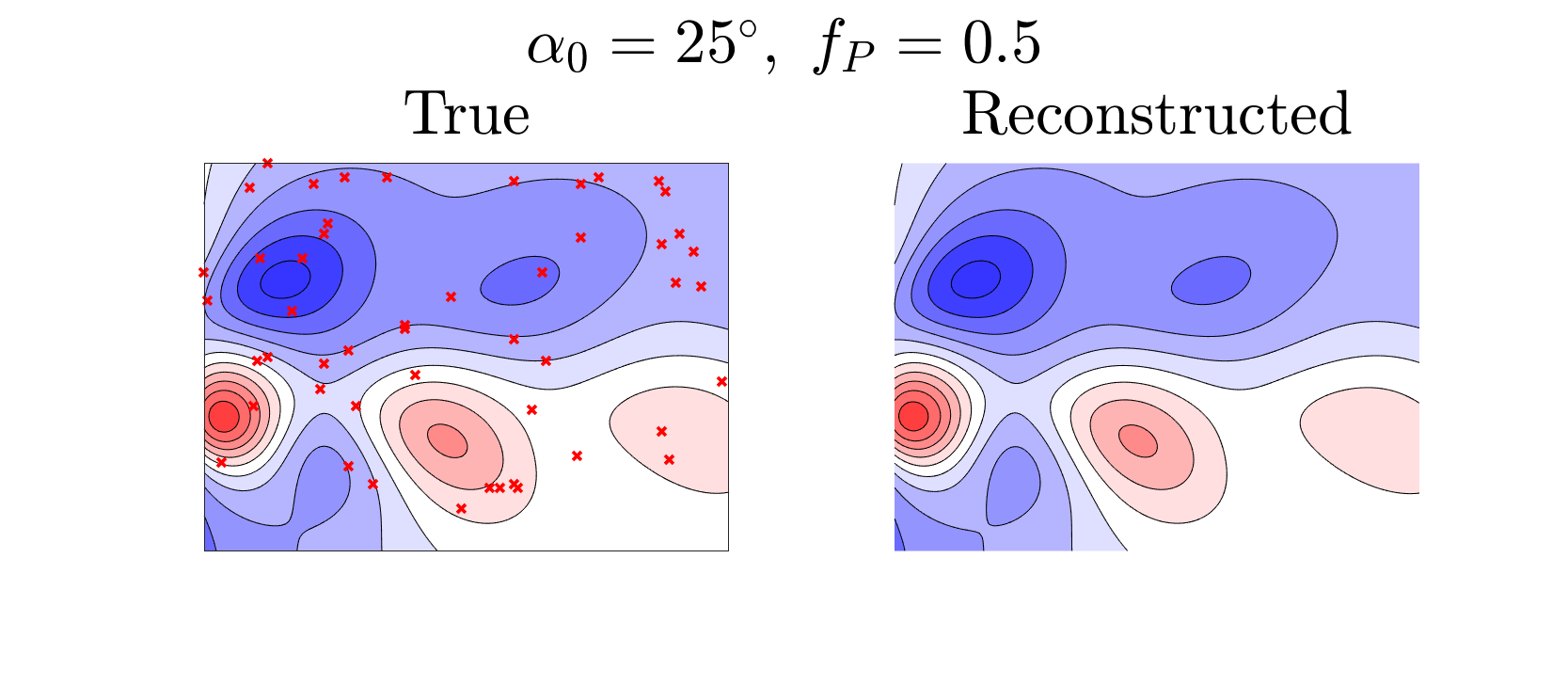}} \\ 
 \sidesubfloat[]{\includegraphics[width= 0.45\textwidth]{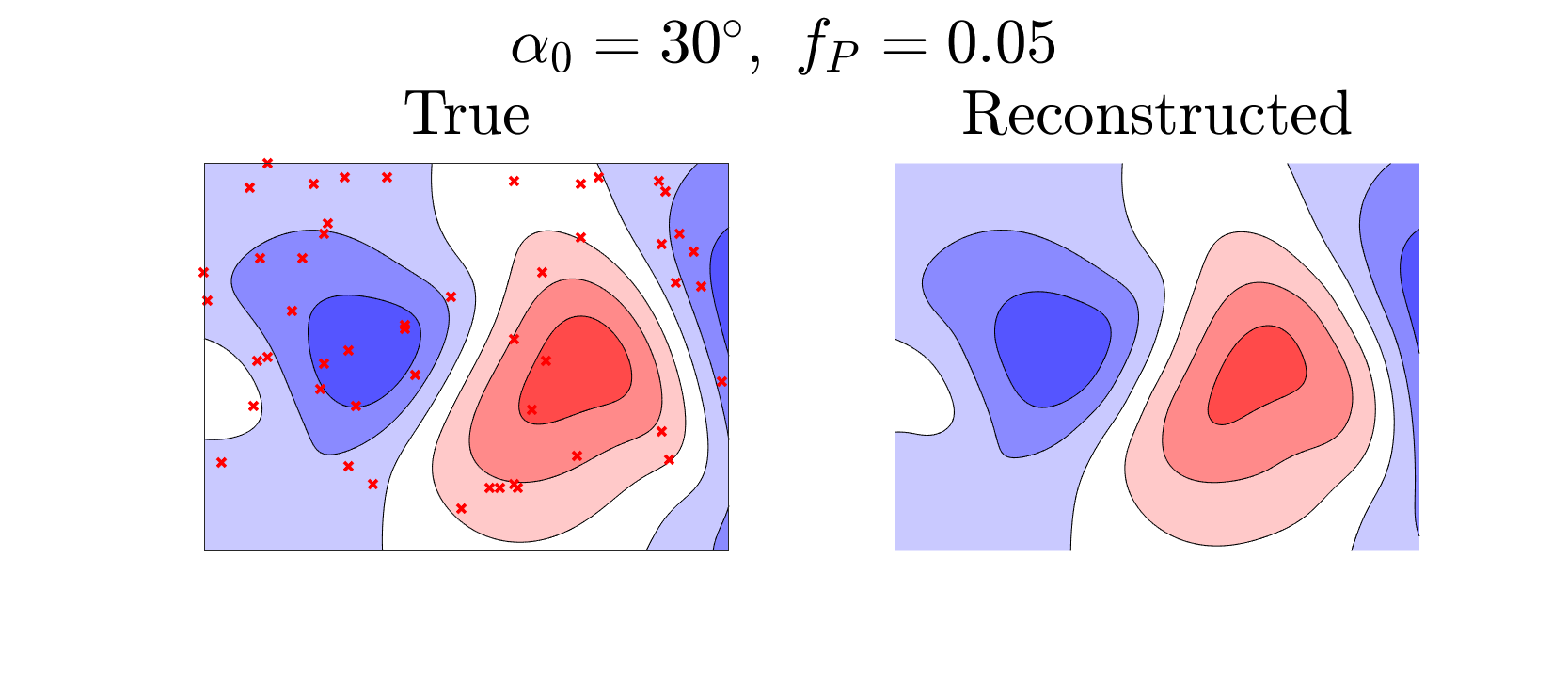}} 
 \sidesubfloat[]{\includegraphics[width= 0.45\textwidth]{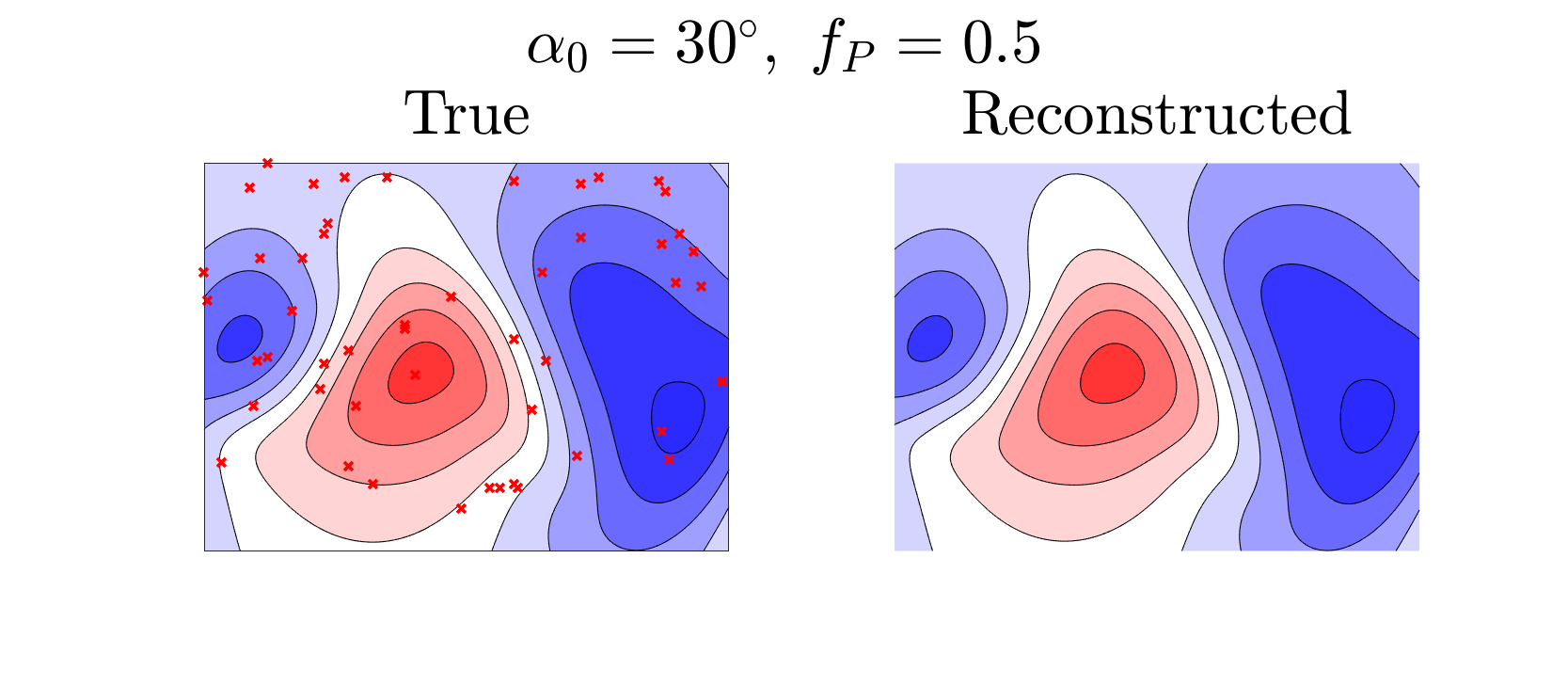}}
}
\caption{True and reconstructed instantaneous flow fields ($u_y$ component) for various cases. Red crosses indicate the randomly-chosen locations where data is collected.}
\label{fig:compressed}
\end{figure}

\begin{figure}
\centering{\includegraphics[width= 0.6\textwidth]{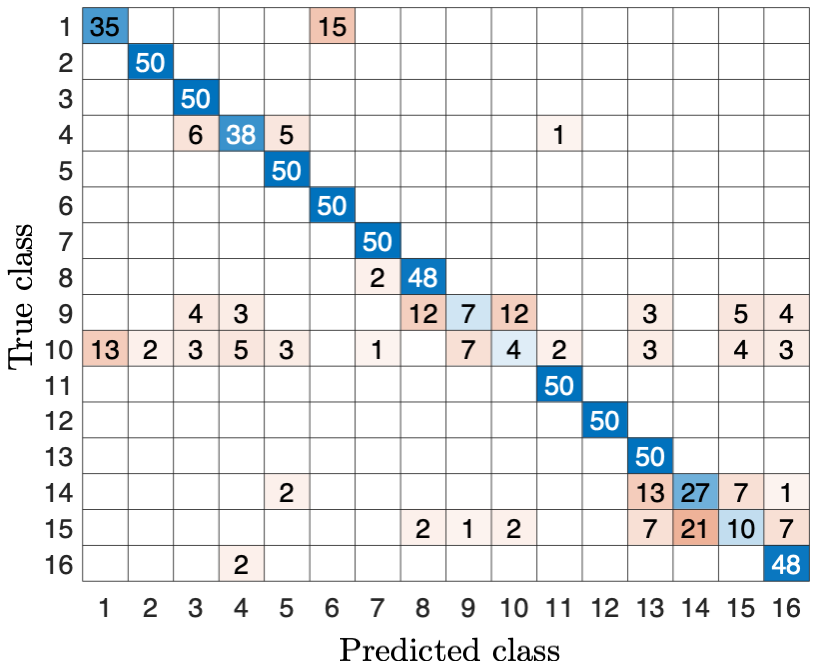}} 
\caption{Confusion matrix showing the number of cases that are correctly and incorrectly classified (with 50 total cases tested for each class). Class indices from 1-16 correspond to $(\alpha_0,f_P) = (25^\circ,0.05), (25^\circ,0.1),\dots,  (25^\circ,0.5),  (30^\circ,0.05),\dots, (30^\circ,0.5)$ respectively.}
\label{fig:confusion}
\end{figure}

Figure~\ref{fig:compressed} shows the true and reconstructed flow fields for several cases.  For determining the airfoil kinematics via Eq.~\eqref{eq:kinRecon}, we achieve an overall accuracy of 78\%. The confusion matrix, which shows how each of the 50 testing snapshots for each simulation are classified is shown in Fig.~\ref{fig:confusion}. It is observed that the $\alpha_0 = 25^\circ$ cases are more accurately classified ( 92.75\% accuracy) than the $\alpha_0 = 30^\circ$ cases (61.50\%), with the $\alpha_0 = 30^\circ$ and $f_P = 0.05$ and $0.1$ being particularly difficult to correctly identify.  This is likely due to the presence of natural vortex shedding in these cases, which obscures the effect of pitching on the wake dynamics. 
It is possible that improved performance in this task could be achieved with alternative machine learning methods for classification, such as $k$-nearest neighbors, mixture models, or neural network models.

%%%%%%%%%%%%%%%%%%%%%%%%%%%%%%%%%%%%%%%%%%%%%%%%%%%%%%%%%%%
% -- Airfoil EXP Section (Anya) ---------------------------

\section{Large-amplitude transverse gust encounter experiments}  % e.g., "Turbulent jet large eddy simulation"
\label{sec:airfoil_EXP}

Fifth, we consider a flat-plate wing at a constant towing speed $U$ and angle of attack $\alpha$ passing through a large-amplitude transverse gust at several gust ratios $V_\mathrm{max}/U_\infty$. The wing-gust encounter provides an example of a transient, separated flow containing large amounts of vorticity, with unsteady effects leading to large and rapidly changing aerodynamic forces. Advances in understanding, predicting, and controlling such systems can lead to improved efficiency and reliability of aerodynamic vehicles susceptible to such gust encounters.  The dataset contains time-resolved force and flow field measurements carried out in a water-filled towing tank, where the large-amplitude gust is created using a water jet system. Key features in the flow fields include the free-jet gust flow, a strong leading-edge vortex near the wing, and the wing wake. The force history is characterized by a large transient as the wing passes through the gust, followed by an eventual relaxation to steady-state. The data and figures shown in this section are taken from Refs.~\cite{bilerthesis, biler2021experimental}.

% -- Airfoil EXP setup  -----------------------------------------
\subsection{Setup}

% Describe the flow and where the data came from (simulation/exp setup, etc.)

\subsubsection{Facility, wing, and kinematics}

\begin{figure}
    \centering
    \includegraphics[width=0.75\textwidth]{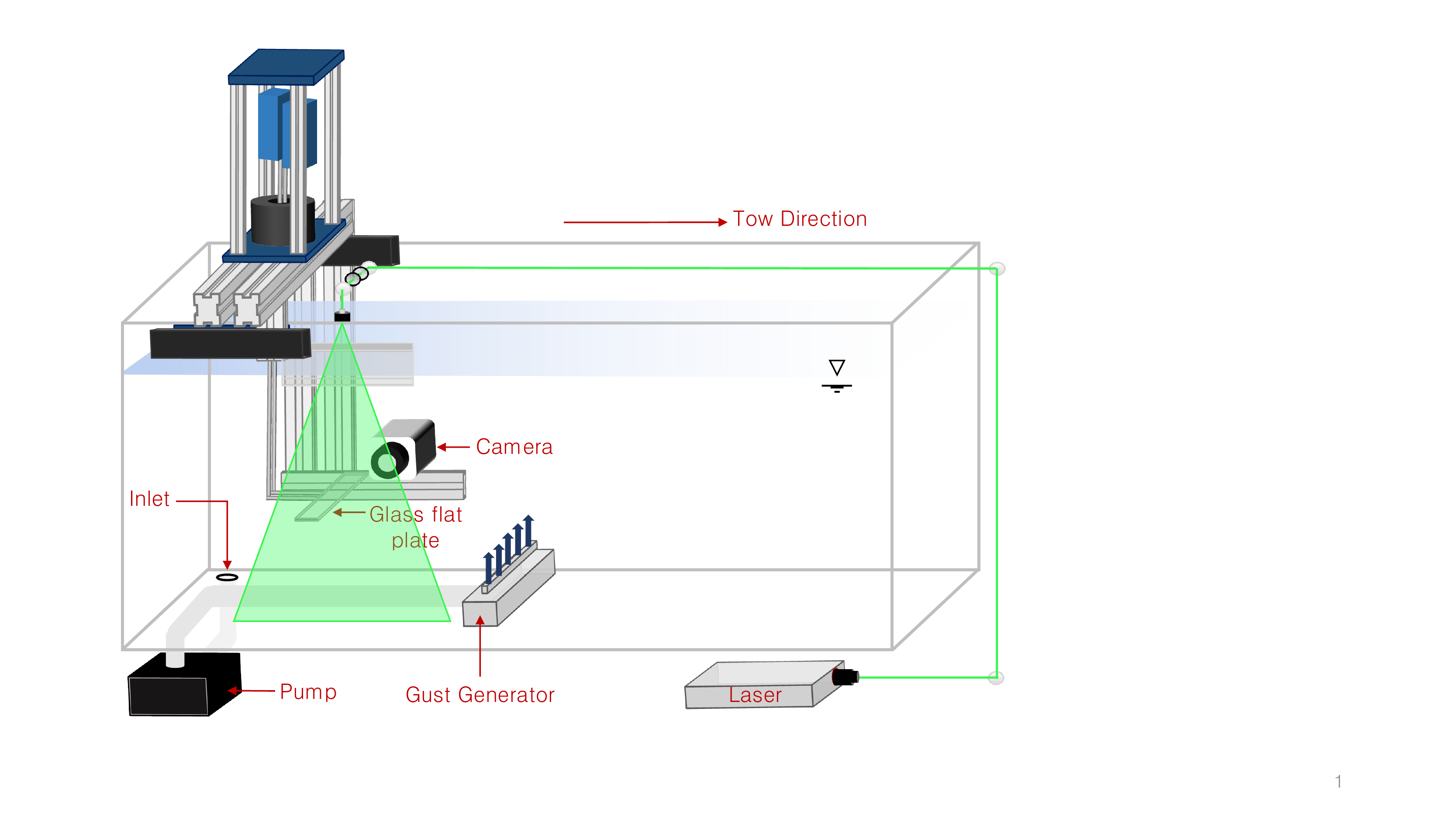}
    \caption{Towing tank with planar gust and PIV setup.}
    \label{fig:UMDsetup}
\end{figure}

This dataset was acquired during experiments performed in the University of Maryland free-surface water towing tank. The towing tank is 7\,m long, 1.5\,m wide, and 1\,m deep with a 4 degree-of-freedom (DOF) model motion control system that allows for streamwise, stream-normal, rotating, and pitching wing motions. A 4.3\% thick flat plate wing of aspect ratio 4 (3\,inch chord, 12\,inch span) was mounted on the towing carriage via two rods as shown in Fig.~\ref{fig:UMDsetup}. A glass flat plate wing was used for PIV flow field measurements to allow both sides of the wing to be imaged simultaneously. An aluminum plate of the same dimensions was used for force and moment measurements. 

For the experiments described here, the wing was set to a fixed angle of attack of $\alpha=0^\circ$ and towed along the length of the tank at a fixed height through a planar gust. The wing was accelerated to a constant velocity, traveled at this constant speed through the gust, and finally decelerated to a stop past the gust. Sufficient travel distances were provided before and after the wing enters and exits the gust to ensure that the flow around the wing reaches a steady state before the flow disturbance (i.e., gust encounter or deceleration). In the results that follow, $t^* = 0$ was defined as the time at which the leading edge of the wing enters the gust, where $t^* = t U_\infty$ denotes convective time. More details of this nominally trapezoidal towing velocity profile can be found in Refs.~\cite{biler2021experimental} and \cite{bilerthesis}.

The data provided here focus on the effects of large-amplitude gust encounters on a wing at a constant angle of attack of $0^{\circ}$ and  Reynolds number ($Re= U_\infty c/\nu$, where $U_\infty$ is the freestream speed, $c$ is the chord of the test model, and $\nu$ is the kinematic viscosity) of 20,000 ($U=0.2635$\,m/s). Force and PIV measurements were obtained for gust ratios $GR=V_\mathrm{max}/U_\infty={0.5, 0.75, 1.0, 1.5}$. Many more cases were also run and are detailed in Refs.~\cite{biler2018experimental, bilerthesis, biler2021experimental, andreu2020effect}. %All cases were repeated 5 times with 10 minutes between each run to ensure that flow disturbances from the previous run did not persist.

\subsubsection{Gust system}

The transverse gust flow was created in the water-filled towing tank using a vertically oriented jet system that produces a gust approximately 3 times as wide as the wingspan. %, made up of a pump (Hayward TriStar SP3202VSP), PVC pipes, plenum, and a honeycomb flow straightener (36\,$\times$\,0.5\,in) with an individual cell width of 1/8\,in. The gust system manifold is approximately 3 times as wide as the wing span in the cross-stream direction. 
To generate the gust flow, water is drawn from an inlet located at the bottom of the tank to the pump, which forces water through the plumbing system and flow straightener. The water pump can operate over a range of different speeds, which allows for a wide range of gust velocities. Gust ratio (i.e., the ratio of the gust velocity to the towed speed of the wing, $GR=V_\mathrm{max}/U_\infty$) can be controlled via the gust velocity, so the towing speed can be selected to obtain a better signal-to-noise ratio for the force measurements, thereby allowing the gust ratio variations to be independent of Reynolds number. Further details of the gust system and resulting flow can be found in Refs.~\cite{biler2018experimental, bilerthesis, bilersedky}. %The resulting gust flow was characterized for 13 pump speeds and 4 spanwise locations using time-resolved PIV as detailed in Ref.~\cite{biler2018experimental, bilerthesis}.

\begin{figure}
 \centering {
 \sidesubfloat[]{\label{fig:UMD_GustShearLayer}\includegraphics[width= 0.42\textwidth]{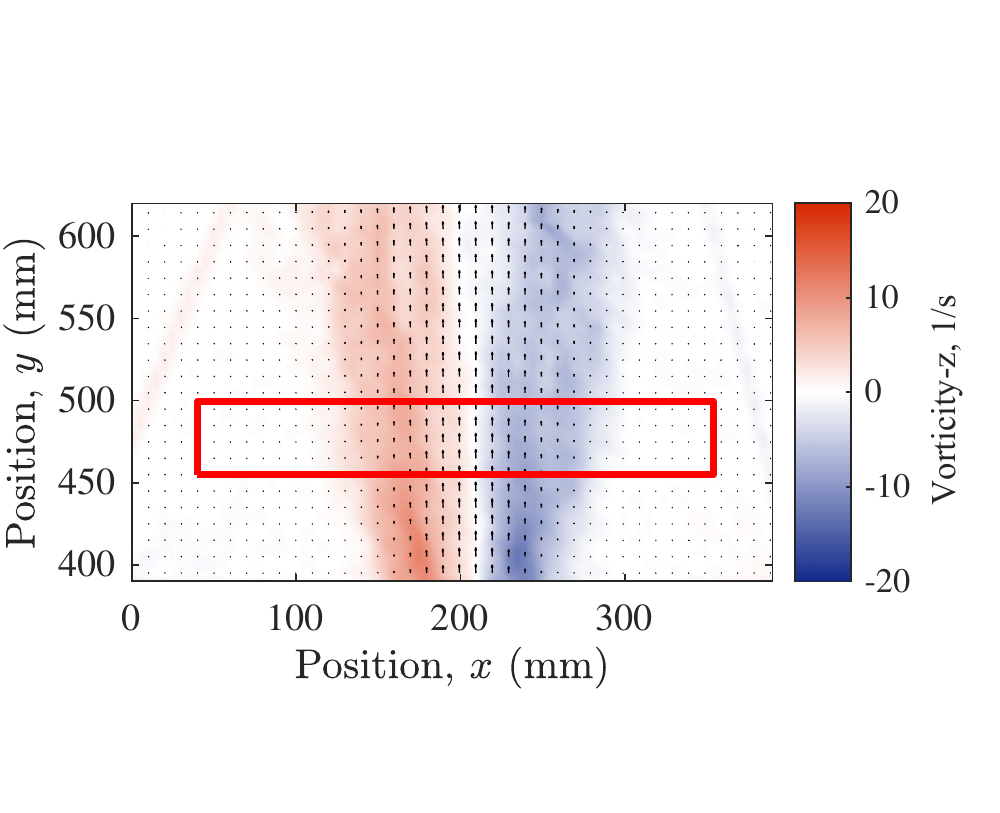}} \ \ 
 \sidesubfloat[]{\label{fig:UMD_Fit}\includegraphics[width= 0.42\textwidth]{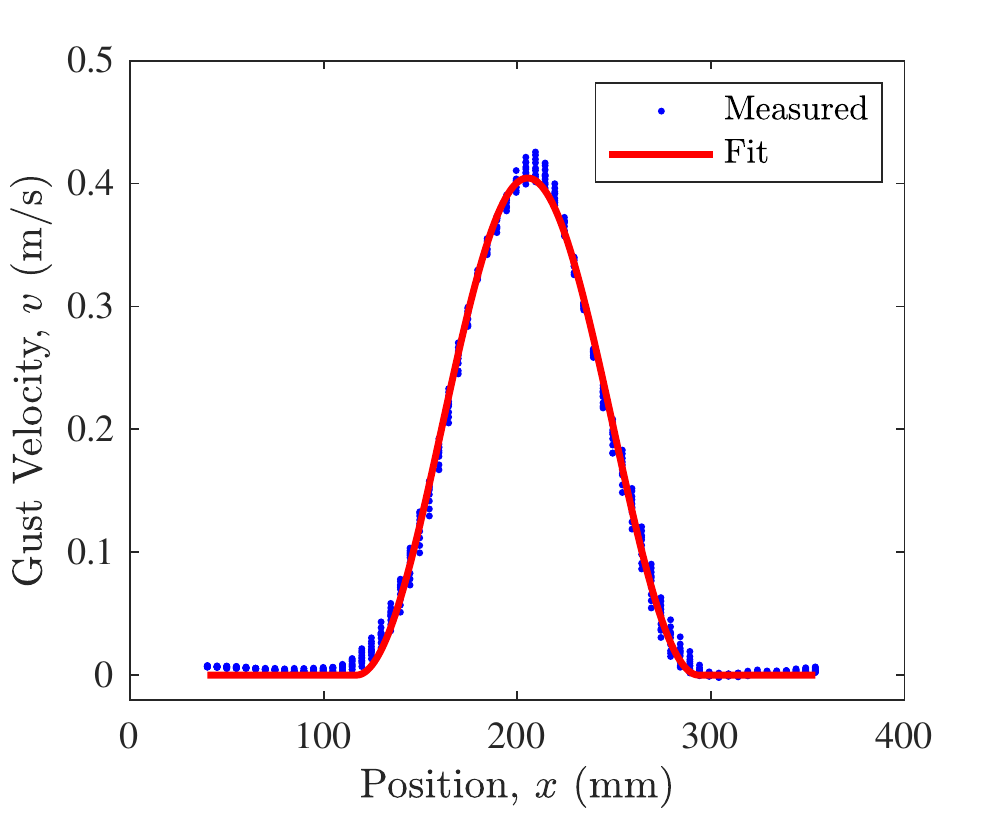}}
}
\caption{Sine-squared gust characterization for the highest pump speed at the centerline of the wing: (a) time-averaged gust vorticity and (b) measured gust velocities and sine-squared fit.}
\label{fig:UMD_streamwise}
\end{figure}

Figure~\ref{fig:UMD_GustShearLayer} shows the time-averaged gust vorticity for the highest pump speed at the center of the gust. As can be seen from this figure, the shear layer covers the entire width of the gust and thus the bulk of the gust flow is rotational. Here, the $x$ and $y$ axes represent the horizontal (streamwise) and vertical (cross-stream) positions, respectively. The red rectangle in this figure represents the region of interest through which the wing passes. The mean measured velocities in the direction normal to the freestream (i.e., vertical velocities) were extracted from this region and are shown in Fig.~\ref{fig:UMD_Fit} along with the best fit to the experimental data, which is found to be a sine-squared profile 
\begin{equation}
\label{eq:gust}
V(x)=V_\mathrm{max} \sin^2\left(\frac{\pi x}{w}\right), 
\end{equation}
where $V_\mathrm{max}$ represents the peak gust velocity, $x$ represents the streamwise position, and $w$ represents the gust width in the streamwise direction. Due to unsteadiness in the gust flow, there was some fluctuation in the width of the gust but the mean gust width was found to be 198\,mm with a standard deviation of 22\,mm or coefficient of variation of 11\%. The ratio of the width of the gust to the wing chord ($w/c$) was thus taken to be 2.6~\cite{biler2018experimental, bilerthesis}. 
The duration of the gust encounter ($T_\mathrm{gust} = w/U_\infty$) was found to be 0.75\,s, corresponding to a frequency of 1.3\,Hz.

%The database provided from this study includes ensemble averaged force coefficents (both lift and drag) and flow fields obtained using time-resolved high-speed planar PIV. This section describes the methods used to obtained the measurements. The reader is referred to Ref.~\cite{bilersedky} for more details.

\subsubsection{Flow field measurements}

Time-resolved flow fields of the gust encounter were measured using high-speed PIV. To acquire the images, a 4~MP (Phantom v641) camera was used with an 85~mm Nikon lens. The water in the towing tank was seeded with class IV soda lime glass spheres (diameter of 37~$\mu$m), and the flow was illuminated by an Nd:YLF laser (Litron LDY 304, 30~mJ/pulse, 10~kHz max). A sketch of the setup is shown in Fig.~\ref{fig:UMDsetup}. During these experiments, both the laser sheet and the camera were towed with the wing to allow for long time histories to be acquired in a wing-fixed reference frame. The laser sheet illuminated a plane 1 chord-length off mid-span, where the images were captured to provide nominally two-dimensional flow fields for the aspect-ratio-4 wing. Images could not be acquired at the mid-span since the support structure does not allow for such a configuration of the optics. 
%4250 images were recorded for each case. DaVis v8.1 was used as the processing software. A sliding background was first subtracted from the acquired images to increase the signal-to-noise ratio of the measurements. Images were then cross-correlated using a decreasing interrogation window size from 24$\times$24 to 16$\times$16\,px. 50\% and 75\% region overlaps were allowed for the first and second passes. The resulting field of view was 24$\times$16\,cm, and the wing was positioned in the image such that approximately 1.1 and 1.2 chords were allowed upstream and downstream of it, respectively. 
The maximum uncertainty in the velocity measurements was found to be 0.2\%.

\subsubsection{Force measurements}

\begin{figure}
 \centering {
\sidesubfloat[]{\label{fig:WingMount}\includegraphics[width=0.5\textwidth]{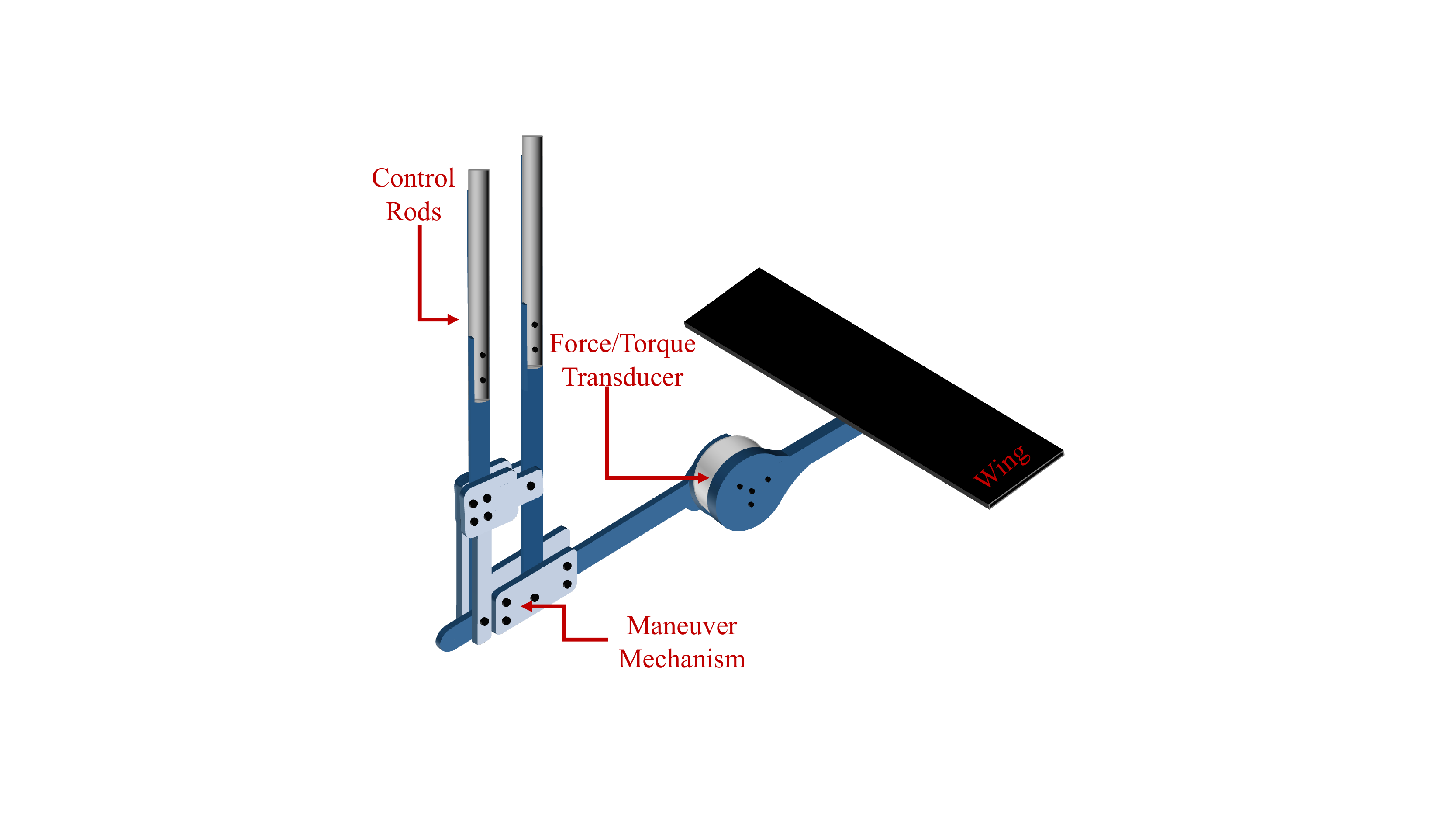}} \ \ 
\sidesubfloat[]{\label{fig:UMD_Filter}\includegraphics[width=0.41\textwidth]{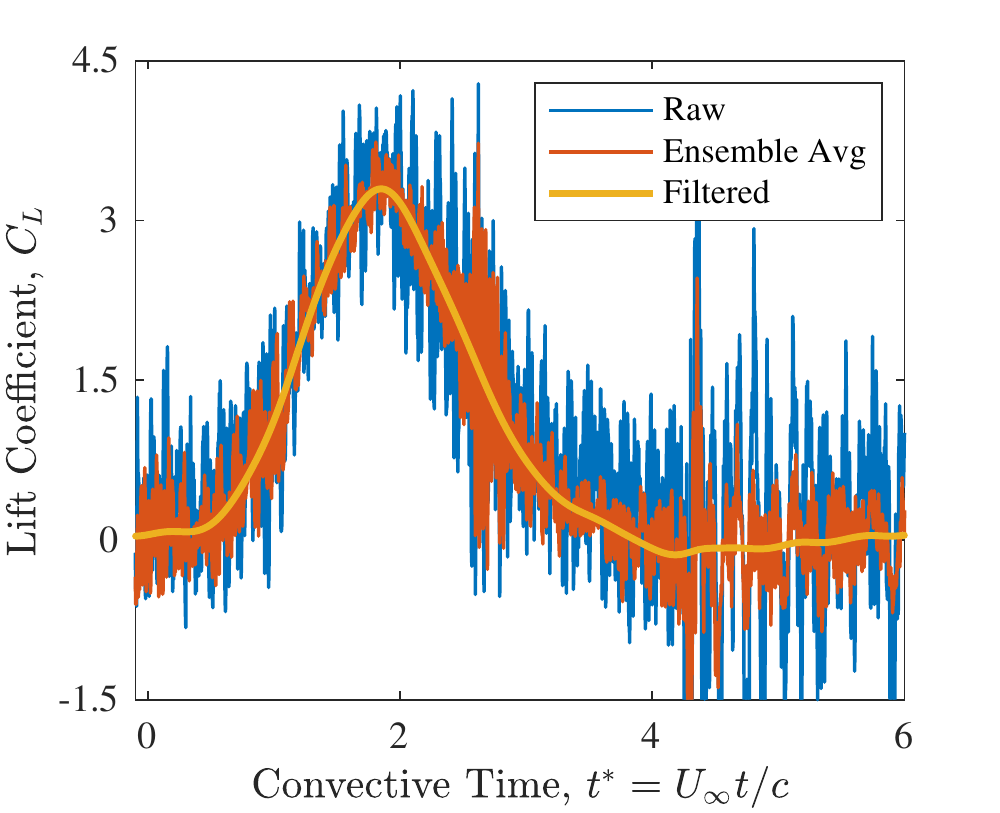}}
}
\caption{(a) Wing installation mechanism. (b) Raw, ensemble averaged, and filtered lift coefficients.}
\label{fig:UMD_ExpSetup2}
\end{figure}

A 6-degree-of-freedom force balance (submergible ATI Mini-40 with a range of $\pm 40$~N and a resolution of $0.01$~N) was used to measure forces on the test model at a sampling rate of 1~kHz. The force balance was mounted to the setup as shown in Fig.~\ref{fig:WingMount}. All of the cases were repeated 5 times, synchronized, filtered, and ensemble-averaged. The average forces were calculated when the wing was stationary and subtracted from the measurements to remove the contributions from gravity and buoyancy. The mean normal and axial forces were then converted to lift and drag components and non-dimensionalized by the dynamic pressure and the wing area. A Fast Fourier Transform was performed for all cases to reveal the frequency content of the measurements and the dominant frequencies were found to be 10, 35, and 110~Hz.
%The period of the gust encounter ($T_\mathrm{gust} = w/U_\infty$) was found to be 0.75\,s, corresponding to a frequency of 1.3\,Hz.
%The period of the gust encounter ($T_\mathrm{gust} = w/U_\infty$) for low-amplitude and large-amplitude cases were found to be 0.2s (corresponding to a frequency of 4\,Hz) and 0.75\,s (corresponding to a frequency of 1.3\,Hz), respectively. 
To remove mechanical and electrical vibrations and measurement noise in the data, a 5 Hz low-pass Butterworth filter was applied. Figure~\ref{fig:UMD_Filter} shows the raw, ensemble-averaged, and filtered lift coefficients for $GR = 1$ case. As can be seen from this figure, the low-pass filter reduces the vibrations and noise without truncating the peak and distorting the response.

\subsection{Data}

The dataset provided includes ensemble-averaged time-resolved flow fields and force coefficients for a flat plate wing at $0^{\circ}$ angle of attack encountering gusts of various strengths, $GR=V_\mathrm{max}/U=\{0.5, 0.75, 1.0, 1.5\}$.

\subsubsection{Time-resolved velocity fields}

\begin{figure}
 \centering {
\sidesubfloat[]{\label{fig:SamplePIV}\includegraphics[width=0.5\textwidth]{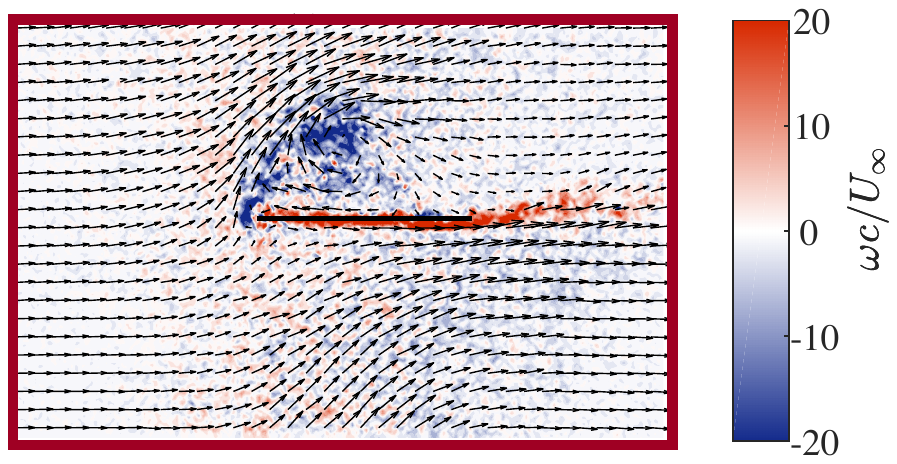}} \ \ 
\sidesubfloat[]{\label{fig:UMD_PIVSensitivity}\includegraphics[width=0.4\textwidth]{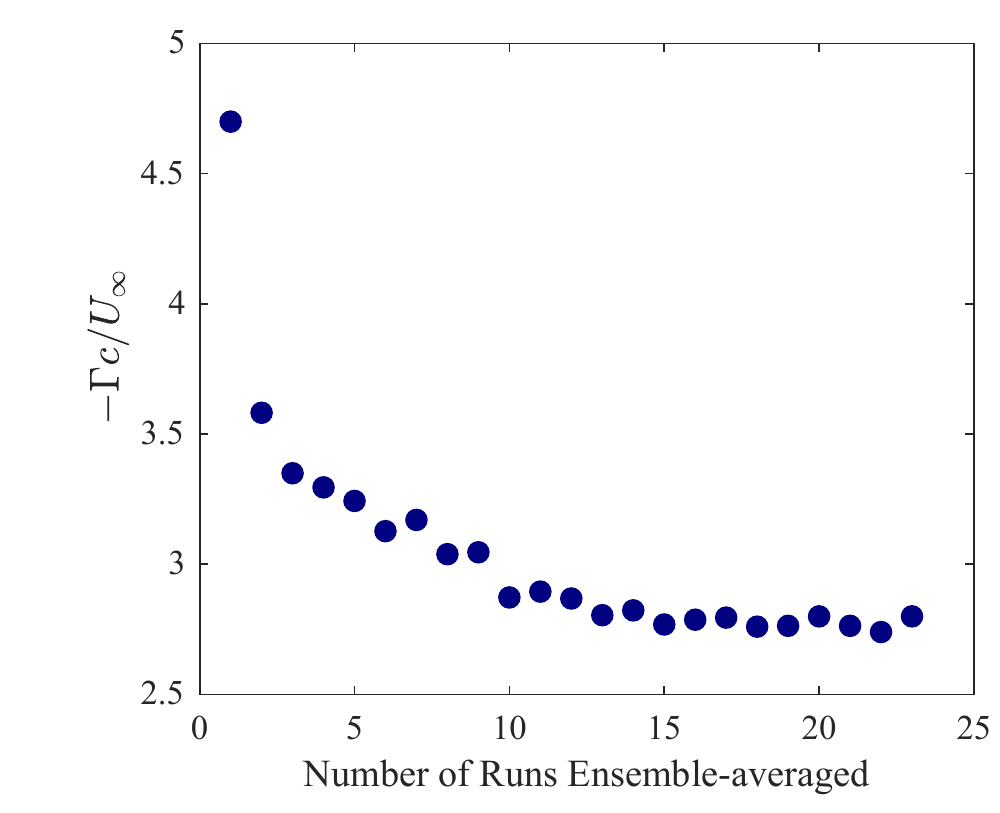}}
}
\caption{(a) A sample PIV flow field and (b) PIV sensitivity sensitivity study based on the non-dimensional leading-edge vortex strength for $GR=1$.}
\label{fig:UMD_PIVSens}
\end{figure}

The dataset includes 1284 time-resolved planar velocity fields for $0 \leq t^*\leq 6.33$ obtained using using high-speed planar PIV. %, where $t^* = t U_\infty$ denotes convective time, and $t^* = 0$ corresponds to the start of the gust encounter. 
The resulting field of view was 24$\times$16\,cm, and the wing was positioned in the image such that approximately 1.1 and 1.2 chords are visible upstream and downstream of it, respectively. The final spatial resolution of the data was 0.1\,mm per pixel (772 pixels per chord-length), and the temporal resolution was 202 images per convective time (giving 1284 snapshots sampled at a dimensionless time step of $\Delta t^* = 0.00345$). After processing, the final vector spacing was 0.79\,mm (96.46 vectors per chord-length). See Ref.~\cite{bilerthesis} for more details.  A sample PIV flow field at $t^{*} = 1.8$ is shown in Fig.~\ref{fig:SamplePIV} for the $GR=1$ case.  

A sensitivity study was performed to determine the number of runs to ensemble average. 23 runs of $GR = 1$ case were acquired. Since the gust meanders, the sensitivity study was carried out for the leading-edge vortex strength rather than the total circulation in the PIV flow field. The procedure used to calculate the leading-edge vortex strength is detailed below and in Ref.~\cite{bilerthesis}. Figure~\ref{fig:UMD_PIVSensitivity} shows the non-dimensional circulation of the leading-edge vortex at $t^{*} = 1.8$ for various numbers of runs ensemble-averaged. As can be seen from this figure, the circulation approaches a constant value as the number of runs increases. To make the work feasible, the cut-off value for number of runs to be ensemble-averaged was selected to be 8 which corresponds to 8\% error for the $GR = 1$ case shown here.

\subsubsection{Time-resolved force measurements}

The force data provided consist of normal and axial forces for $-5 \leq t^*\leq 10$, with 4337 samples at a dimensionless time step of $\Delta t^* = 0.00346$. Since $\alpha=0^{\circ}$, these correspond to lift and drag forces.  As mentioned above, the raw force data is ensemble averaged (over five realizations) and filtered to improve the signal-to-noise ratio.  The database includes the individual filtered realizations as well as the ensemble-averaged data.  Figure~\ref{fig:UMD_Filter} shows the raw, ensemble-averaged, and filtered lift coefficients for the $GR = 1$ case. The low-pass filter used here reduces the vibrations and noise without truncating the peak or distorting the response.

%The database provided from this study includes ensemble-averaged force coefficients and flow fields obtained using time-resolved high-speed planar particle image velocimetry for a flat plate wing at $0^{\circ}$ angle of attack encountering gusts of various strengths ($GR=V_\mathrm{max}/U={0.5, 0.75, 1.0, 1.5}$). Force data provided consists of normal and axial forces for $-5 \leq t^*\leq 10$, which corresponds to lift and drag forces for $\alpha=0^{\circ}$. The database also contains 1284 velocity fields obtained using PIV for $0 \leq t^*\leq 6.33$. The resulting field of view was 24$\times$16\,cm, and the wing was positioned in the image such that approximately 1.1 and 1.2 chords were allowed upstream and downstream of it, respectively. The final spatial resolution of the data were 0.1\,mm per pixel (772 pixels per chord-length), and the temporal resolution was 202 images per convective time. The final vector spacing was 0.79\,mm (96.46 vectors per chord length). The reader is referred to Ref.~\cite{bilerthesis} for more details.

\subsection{Example analysis}

The forces and flow fields included here establish an experimental database for the validation of computational study, development of low-order models, and comparison of transverse gusts. From these data it is possible to both compute the strength and trajectory of vortices that form on the wing and in the wake and to relate the resulting wing forcing to canonical models, thereby highlighting the underlying physics of unsteady force production in a wing-gust encounter.

\subsubsection{Circulation measurements}
One of the key characteristics of a large-amplitude gust encounter is large-scale flow separation and the formation of a leading-edge vortex. 
The presence, strength, and evolution of this vortex can have a significant effect on aerodynamic forces, so measuring its strength at a given instance in time from experimental data can be valuable both for understanding flow physics and to provide a circulation parameter that can be used to test and/or calibrate reduced-complexity models.

\begin{figure}
 \centering {
\sidesubfloat[]{\label{fig:UMD_Threshold}\includegraphics[width=0.45\textwidth]{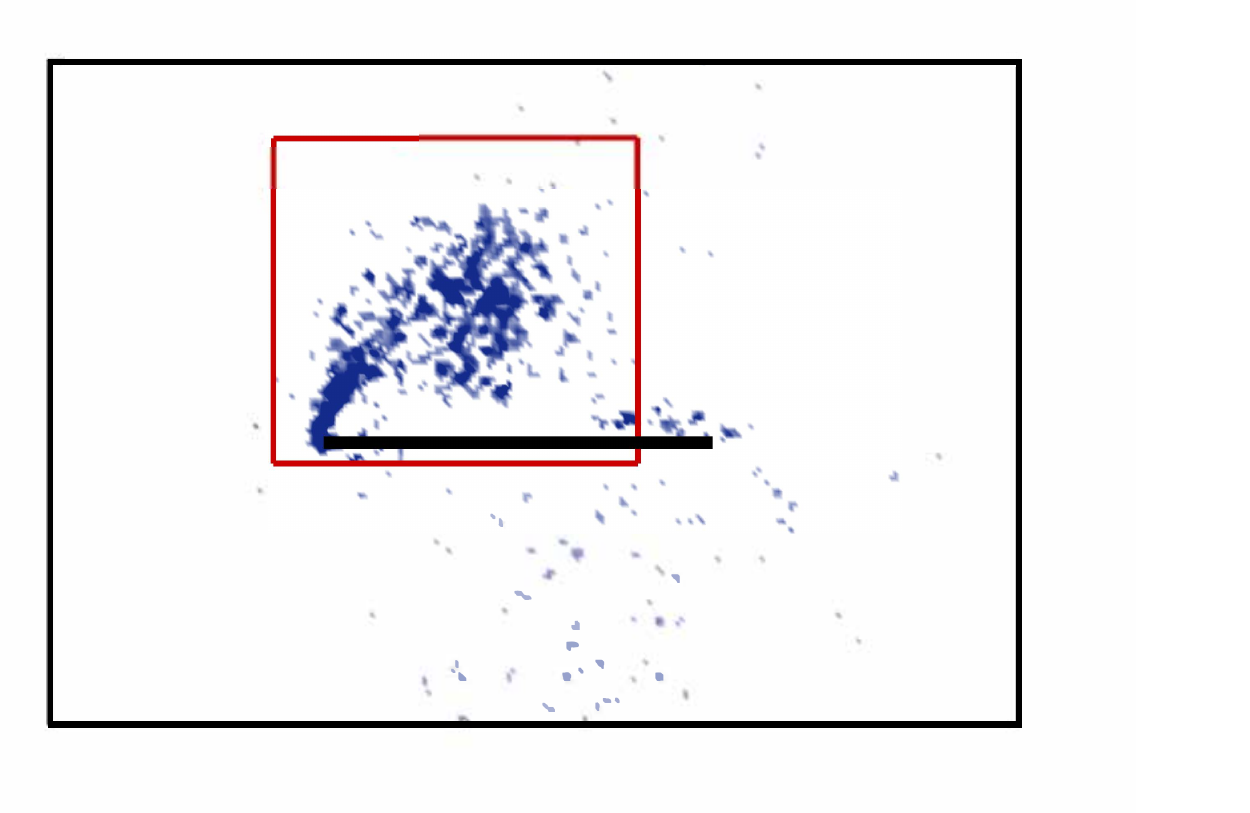}} \ \ 
\sidesubfloat[]{\label{fig:UMD_Polygon}\includegraphics[width=0.45\textwidth]{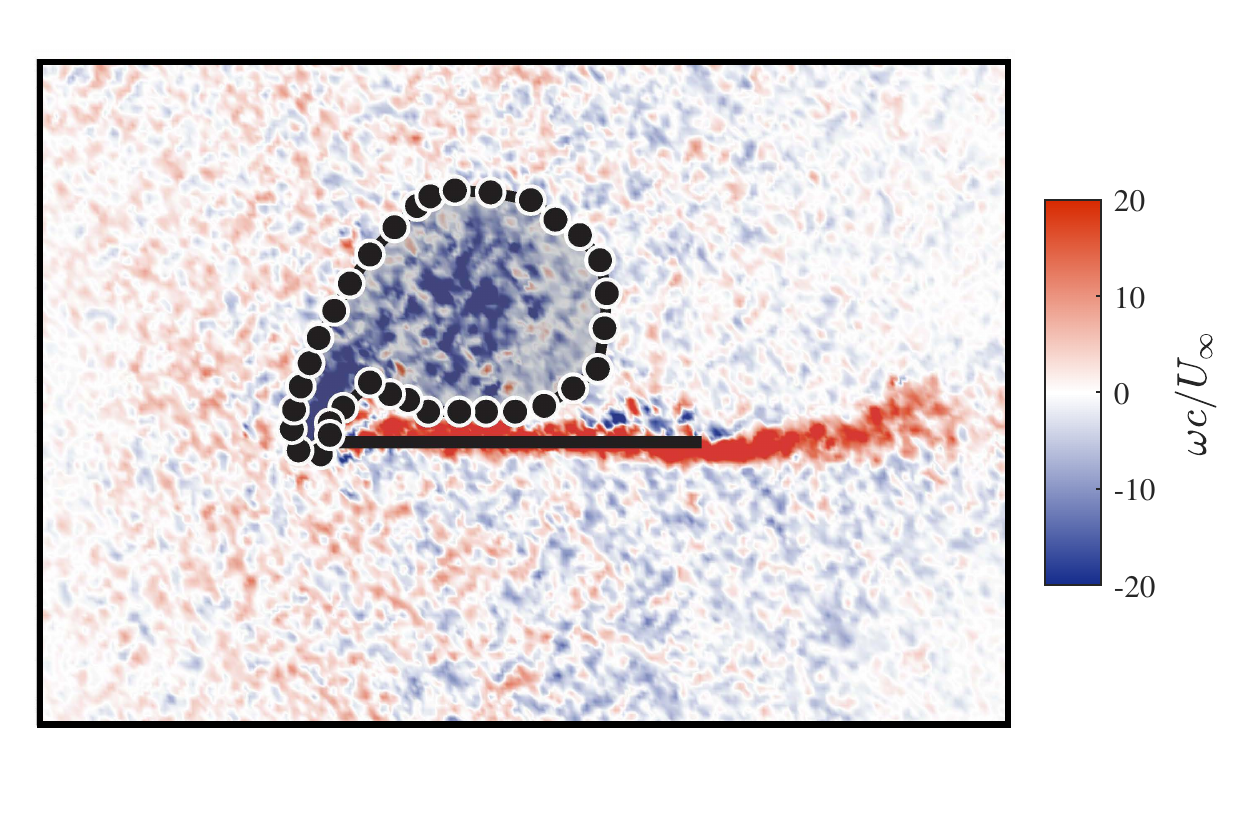}}
}
\caption{Vorticity contours at $t^*=1.8$ for $GR=1, \alpha=0^\circ$ case showing (a) threshold method and (b) segmentation method.}
\label{fig:CircMeasVort}
\end{figure}

\begin{figure}[t]
	\centering
 \includegraphics[width=0.4\textwidth]{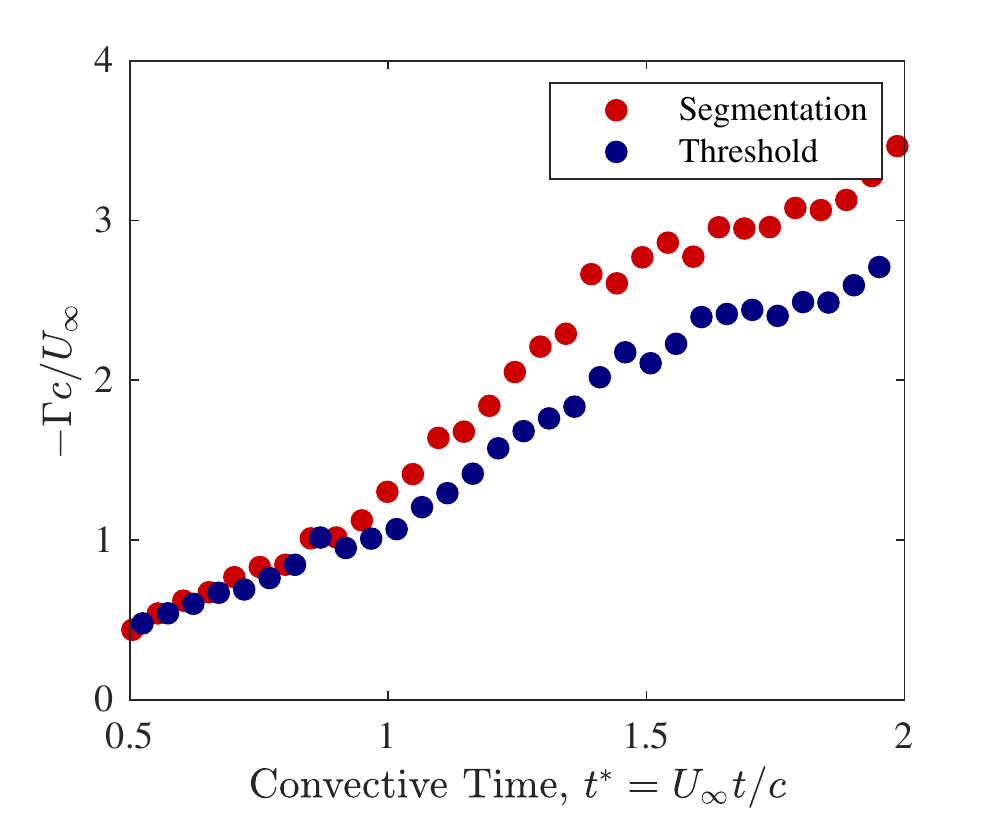}
 \caption{Nondimensional circulation calculated from both the segmentation and the threshold methods.}
 \label{fig:UMD_CircComparison}
\end{figure}

The strength of this vortex can be computed from experimental flow fields by computing the vorticity field and integrating to obtain the total circulation of the vortex, $\Gamma = \int \omega \, \mathrm{d}A$. For a sine-squared gust, this is somewhat challenging since the gust flow introduces a continuous sheet of vorticity due to the spatial variations in the velocity profile of the gust. Two methods that are appropriate for this task are (1) a threshold method (as shown in Fig.~\ref{fig:UMD_Threshold}) and (2) a segmentation method (as shown in Fig.~\ref{fig:UMD_Polygon}). In the threshold method, a stationary rectangular box is drawn near the wing and the vorticity field calculated within it. A threshold is set below which any vorticity is removed to avoid including effects from the gust shear layer, thereby isolating the leading-edge vortex. In the segmentation method, the leading-edge vortex is manually isolated from the gust shear layer using an arbitrary polygon. Since the leading-edge vortex contains negative vorticity and a part of the gust shear layer is positive, a threshold was also used to remove any positive vorticity due to the gust shear layer. The vorticity was then integrated around the polygon area, and the strength of the leading-edge vortex found. The resulting circulation from both the threshold and the segmentation methods was nondimensionalized by the freestream velocity and chord. Figure~\ref{fig:UMD_CircComparison} shows these nondimensional circulations for the $GR = 1$ case. As can be seen from this figure, these methods yield similar results early on. However, as the wing moves towards the center of the gust, the circulations calculated from these two methods start to differ. These differences are attributed to the fact that the vortex diffuses and at some point the threshold method eliminates some of the vorticity that originally belonged to the leading-edge vortex. This issue can be addressed by increasing the threshold value. However, the increased threshold value causes the gust shear layer to reappear in the box and thus the method to fail. %To conclude, the best method was determined to be the segmentation method, and it was used to calculate the leading-edge vortex strength for the sine-squared gust encounters. Please note that the measurements were down-sampled to 70 Hz to make the manual work feasible.

\subsubsection{K\"{u}ssner's model}

K\"{u}ssner's model is a linear analytical model for the lift force on a flat plate encountering a sharp-edged transverse gust of infinite width. Although the assumptions in this model are quite strict, the model can be used to predict lift for a wide variety of small gust velocity disturbances. It does, however, fail to do so for moderate to large gusts, where flow separation occurs. K\"{u}ssner's \cite{kussner1936zusammenfassender} lift equation for a wing passing through a sharp-edged gust is

\begin{equation}  \label{eqnA}
\centering
C_l= 2 \pi \frac{V}{U_{\infty}} \psi\left(t^{*} \right),
\end{equation}
where the dimensionless time $t^{*}$ also represents the number of chord lengths the wing has traveled, $V$ is the gust velocity, and $\psi(t^{*})$ is the K\"{u}ssner function which can be calculated iteratively. The K\"{u}ssner function, however, is generally replaced by the exponential approximation given by \cite{aeroelasticity}
\begin{equation}  \label{eqnB}
\centering
\psi(t^{*})\approx 1-0.5 e^{-0.26 t^{*}}-0.5 e^{-2 t^{*}},
\end{equation}

K\"{u}ssner's model can be used to find the lift response to any arbitrary transverse gust profile when used with the Duhamel superposition integral \cite{leishman2006principles}

\begin{equation}  \label{eqnC}
\centering
C_l= \frac{2 \pi}{U_{\infty}}\left ( V_{g}(0)\, \psi(t^{*}) + \int_{0}^{t^{*}} \frac{\mathrm{d}V_{g}(\sigma) }{\mathrm{d}t} \psi(t^{*}-\sigma)\, \mathrm{d}\sigma  \right ),
\end{equation}
where $\sigma$ is a variable of integration, $\psi(t^{*})$ is taken as in Eq.~\eqref{eqnB} and $V_{g}$ is the forcing function, i.e., the gust velocity profile. In order to provide a direct comparison between K\"{u}ssner's analytical solution and the current experiments and simulations, the forcing function was defined as a sine-squared gust velocity profile given by Eq.~\ref{eq:gust} with $x=t^* c$.

\begin{figure}[t]
	\centering
 \includegraphics[width=0.4\textwidth]{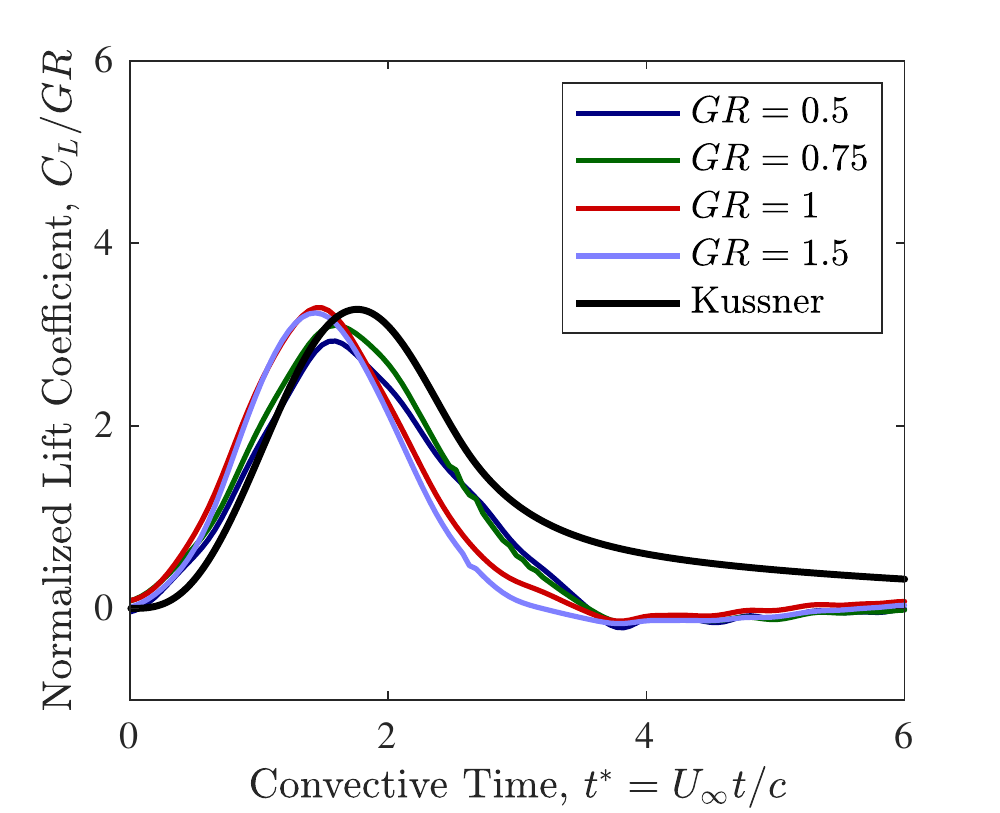}
 \caption{Measured and convoluted gust ratio normalized lift coefficients for a flat plate at $0^{\circ}$ angle of attack encountering sine-squared gust of various strengths.}
 \label{fig:UMD_Kussner}
\end{figure}

Figure~\ref{fig:UMD_Kussner} shows the lift histories normalized by gust ratio from the convolution as well as experimental measurements for the various strength sine-squared gust encounters. As can be seen from this figure, K\"{u}ssner's model does a good job predicting the lift maximum and the transients up to that point in time despite violation of the small-amplitude disturbance assumption inherent in the model. However, it also predicts a much slower recovery than the measurements.  The reader is referred to Refs.~\cite{bilerthesis,andreu2020effect,bilersedky} for more details on K\"{u}ssner's model and its application to sine-squared transverse gusts. It should also be noted that unsteady potential flow models can be extended to explicitly account for some of the phenomena neglected in K\"{u}ssner's model, e.g., the presence of leading-edge vortices \cite{ramesh2014discrete}.

%%%%%%%%%%%%%%%%%%%%%%%%%%%%%%%%%%%%%%%%%%%%%%%%%%%%%%%%%%%
% -- Airfoil LES Section (Chi-An & Sam) -------------------

\section{Turbulent airfoil wake large eddy simulation}
\label{sec:airfoil_LES}

Sixth, we present a turbulent wake flow downstream of a spanwise-periodic NACA 0012 airfoil at a moderate Reynolds number of $Re = 23{,}000$.  The flow features coherent structures associated with the Kelvin-Helmholtz instability over the separation bubble and von K\'arm\'an vortex shedding in the wake while retaining the broadband complexity of the turbulent flow, making it an ideal platform for CFD validation, flow analysis, and exploring reduced-complexity models.  The dataset contains 16,000 time-resolved snapshots of the mid-span and spanwise-averaged velocity fields obtained with LES.

% -- Airfoil LES setup  ---------------------------------------
\subsection{Setup}
% Describe the flow and where the data came from (simulation/exp setup, etc.)

\begin{figure}[t]
\begin{center}
\begin{overpic}[width=6.5in]{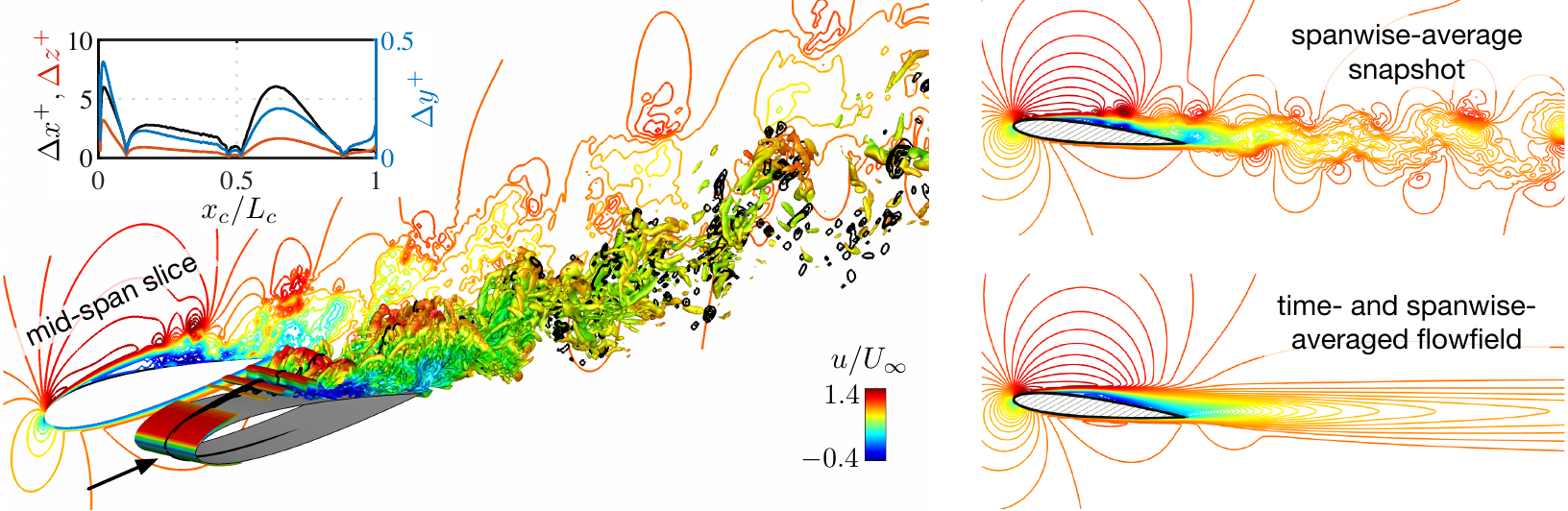}
	\put(00, 31){\small (a)}
	\put(63, 31){\small (b)}
	\put(63, 13.5){\small (c)}
\end{overpic}
\end{center}
\caption{\label{fig:airfoilLES_3DVis}Visualizations of the turbulent airfoil wake at a chord-based Reynolds number $Re = 23,000$ and free stream Mach number $M_\infty = 0.3$.  The wall-adjacent cell size of the computational mesh is superposed on the top-left, with respect to $\Delta x^+$, $\Delta y^+$, and $\Delta z^+$ over the suction side of the wing. }
\end{figure}

We consider a turbulent wake flow downstream of a NACA 0012 airfoil at a Reynolds number $Re = \rho_\infty U_\infty L_c / \mu_\infty = 23{,}000$, a freestream Mach number $M_\infty = U_\infty /a_\infty = 0.3$, and an angle of attack of $6^\circ$ from \citet{Yeh:JFM2019}.  Here, $L_c$ is the chord length, and $\rho_\infty$, $U_\infty$, $\mu_\infty$ and $a_\infty$ are the freestream density, velocity, dynamic viscosity, and speed of sound, respectively.  A visualization of an instantaneous flow field using the iso-surface of $Q$-criterion \citep{Hunt:CTR1988} is shown in Fig.~\ref{fig:airfoilLES_3DVis}a.   

The dataset is generated from the wall-resolved large-eddy simulation (LES) of a spanwise-periodic flow over the airfoil using the finite-volume compressible flow solver {\it{CharLES}} \citep{Khalighi:AIAA11,BresAIAAJ2017,BresJFM2018} with Vremen's sub-grid scale model \citep{Vreman:2004p754}.  The computational mesh is designed based on a C-shape structured grid adaptively refined in the wake region, resulting in a hybrid topology with structured sub-domains and unstructured regions at the interface of adaptive refinement.  The domain size for the LES is $x/L_c \in [-19, 26]$, $y/L_c \in [-20, 20]$ and $z/L_c \in [-0.1, 0.1]$, respectively, in the streamwise, transverse, and spanwise directions, with the leading edge of the airfoil positioned at the origin.  The wall-adjacent cell sizes  $\Delta \boldsymbol{x}^+ \equiv \Delta (x, y, z)^+ \sqrt{\bar{\tau}_w/\rho_\infty}/\nu_\infty$ over the suction surface are inserted in Fig.~\ref{fig:airfoilLES_3DVis}a.  These values are evaluated using the time- and spanwise-averaged wall shear stress $\bar{\tau}_w$ obtained from the LES, suggesting sufficient near-wall resolution for the wall-resolved LES \cite{Georgiadis:AIAAJ2010,ChoiMoin:PoF2012}.  Higher values for $\Delta \boldsymbol{x}^+$ are observed over the first $10\%$ of the chord compared to the rest of the suction surface, but this is deemed to be of no concern since the flow is laminar in this region.  The resulting computational mesh contains approximately 35 million cell volumes.   The time integration is performed at a constant time step of $\Delta t U_\infty/L_c = 4.14 \times 10^{-5}$, corresponding to a maximum Courant-Friedrichs-Lewy (CFL) number of $0.86$.  Additional information on the boundary conditions, grid convergence, and validations with respect to aerodynamic forces and pressure distribution can be found in \citet{Yeh:JFM2019}.

% -- Airfoil LES data  ---------------------------------------
\subsection{Data}
% Describe what data is available in the database

\subsubsection{Time-resolved midspan and spanwise-averaged velocity fields}

The dataset is comprised of $n_t = 16{,}000$ time-resolved snapshots along the mid-span slice and the spanwise-averaged velocity fields extracted over a near-wake region of $x/L_c \in [-0.5, 6]$ and $y/L_c \in [-2.5, 2.5]$.  Two representative snapshots for the mid-span and spanwise-averaged velocity fields are shown in Figs.~\ref{fig:airfoilLES_3DVis}a and \ref{fig:airfoilLES_3DVis}b.  These snapshots are collected at a constant convective time increment of $\Delta tU_\infty/L_c = 0.0104$, spanning a time window of $tU_\infty/L_c \in [0, 165]$.  This time span is equivalent to approximately $248$ shedding cycles, according to the dominant vortex-shedding frequency $St \equiv fU_\infty /L_c = 1.43$, identified from the lift spectrum. 

For each snapshot, all velocity components ($u_x$, $u_y$, and $u_z$) are provided in both mid-span and spanwise-averaged settings. Due to the use of an unstructured grid, each velocity component is individually formatted as a column vector $\boldsymbol{u} \in \mathbb{R}^{n}$ for each snapshot, where $n = 222,210$ is the number of grid points inside the chosen near-wake region.   Due to the non-uniform unstructured grid, a vector $\boldsymbol{w}\in \mathbb{R}^{n}$ for the cell sizes is also provided to appropriately weight each element in the velocity vectors when performing modal analyses.  %While it is not recommended to form the large data matrix $\boldsymbol{Q} \equiv [\boldsymbol{q}_1, \boldsymbol{q}_2, \dots, \boldsymbol{q}_{n_t}]$, its size is $\mathbb{R}^{3n\times n_t}$, where $\boldsymbol{q}$ is the state vector constructed for each snapshot by stacking $u_x$, $u_y$, and $u_z$ into a single column vector.

\subsubsection{Mean velocity fields}
In addition to the time-resolved snapshots, the time- and spanwise-averaged velocity field is also provided, as shown in Fig.~\ref{fig:airfoilLES_3DVis}c. These mean data can be used as input to linear analyses.

\subsection{Example analyses}

This section describes a range of analysis methods that can be applied to study this airfoil wake configuration. After first describing pertinent flow physics present in this system, we summarize the results of applying a variety of modal analysis, including operator-based decompositions (stability and resolvent analyses, which utilize only mean field data), and data-driven decompositions (POD and DMD). These examples demonstrate the range of methods that can be applied to such data, and how applying numerous methods can give complementary insights into the dynamics of the system.

\subsubsection{Flow physics}

\begin{figure}[t]
\begin{center}
\begin{overpic}[width=6.5in]{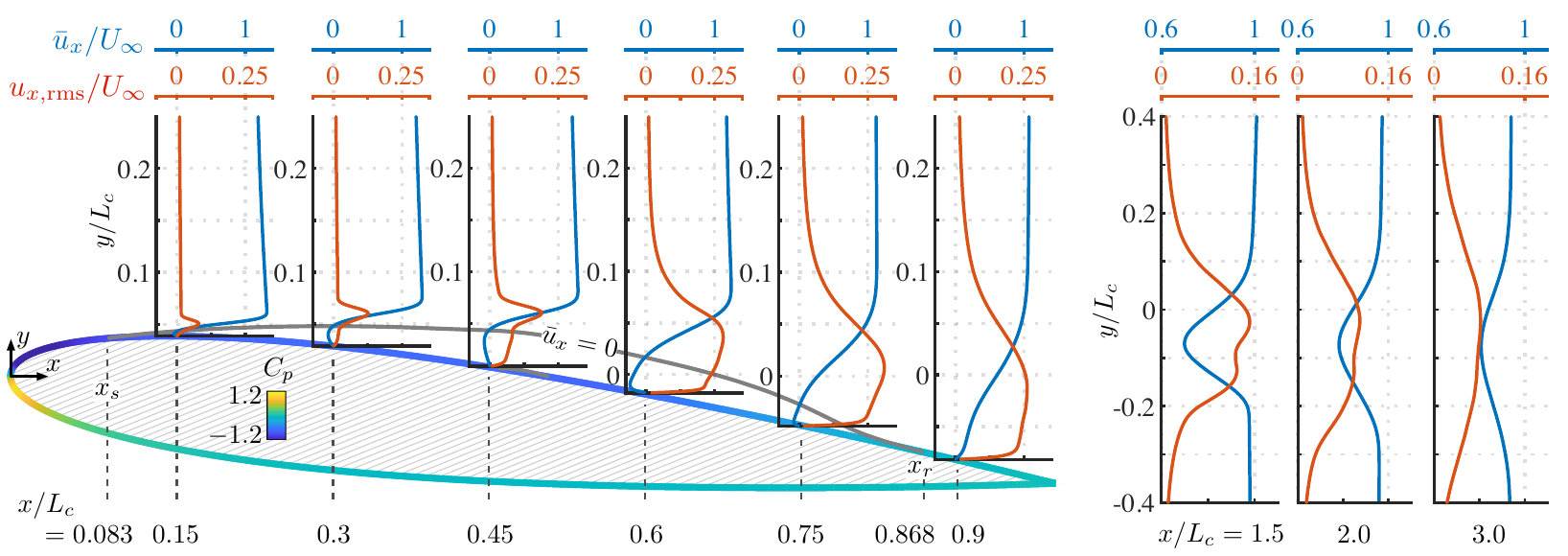}
\end{overpic}
\end{center}
\caption{\label{fig:U_Profiles}Time- and spanwise averaged streamwise velocity profiles and the fluctuation magnitudes over the airfoil suction side and the wake. The contour line for $\bar{u}_x = 0$ is used to identify the location of the separation bubble.  Pressure coefficients are also shown over the surface of the airfoil. }
\end{figure}

Streamwise velocity profiles and the fluctuation magnitudes at a series of streamwise stations over the suction surface are shown in Fig.~\ref{fig:U_Profiles}.  In addition to the velocity profiles, we use the $\bar{u}_x = 0$ contour line to identify the location of the separation bubble.  According to the contour line, laminar separation occurs at $x/L_c = 0.083$ and the flow reattaches at $x/L_c = 0.868$ over the suction surface.  Moving downstream along the shear layer over the separation bubble, the level of velocity fluctuation increases due to the Kelvin-Helmholtz instability, accompanied by the increasing deficit in the streamwise velocity profiles.  The amplified perturbations trigger the roll-up of this shear layer.  The formation of spanwise vortices from the roll-up process is observed near $x/L_c \approx 0.4$, which can be seen in Fig.~\ref{fig:airfoilLES_3DVis}a.   These roller vortices induced by the Kelvin-Helmholtz instability merge near the mid-chord and transition to turbulence.  Turbulent flow is observed over the latter half of the chord.  The merged vortices advect along the suction surface and shed into the wake, followed by vortices shed from the pressure side of the wing.   Moving into the wake, we observe that the streamwise velocity gradually recovers to the freestream level, and the maximum level of fluctuation also decreases along the streamwise direction.  The vortex merging and the von K\'arm\'an vortex shedding, taking place at $St = 1.43$, are identified by the spectral, POD, resolvent, and DMD analyses described in the following subsections as the dominant energetic dynamics of the present turbulent wake flow. 

\subsubsection{Global stability and Fourier spectral analyses}

\begin{figure}[t]
\hspace{0.1in}
\begin{center}
\begin{overpic}[width=1\textwidth]{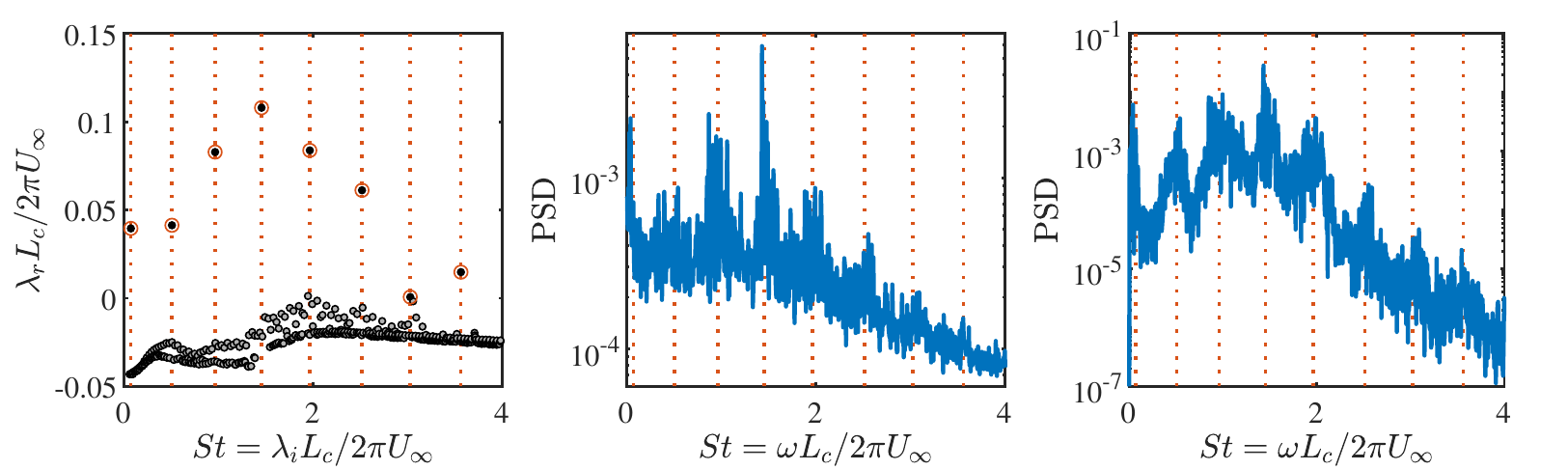}
	\put(09, 29){\small (a)}
	\put(41, 29){\small (b)}
	\put(73, 29){\small (c)}
	
	\put(50, 24){\small $\sum_{i = 1}^{n_x} |\hat{\boldsymbol{u}}(\boldsymbol{x}_i, \omega)|^2$ }
	\put(83, 24){\small $\sum_{i = 1}^{16} |\hat{\boldsymbol{\psi}}_i(\omega)|^2$ }
\end{overpic}
\end{center}
\caption{\label{fig:BiGSpectral}(a) Eigenvalues from the global stability analysis \cite{Yeh:JFM2019}, where the frequencies of the dominant modes are marked with red dashed lines; (b) the PSD summed over all spatial points, $\sum_{i = 1}^{n} |\hat{\boldsymbol{u}}(\boldsymbol{x}_i, \omega)|^2$; (c) the PSD for the Fourier spectra of the POD time coefficients, $\hat{\boldsymbol{\psi}}(\omega)$, summed from modes 1 to 16, i.e., $\sum_{i = 1}^{16} |\hat{\boldsymbol{\psi}}_i(\omega)|^2$.}
\end{figure}
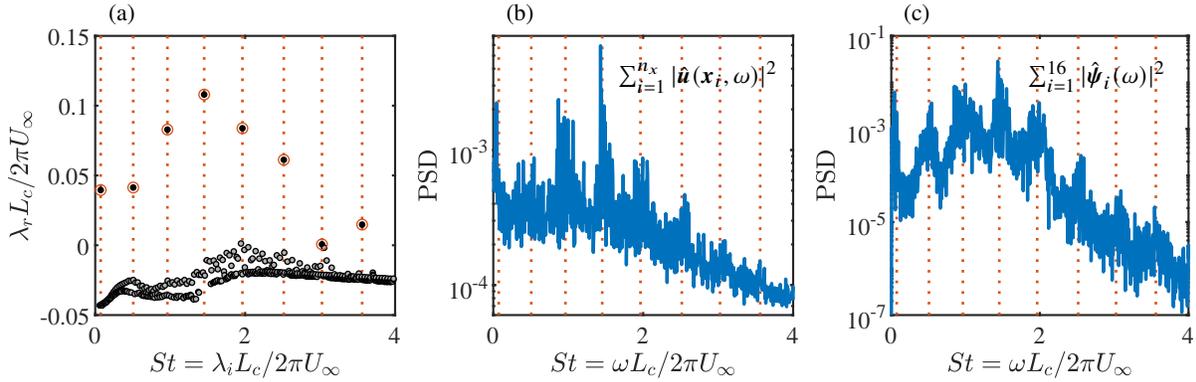

The global stability analysis performs an eigenvalue decomposition $\boldsymbol{L}_{\bar{\boldsymbol{q}}}\boldsymbol{v} = \lambda\boldsymbol{v}$, where $\boldsymbol{L}_{\bar{\boldsymbol{q}}}$ is the linearized Navier-Stokes operator about the time- and spanwise-averaged flow $\bar{\boldsymbol{q}}$, as shown in Fig.~\ref{fig:airfoilLES_3DVis}c.  Here, we consider the bi-global linearized Navier-Stokes operator where a spatial Fourier expansion is performed in the homogeneous spanwise direction, such that the mean-flow-based linear operator $\boldsymbol{L}_{\bar{\boldsymbol{q}}}$ depends on the spanwise wavenumber, i.e. $\boldsymbol{L}_{\bar{\boldsymbol{q}}} = \boldsymbol{L}_{\bar{\boldsymbol{q}}}(k_z)$.  Fig.~\ref{fig:BiGSpectral}a shows the eigenvalue spectrum of the linearized Navier-Stokes operator $\boldsymbol{L}_{\bar{\boldsymbol{q}}}$ with $k_z = 0$.   From the spectrum, we identify a number of eigenmodes with dominant growth rates and mark their frequencies with red dashed lines.  Based on the structures of the eigenvectors, \citet{Yeh:JFM2019} found that the eigenmodes of frequencies lower than $St = 2$ are associated with the von K\'arm\'an vortex shedding.  Increasing the frequency above $St > 2$, the shear-layer structures associated with the Kelvin-Helmholtz instability become dominant.   In Fig.~\ref{fig:BiGSpectral}b, we show the PSD summed over all spatial points.  We observe that the peak frequencies align well with those of the dominant eigenmodes revealed by the global stability analysis.  The eigenmode with the dominant growth rate at $St = 1.43$ also agrees with the frequency of the highest amplitude from the spectral analysis in Fig.~\ref{fig:BiGSpectral}b.  This analysis is an example of a global stability analysis performed on a mean turbulent flow which is not a solution to the governing equation.  The objective here is to uncover the structures and frequencies for large-scale coherent structures but not the stability characteristics of the base flow \cite{Sipp:JFM2007, Beneddine:JFM2016, Edstrand:JFM2016, Sun:TCFD2017}.

\subsubsection{Resolvent analysis}
\begin{figure}[t]
\begin{center}
\begin{overpic}[width=6.5in]{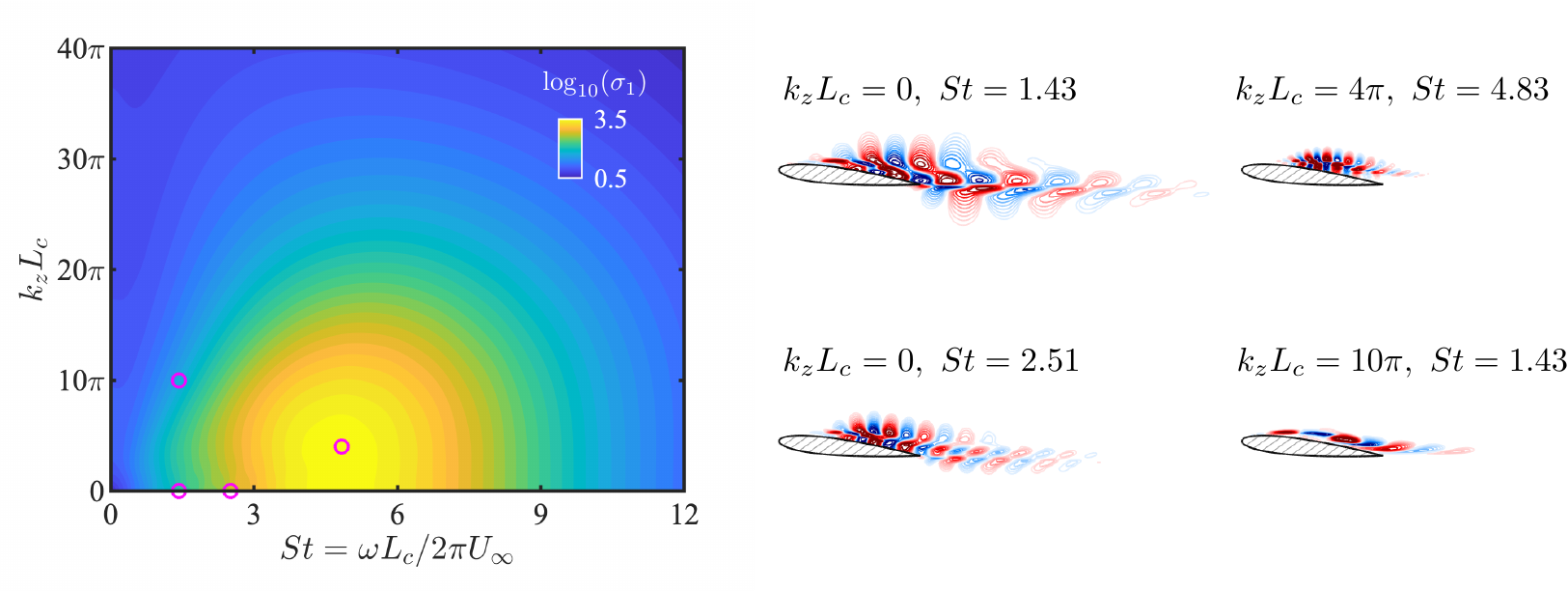}
	\put(01, 36){(a)}
	\put(48, 36){(b)}
	\put(48, 20){(c)}
	\put(77, 36){(d)}
	\put(77, 20){(e)}
\end{overpic}
\end{center}
\caption{\label{fig:Resolvent}Resolvent analysis of the turbulent airfoil wake: (a) Optimal amplification (gain) as a function of frequency and spanwise wavenumber \cite{Yeh:JFM2019}; (b-e) representative resolvent response modes visualized by the $\hat{u}_x$-components.  These representative modes are highlighted by magenta circles in (a).  Note that at $[k_zL_c, St] = [4\pi, 4.83]$, the gain reaches the global maximum over the $St$-$k_z$ plane.  }
\end{figure}
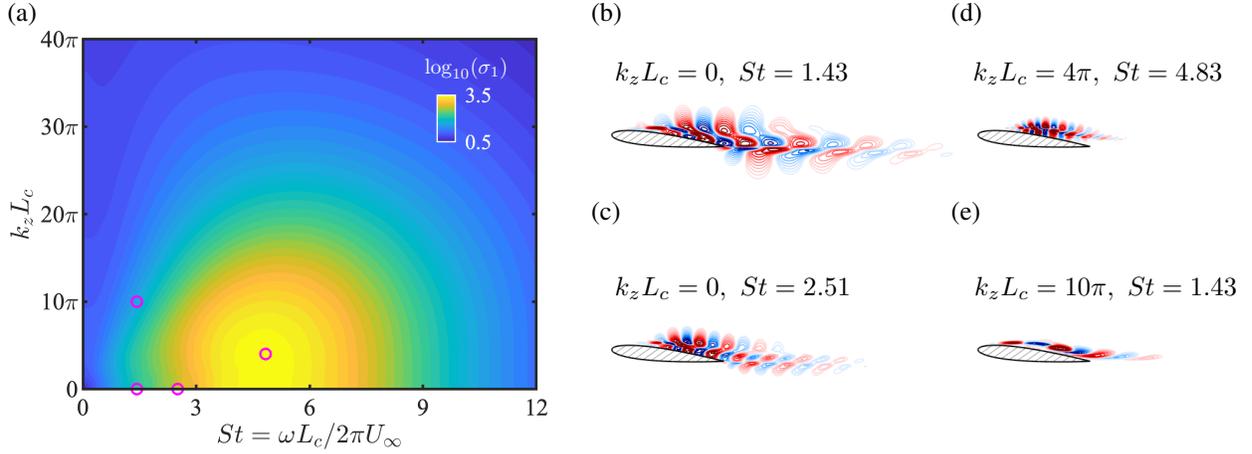

Using the mean-flow-based bi-global linear operator $\boldsymbol{L}_{\bar{\boldsymbol{q}}}(k_z)$, resolvent analysis, or pseudospectral analysis \cite{TrefethenEmbree:2005}, performs a singular value decomposition of the resolvent operator, $\boldsymbol{R}(\omega, k_z) \equiv [i\omega\boldsymbol{I} - \boldsymbol{L}_{\bar{\boldsymbol{q}}}(k_z)]^{-1}$ for a prescribed frequency-wavenumber pair.  In particular, we consider the use of exponential discounting \cite{Jovanovic:Thesis2004,Yeh:JFM2019,Yeh:PRF2020} in the analysis, since $\boldsymbol{L}_{\bar{\boldsymbol{q}}}$ has eigenvalues with positive real parts, as shown in Fig.~\ref{fig:BiGSpectral}a.  The resolvent gain (leading singular values) over the $k_z$-$\omega$ plane and four representative response modes (left singular vectors) are shown in Fig.~\ref{fig:Resolvent}.   We observe that the maximum resolvent gain appears in the regions near $St \approx 4.8$ at low spanwise wavenumbers.  Taking advantage of this $St$-$k_z$ range of high energy amplification for active flow control, \citet{Yeh:JFM2019} showed that the resolvent-analysis informed actuation is capable of suppressing flow separation and delivers the highest level of aerodynamic performance enhancement.  The response modes over this $St$-$k_z$ range, as exemplified by the mode shown in Fig.~\ref{fig:Resolvent}d for the location where the gain reaches the global maximum, also suggest that the actuation can trigger the Kelvin-Helmholtz instability over the separation bubble.  While the intrinsic shear-layer frequency in the baseline flow is near $St = 2.5$, the resolvent analysis shows that higher amplification can be achieved near the $St \approx 4.8$.  This example shows the capability of resolvent analysis in providing insights into the optimal frequency range for forcing amplification, which may not be generally accessible by targeting the baseline shear-layer frequencies. 

\subsubsection{Proper orthogonal decomposition}

POD analysis is performed with 16,000 snapshots of both mid-span and spanwise-averaged velocity fields. 
While POD can be formulated as the singular value decomposition of the (cell-volume weighted) data matrix $ \boldsymbol{Q}= \boldsymbol{\Phi}\boldsymbol{\Sigma}\boldsymbol{\Psi}^T$, for large datasets the method of snapshots \cite{sirovich1987turbulence} is more computationally tractable. In this method, the eigendecomposition of the snapshot covariance matrix $\boldsymbol{C} \in \mathbb{R}^{n_t \times n_t}$ is found, where each entry $c_{ij}$ in $\boldsymbol{C}$ is the (appropriately-weighted) inner product between the state vectors from snapshots $i$ and $j$. The eigenvalues $\boldsymbol{\Lambda} = \boldsymbol{\Sigma}^2$ of $\boldsymbol{C}$ correspond to modal energies, and the eigenvectors $\boldsymbol{\Psi}$ correspond to the time-varying coefficients of the POD modes. The POD modes themselves can be recovered via $\boldsymbol{\Phi} = \boldsymbol{Q}\boldsymbol{\Psi}\boldsymbol{\Sigma}^{-1}$, which can be computed while loading only one snapshot at a time.  

%It considers the singular value decomposition of the cell-volume weighted data matrix, $\boldsymbol{Q}_{\boldsymbol{W}} = \boldsymbol{\Phi}_{\boldsymbol{W}}\boldsymbol{\Sigma}\boldsymbol{\Psi}^T$, where $\boldsymbol{\Phi}_{\boldsymbol{W}} = [\boldsymbol{\phi}_1, \boldsymbol{\phi}_2, \dots, \boldsymbol{\phi}_{n_t}]$ are the weighted POD modes and $\boldsymbol{\Psi} = [\boldsymbol{\psi}_1, \boldsymbol{\psi}_2, \dots, \boldsymbol{\psi}_{n_t}]$ are the time coefficients.  Due to the large data size, the POD is initiated by constructing of covariance matrix $\boldsymbol{C} \in \mathbb{R}^{n_t \times n_t}$, where each entry $c_{ij}$ in $\boldsymbol{C}$ is the weighted inner product between the state vectors from snapshot $i$ and $j$, $c_{ij} = \boldsymbol{q}_i^T \boldsymbol{W} \boldsymbol{q}_j$, where $\boldsymbol{W} = \text{diag}(\boldsymbol{w}, \boldsymbol{w}, \boldsymbol{w})$.  With the covariance matrix $\boldsymbol{C}$, the modal energies $\boldsymbol{\Lambda} = \boldsymbol{\Sigma}^2$ and time coefficients $\boldsymbol{\Psi}$ are found from the eigenvalue decomposition $\boldsymbol{C}\boldsymbol{\Psi} = \boldsymbol{\Lambda}\boldsymbol{\Psi}$, and POD modes can be recovered via $\boldsymbol{\Phi} = \boldsymbol{Q}\boldsymbol{\Psi}\boldsymbol{\Sigma}^{-1}$ by loading one snapshot at a time.  

\begin{figure}[t]
\begin{center}
\begin{overpic}[width=3in]{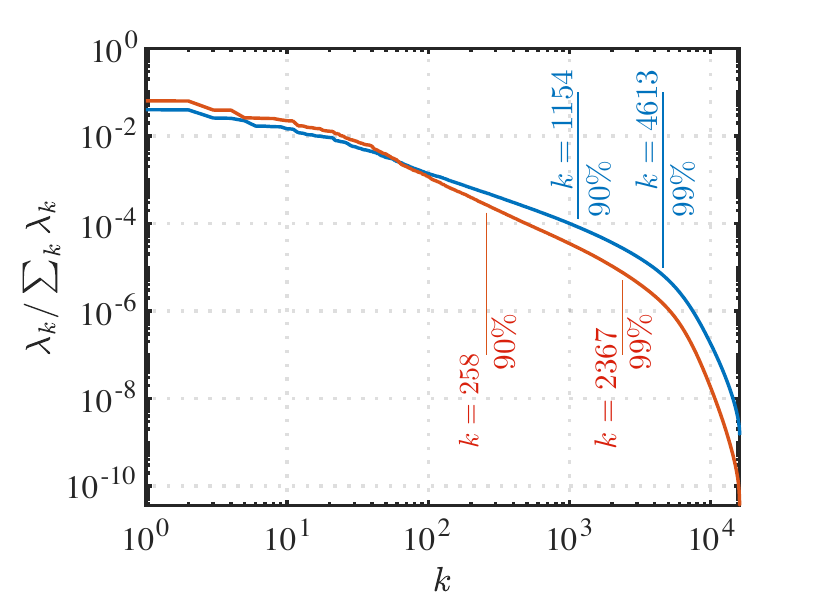}
	\put(30, 40){\definecolor{red2}{rgb}{0.8500, 0.1250, 0.0480}\footnotesize \color{red2} spanwise-averaged}
	\put(43, 62){\definecolor{blue2}{rgb}{0, 0.4470, 0.7410}\footnotesize \color{blue2} mid-span slice}
\end{overpic}
\end{center}
\caption{\label{fig:POD_Energy}POD modal energy distributions for mid-span slice and spanwise-averaged velocity fields. }
\end{figure}

The distributions of modal energy $\lambda_k$ obtained from both the mid-span slice and spanwise-averaged velocity fields are shown in Fig.~\ref{fig:POD_Energy}.  The first POD modes of the mid-span and spanwise averaged data contain $3.98\%$ and $6.33\%$ of the total energy, respectively.  The spanwise-averaged POD modes capture $90\%$ of the total energy at $r = 258$, which is only $1.6\%$ of the data size.  Compared to the spanwise-averaged case, the mid-span case requires $1154$ modes to capture $90\%$ of the total energy.  The modal energies for rank $k > 40$ from the spanwise-averaged case show a higher roll-off rate compared to that from the mid-span data.  In fact, for both cases the modal structures for $k > 40$ exhibit only small-scale turbulent fluctuations, which are expected to be averaged out over the spanwise direction.  

\begin{figure}[t]
\begin{center}
\vspace{0.1in}
\begin{overpic}[trim=0.2in 0 0.2in 0, clip, width=1.0\textwidth]{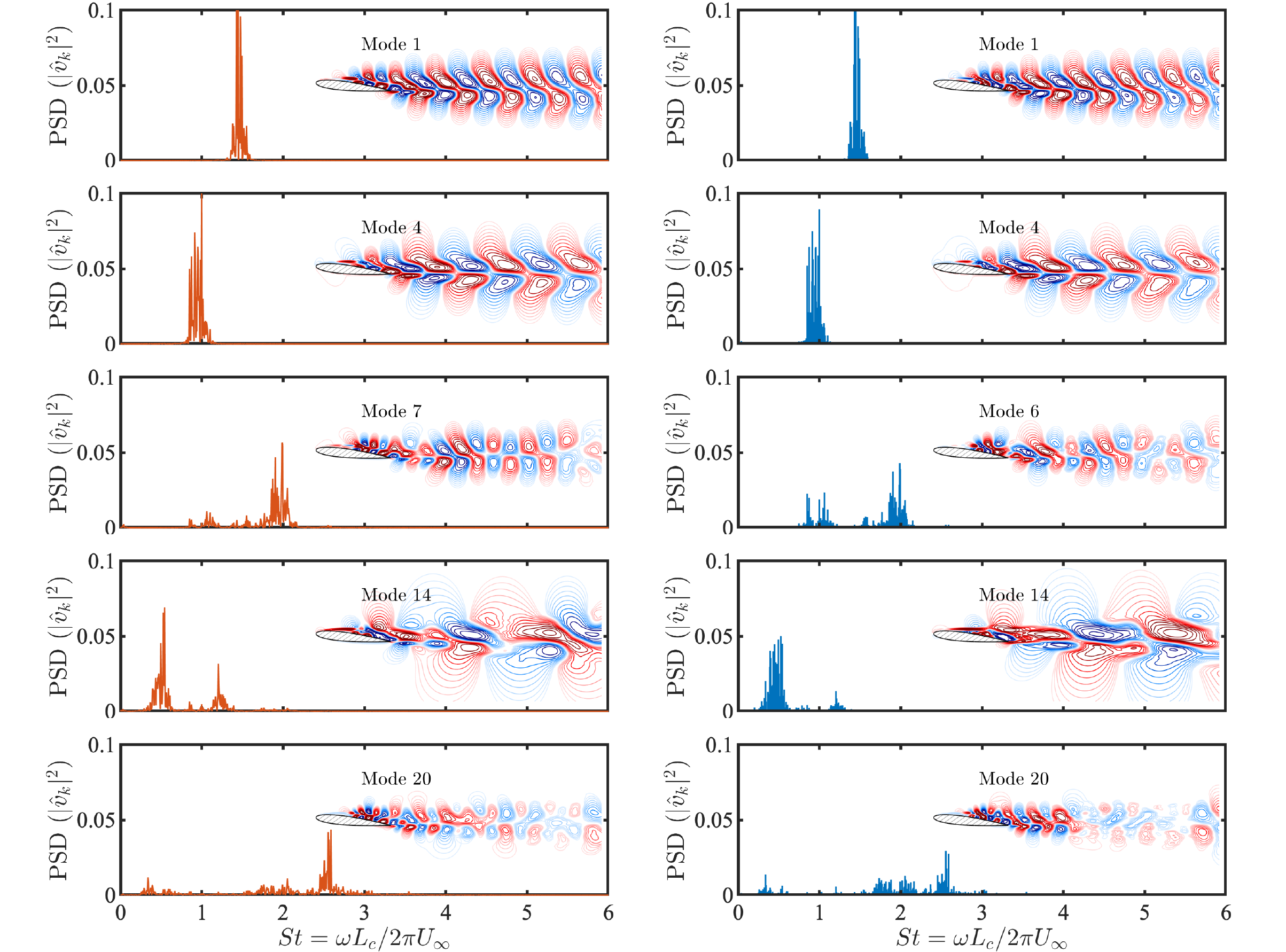}
	\put(20, 78){\definecolor{red2}{rgb}{0.8500, 0.1250, 0.0480}\small \color{red2} spanwise-averaged}
	\put(70, 78){\definecolor{blue2}{rgb}{0, 0.4470, 0.7410}\small \color{blue2} mid-span slice}
\end{overpic}
\end{center}
\caption{\label{fig:POD_Modes_Spectra} Modal structures and spectral contents of POD modes.  Showing for both spanwise-averaged (left column) and mid-span (right) cases.  POD modes are visualized by the $\hat{u}_x$ components. }
\end{figure}

The frequency contents for the POD time coefficients $\boldsymbol{\psi}$ are also analyzed in Fig.~\ref{fig:BiGSpectral}c and Fig.~\ref{fig:POD_Modes_Spectra}.  The frequency contents of the first 16 modes capture the dominant peak frequencies from the global stability and the Fourier analyses in Fig.~\ref{fig:BiGSpectral}c.  In Fig.~\ref{fig:POD_Modes_Spectra} we show representative POD modes and the frequency spectra of their time coefficients for both spanwise-averaged and mid-span cases.  We observed that the ordering of the modes according to their spectral contents as well as the modal structures are well aligned.  This suggests the dynamics of the present turbulent wake flow are energetically dominated by two-dimensional spanwise structures (at spanwise wavenumber $k_z = 0$.)  For both cases, mode 1 shows dominant frequency at $St = 1.43$.  The modal structure exhibits the development of Kelvin-Helmholtz instability over the separation bubble and the latter half of the chord.  The wake is receptive to these perturbations at $St = 1.43$, allowing for the structures to continue advecting into the wake.  The first shear-layer dominated mode appears in the third conjugate pair for both cases, exhibiting a dominant frequency near $St \approx 2$.  Mode 14 is the first mode that shows a strong subharmonic coherence near $St \approx 0.5$.  Mode 20 shows a higher shear-layer frequency near $St \approx 2.5$.  In this frequency range, the small Kelvin-Helmholtz structures are only observed over the airfoil.  Beyond one chord downstream of the trailing edge, the wake no longer shows a coherent pattern for $St \approx 2.5$.  We also note all of these dominant frequencies are observed in the spectral analyses presented in Fig.~\ref{fig:BiGSpectral}.

\subsubsection{Dynamic mode decomposition}
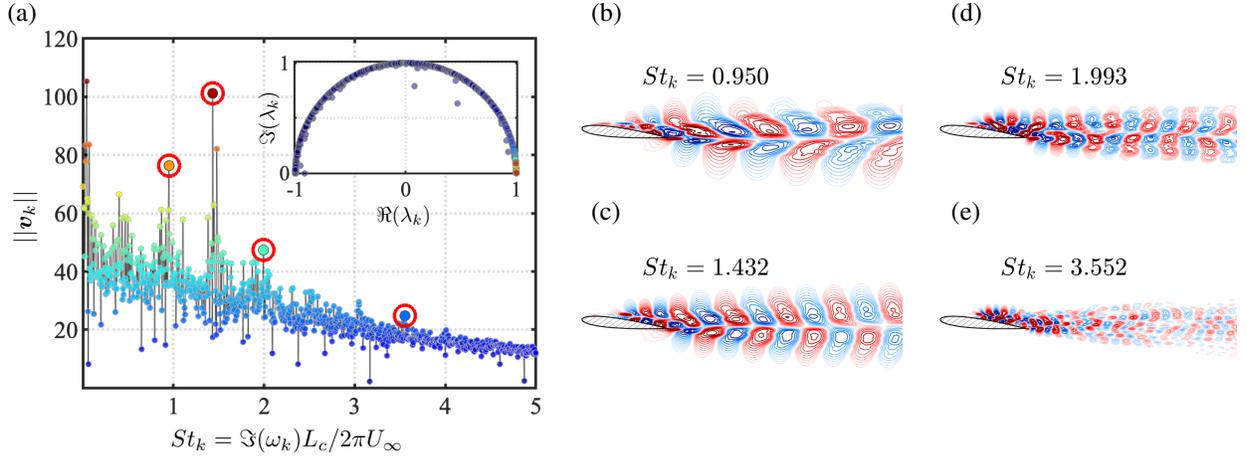
\begin{figure}[t]
\begin{center}
\begin{overpic}[width=6.5in]{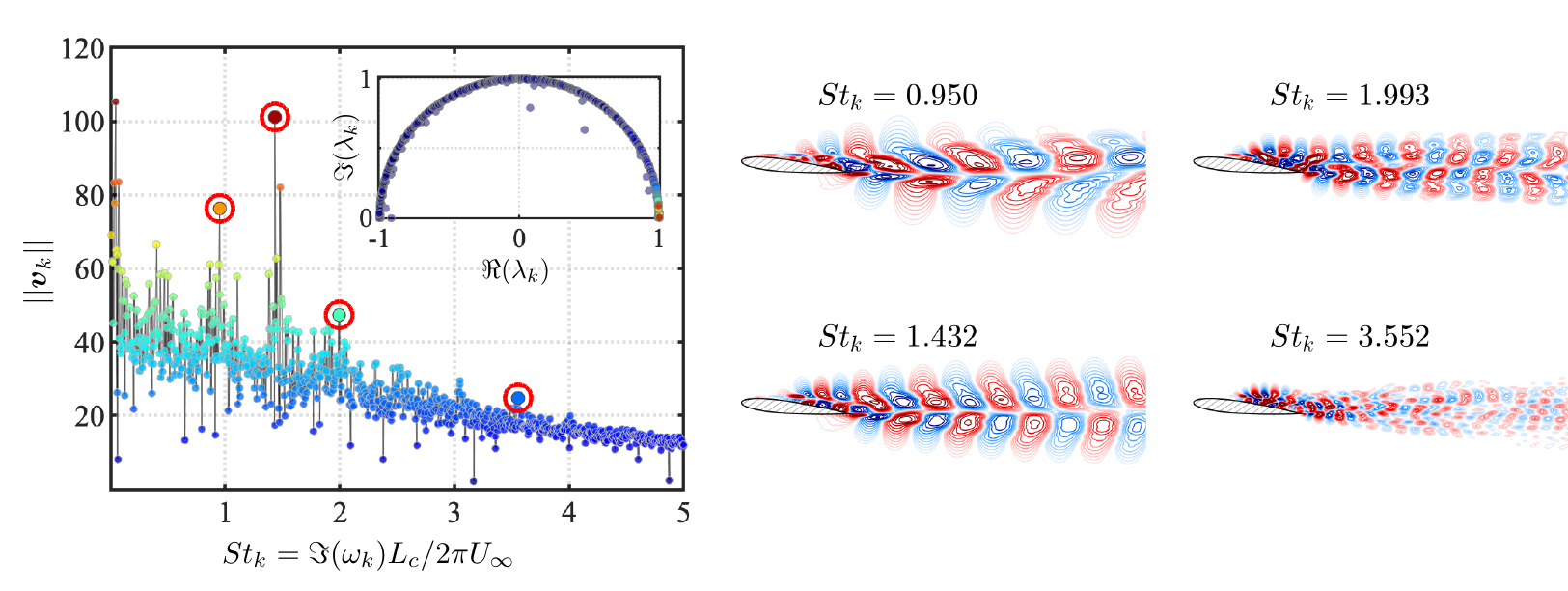}
	\put(01, 36){(a)}
	\put(48, 36){(b)}
	\put(48, 20){(c)}
	\put(77, 36){(d)}
	\put(77, 20){(e)}
\end{overpic}
\end{center}
\caption{\label{fig:DMD}Dynamic mode decomposition for snapshots collected over the mid-span slice. (a) DMD modal amplitudes and the Ritz values with positive frequencies; (b-e) Representative DMD modes ($\boldsymbol{v}_k$) visualized by the $\hat{u}_x$-component at four peak frequencies highlighted by red circles in (a).  The DMD modal frequencies are computed using $\Im(\omega_k)$, where $\omega_k = \log(\lambda_k)/\Delta t$ and $\lambda_k$ are the eigenvalues of the companion matrix $\boldsymbol{S}$.  }
\end{figure}

Dynamic mode decomposition (DMD) is performed for the dataset over the mid-span slice.  As commented in \citet{Schmid:JFM2010}, a dataset extracted from such a subdomain can be considered as projecting the three-dimensional flow field onto a two-dimensional space.  The modal structures can be influenced by such a projection, but the temporal dynamics (frequencies) of the full three-dimensional flow can still be captured by processing the two-dimensional slices of the flow field.  Here, the DMD analysis considers all 16,000 snapshots and adopts the algorithm presented in \citet{Rowley:JFM2009DMD} where the eigenvalue decomposition is performed for the full companion matrix $\boldsymbol{S} \in \mathbb{R}^{(n_t-1) \times (n_t-1)}$.  While the present approach appears to uncover physical DMD modes at high amplitudes, we note that DMD exhibits better numerical stability by projecting the discrete-time linear operator onto an orthogonal basis of lower dimension, such as the POD modes \cite{Schmid:JFM2010, Tu:JCD2014}.  In Fig.~\ref{fig:DMD}, we show the amplitudes and modal structures obtained from the DMD analysis.  The peak frequency appears at $St = 1.43$, agreeing with the linear stability and Fourier spectral analyses.  The structure of this DMD mode exhibits strong coupling between the shear-layer and wake instabilities at this frequency, as opposed to the modal structures for $St_k = 1.99$ and $3.55$ where the fluctuations predominantly present along the shear layer over the separation bubble.

%%%%%%%%%%%%%%%%%%%%%%%%%%%%%%%%%%%%%%%%%%%%%%%%%%%%%%%%%%%
% -- Conclusions ------------------------------------------

\section{Conclusions}
\label{sec:conclusions}

We have presented a curated database designed to aid in the conception, training, demonstration, evaluation, and comparison of reduced-complexity models for fluid mechanics.  The data are publicly accessible via a web browser interface or Globus at \url{ http://deepblue.lib.umich.edu/data/collections/kk91fk98z}.  The database contains six distinct datasets selected to span a wide range of flow conditions and data types.  

The first dataset (\S\ref{sec:jet}) comprises a LES of a turbulent jet at Mach number 0.9.  The dataset contains 10,000 time-resolved snapshots of three-dimensional velocity, density, and pressure fields spanning 2000 acoustic time units.  Also included are preprocessed azimuthal Fourier modes for each snapshot and the mean flow.  The jet offers an example of a canonical, statistically stationary turbulent free-shear flow containing both broadband (in the turbulent plume) and tonal (in the potential core) spectral content.  The use of these data was exemplified via comparisons between SPOD and resolvent analysis and validation of a linear-stability-based theory of acoustic waves trapped in the jet core.

The second dataset (\S\ref{sec:BL_DNS}) comprises a pair of DNS calculations of a turbulent zero-pressure-gradient flat-plate boundary layer spanning $292 \leq Re_{\tau} \leq 729$ and $481 \leq Re_{\tau} \leq 1024$.  The dataset contains roughly 10,000 time-resolved snapshots of the three-dimensional velocity and pressure fields spanning more than 20 eddy-turnover times for the first case and 100 snapshots for the second case.  Also included are preprocessed $x$-$t$ and $x$-$z$ correlations at several wall-normal distances, mean and root-mean-squared velocity and vorticity profiles, various boundary-layer metrics such as the mean friction coefficient along the plate, and planar velocity data that can be accessed without downloading the full volumetric data.  The boundary layer offers an example of a canonical, statistically stationary wall-bounded turbulent flow dominated by broadband spectral content, computed numerically.  As an example, the data were used to demonstrate a novel causality analysis, which confirmed that information flows predominantly from the outer-layer large-scale motions to inner-layer small-scale motions.

The third dataset (\S\ref{sec:BL_EXP}) comprises experimental (PIV) measurements of a similar zero-pressure-gradient flat-plate boundary layer at five different Reynolds numbers in the range  $ 605 \leq Re_{\tau} \leq 2227$.  For each Reynolds number, the dataset contains 6000 snapshots of planar ($x$-$y$) velocity fields.  This dataset provides an experimental counterpart to the previous DNS data, and the overlap in Reynolds numbers could be leveraged to assess the impact of realistic experimental uncertainties and limited measurements (planar vs. volumetric) on reduced-complexity model methods.  The data were used to demonstrate a new conditional projection averaging method applied to SPOD modes. 

The fourth dataset (\S\ref{sec:airfoil_DNS}) comprises direct numerical simulations of two-dimensional stationary and pitching flat-plate airfoils at Reynolds number 100.  The dataset contains time-resolved snapshots of the velocity field, lift and drag coefficients, and airfoil kinematics spanning 40 or 100 convective time units, depending on the case.  Cases include a stationary airfoil and eight different pitching frequencies.  The pitching airfoils offer examples of application-oriented, laminar, transient flow exhibiting tonal spectral content.  Three analyses were used to demonstrate the utility of these data for reduced-complexity modeling: a comparison of DMD and SPOD modes, construction of a dynamic data-driven model of the forces on the airfoil, and flow-field reconstruction and parameter estimation using machine learning.

The fifth dataset (\S\ref{sec:airfoil_EXP}) comprises experimental measurements of a flat-plate airfoil passing through a large-amplitude transverse gust.  The dataset is generated from an ensemble of the airfoil-gust encounters, to account for experimental uncertainty and variability in the gust profile. Time-resolved force data are provided from each realization, while time-resolved  planar PIV velocity fields are provided for  ensemble-averaged gust encounters.  This flow offers an example of application-oriented, laminar, transient flow probed via experimental measurements.  These data were used to compare two methods for deducing the strength of the leading-edge vortex and to evaluate an analytical model for the time-varying lift coefficient.   

The sixth dataset (\S\ref{sec:airfoil_LES}) comprises a three-dimensional LES of Mach 0.3 flow over a NACA 0012 airfoil at Reynolds number 23,000, which features a transitional boundary layer, separation over a recirculation bubble, and a turbulent wake.  The dataset contains 16,000 time-resolved snapshots of the mid-span and spanwise-averaged velocity fields.  The airfoil wake offers an example of an application-oriented, statistically stationary turbulent flow with broadband spectral content.  These data were used to perform a range of operator-based and data-driven modal decomposition methods to elucidate complementary information about the underlying flow physics.

This database is uniquely tailored for reduced-complexity modeling.  It spans a wide range of flow conditions, e.g., laminar and turbulent, transient and statistically stationary, and includes both simulation and experimental data.  Both lengthy series of time-resolved data needed for data-driven models and mean data required for linear physics-based methods are included.  

We envision that this database will aid the fluids community by providing a common set of test cases for developing and testing reduced-complexity models.  If widely adopted, the use of common test cases will help users directly compare new and existing methods and understand their generalizability and applicability to different types of flows and data, beyond the overly simplified cases like low-Reynolds-number cylinder flow and Lorenz equations frequently used in the contemporary literature.  We view this effort as an extension of other recent efforts to lower the barrier of entry into the field of reduced-complexity modeling. 

As described above, we have included examples of how the database can support the exploration of reduced-complexity models.  We showed how the data can be used to evaluate and demonstrate new methods, e.g., the causality analysis and the conditional phase averaging applied to the DNS and experimental boundary layer data, respectively.  Several examples showed how the data can be used to compare methods, e.g., the comparisons between SPOD and resolvent analysis for the jet and between DMD and SPOD for the laminar airfoil.  Other examples showed how the data can be used as a point of comparison to validate analytical and/or linearized models, e.g., the linear theory of the trapped acoustic waves in the jet and the lift model for the airfoil-gust encounters. 

Realizing the full potential of the database toward advances in modeling capabilities and understanding will require ongoing contributions from the fluids and reduced-complexity modeling communities.  While we included in this paper a few nominal examples of comparisons between methods, much work remains in establishing best practices for selecting appropriate methods for a given dataset.  Additionally, with a few exceptions, we have focused on traditional analytical, modal decomposition, and linear stability methods.  Systematic comparisons between these methods and alternative machine learning methods, in terms of training requirements, cost, accuracy, generalizability, etc., are urgently needed.  The database has been designed with these and other challenges in view, and it is our hope that it will contribute positively to future advances in the field of reduced-complexity modeling for fluid dynamics.

%%%%%%%%%%%%%%%%%%%%%%%%%%%%%%%%%%%%%%%%%%%%%%%%%%%%%%%%%%%
% -- Back-matter ------------------------------------------

% Funding
\section*{Funding Sources}

The jet LES study was supported in part by a NAVAIR SBIR project, under the supervision of Dr. John T. Spyropoulos. The main calculations were carried out on CRAY XE6 machines at DoD supercomputer facilities in ERDC DSRC.
The boundary layer simulations and causality analysis were supported by the National Science Foundation under Grant No. 2140775.  The authors acknowledge the MIT SuperCloud and Lincoln Laboratory Supercomputing Center for providing HPC resources that have contributed to the research results reported in this paper.
The experimental boundary layer work was made possible by the support of the National Science Foundation Graduate Research Fellowship Program and a National Science Foundation Grant No. 2118209. The support of the Department of Aerospace Engineering and the Grainger College of Engineering at the University of Illinois Urbana-Champaign is also gratefully acknowledged.  
The airfoil DNS simulations were supported in part by the U.S. Air Force Office of Scientific Research under grant FA9550-14-1-028. 
The airfoil gust experiments were supported in part by the U.S. Air Force Office of Scientific Research under grant FA9550-16-1-0508 and the National Science Foundation under grant no.~1553970.
The airfoil LES simulations were supported by the Office of Naval Research under grant number N00014-19-1-2460 and the Department of Defense High Performance Computing Modernization Program.

% Acknowledgments
\section*{Acknowledgments}

The authors gratefully acknowledge the following contributors: Peter Jordan and Tim Colonius for their role in generating the jet data; Elizabeth Torres De Jesus for her efforts in collecting the experimental boundary layer data; and Girguis Sedky for his work on the circulation measurements for the airfoil gust experiments. 

% Bibliography
\bibliography{dataset} % Please name your bib file DATASET.bib, where DATASET = 'jet', etc.  

\end{document}